\documentclass{nature}
\usepackage{graphicx}
\usepackage{grffile}
\usepackage{bm}
\usepackage{amsmath}
\usepackage{color}
\usepackage{float}
\usepackage{wasysym}
\usepackage{amssymb}
\usepackage{float}
\usepackage{dsfont}
\usepackage{bbold}
\usepackage{cancel}
\usepackage{booktabs}
\usepackage{cleveref}
\usepackage{indentfirst}
\usepackage{pdfpages}

\makeatletter
\let\saved@includegraphics\includegraphics
\AtBeginDocument{\let\includegraphics\saved@includegraphics}
\renewenvironment*{figure}{\@float{figure}}{\end@float}
\makeatother
\newcommand{\argmax}[1]{\underset{#1}{\operatorname{arg}\,\operatorname{min}}\;}

\newcommand{\Ac}{\mathcal{A}}
\newcommand{\Cc}{\mathcal{C}}

\def\bal#1\eal{\begin{align*}#1\end{align*}}
\title{Supplementary Information: Extracting truth from science}

\author{Alexander V. Belikov$^1$, Andrey Rzhetsky$^{2, 3}$ \& James Evans$^1$}

\begin{document}



Detecting signal from science:
The structure of research communities and prior knowledge improves prediction of genetic regulatory experiments

Alexander V. Belikov[1,2]
Andrey Rzhetsky[3,4]
James Evans[1,5]

**The explosive growth of scientists, scientific journals, articles and findings in recent years [1,2] exponentially increases the difficulty scientists face in navigating prior knowledge and collectively reasoning over it to drive future advance [3,4]. This challenge is exacerbated by uncertainty about the reproducibility of published findings [5–8]. The availability of massive digital archives, machine reading and extraction tools on the one hand, and automated high-throughput experiments on the other, allow us to evaluate these challenges at scale and identify novel opportunities for accelerating scientific advance [9]. Here we demonstrate a Bayesian calculus that enables the positive prediction of robust, replicable scientific claims with findings automatically extracted from published literature on gene interactions. We matched these findings, filtered by science, with unfiltered gene interactions measured by the massive LINCS L1000 high-throughput experiment to identify and counteract sources of bias. Our calculus is built on easily extracted publication meta-data regarding the position of a scientific claim within the web of prior knowledge, and its breadth of support across institutions, authors and communities, revealing that scientifically focused but socially and institutionally independent research activity is most likely to replicate. This contrasts with the ineffectiveness of alternative strategies like "follow the leader"—trusting top journals and top scientists—which do not predict robust findings. These findings recommend policies that go against the common practice of channeling biomedical research funding into centralized research consortia and institutes rather than dispersing it more broadly. Our results demonstrate that robust scientific findings hinge upon a delicate balance of shared focus and independence, and that this complex pattern can be computationally exploited to decode bias and predict the replicability of published findings. These insights provide guidance for scientists navigating the research literature and for science funders seeking to improve it. Moreover, our project models an entirely machine-driven research pipeline, from machine reading to evaluation, that could be incorporated by intelligent algorithms to augment scientific search, recommending fruitful research hypotheses to accelerate cumulative scientific advance.**

[1] Knowledge Lab & Department of Sociology, University of Chicago; [2]Hello Watt, Paris; [3]Departments of Medicine and Human Genetics, University of Chicago; [4]Institute for Genomic and Systems Biology, University of Chicago; [5]Santa Fe Institute. Correspondence should be addressed to belikov@uchicago.edu or jevans@uchicago.edu.

# Main

Millions of research papers are published globally each year[3,4], with nearly a million biomedical articles published and indexed in MEDLINE, and more than a hundred thousand that discuss genes and their biological interactions. Popular individual genetic pathways receive attention from thousands of papers. This deluge of information makes it very difficult for researchers and other audiences of science to decide what to trust—what insights will replicate and generalize beyond the particular experiment or observation from which they were initially demonstrated. More certainty about what findings to trust will allow us to better select useful signals from science that advance science and accelerate technology. Knowing what findings are robust will help scientists decide what to study next, policy-makers and philanthropists what to fund next, and institutes and technologists what next to develop as life-saving medical diagnostics and therapies. In this investigation, we ask how to predict certain signals from scientific publications. We further consider the implications of our predictions for how scientific institutions might be reformed to improve those signals. Subjective bias is an inevitable reality of published science. Accurate and robust insight about nature is only one of several factors scientists consider as they undertake research and publish findings. They must also strategically think about scientific influence and academic survival. What would be theoretically significant, what will attract attention and what can inspire scientists to build on their work in future? What will journal editors and reviewers allow[10]? What will patrons fund? What will promotion committees accept? Beyond complex motivations, scientists, their experiments and observations are situated in particular positions with respect to their objects of study, which defy detached and universal notions of objectivity[11] and necessarily shape their assessments. Scientists foster expectations inculcated from disciplinary education and prior experience[12], and they rationally incorporate the beliefs of those they trust—respected mentors and colleagues—into their own scientific expectations and certainties[6]. These contextual forces add noise to the signal about what findings will replicate and generalize, above and beyond the complexity involved in deciphering experimental and observational protocols from ambiguous language.

Above the level of the scientist, the modern scientific system promulgates predictable bias. Competition between journals makes it far easier to publish positive findings than neutral or 'negative' ones[1,2], which become underrepresented in the published record[5]. Moreover, favorable conditions for the "wisdom of crowds" phenomenon[13], where collectives produce systematically more accurate estimates than individuals, are widely violated in science. Crowds are wise when their members have access to independent data[14] or utilize independent methods[15] to derive their answers, but they falter when engaged in centralized communication[16,17] and share prior experience, knowledge, and methods[9]. By contrast, the modern life sciences are characterized by intensive and repeated collaboration[18–20], increasingly large[21,22] and distributed teams[23], star scientists[24,25], canonical citations[26,27] and expensive shared equipment[18,20].

These forces have led to widespread concerns regarding the reliability and reproducibility of findings in fields ranging from pharmacology[7,28,29 5] and genetics[5,30–32 3,4] to psychology[33,34] with widespread implications for the accumulation of certainty in science. Some have even feared that distortions associated with publication and confirmation bias could lead to the canonization of false facts[10]. This is a problem for scientists, but also science as a system of insights on which future innovation relies. Prior work has attempted to identify the robust replicability of the scientific literature and identify sources of

distortion that might weaken the signal of science through simulation[10,35,36], experiments[33,37], and meta-analysis of prior results[9,38]. Psychologists have called efforts to replicate the robust essence of an experiment with an alternative research design and methods a "conceptual replication" [39]. Here we build on that work by demonstrating an automated research pipeline that (1) extracts gene-gene interaction claims from the biomedical literature, then (2) aligns them for conceptual replication with results from the massively replicated LINCS L1000 high-throughput experiment (Fig. 1), (3) identifies factors that increase and decrease the likelihood of replication and (4) updates our understanding and scientific certainty through a Bayesian calculus (Fig. 2). Finally, we use the results of our investigation (Fig. 3) to identify opportunities that could redesign scientific institutions and investigations for accelerated advance (Fig. 4).

A number of scientific and social factors could influence the likelihood that a claim is robust and generalizes. Two important classes of factors involve (1) how a claim fits into prior knowledge about nature and (2) its breadth of prior support. We may investigate how a claim fits with preexisting knowledge by assessing its position in the complex network of other scientific claims. A claim may be central or peripheral in the network, and entire claim networks may be decentralized or hierarchical, controlled through a small number of central nodes like the CEO of a corporation. New scientific claims about the influence of central nodes, such as genes at the center of a genetic regulatory network, will be more plausible than claims about peripheral genes for which we have no prior signals of influence. A claim's plausibility may also be affected by its position in the macro-structure of the network. If a claim about a genetic influence pathway is embedded within a large, dense cluster of claims about related interactions, the structure and direction of those interactions may logically and physically constrain the direction of the focal claim[6], which might help researchers triangulate the correct, robust claim. The presence of other researchers asking nearby questions might also socially discipline researchers to share their most robust results, as their work will receive scrutiny by contemporaries.

We may examine a scientific claim's breadth of prior support through evaluating the range researchers who have reiterated it and the depth of time over which a claim has been examined. Nevertheless, recent research has shown that the independence underlying a claims' support matters. More dependencies linking research that forwards a claim because they share authors, institutions, methods, or reference the same prior work decrease the replicability of that claim by outsiders [9]. Dense communities become scientific echo-chambers that drive out diverse perspectives[40], ignoring dissent or precisely reproducing fragile experiments unlikely to generalize beyond the context of initial demonstration. In biomedicine, findings from dense social and methodological communities are less likely to become relevant to human health and enter the clinic as practical diagnostics or therapies. They have not yet received independent support. By contrast, if more, independent research communities support a scientific claim, it will more likely be robust and generalize because it already has.

Other features of apparent support also exist, most notably the authority of whether a finding was published in elite, high-impact journals, or was authored by scientists from elite, strong reputation schools. As our analysis below reveals, however, these status signals deceive and are not associated with robust replication.

**Automated Validation Pipelines**

In order to predict robust signals from published science, we must first extract claims from published literature. Many prominent efforts have manually extracted statements regarding specific biological and chemical interactions from the literature and aggregate them into publicly available databases, which include the Gene Ontology[41], Comparative Toxicogenomics Database[6], the American Chemical Society's CASREACT, and others. These projects take a crisp view of logical inference that does not directly account for history and uncertainty. In contrast, we deploy two algorithmic approaches built on distinct architectures, GeneWays[7] and Literome[8]. The GeneWays system (portmanteau for genetics and pathways) semantically parses the biomedical literature. It first identifies biological substances and processes—nouns and verbs—then parses them with a context-free grammar tuned to the sublanguage of biomedicine[42] that extracts direct and indirect relations yielding a graph with directed links from source to target. Proteins may bind, activate, inhibit, or unleash more specific properties (acetylate, methylate, phosphorylate) upon their targets[43]. We simplify these interactions into positive (e.g., 'activate', 'enhance', 'increase', 'promote', 'stimulate', '[over]produce', 'upregulate', etc.) and negative (e.g., 'inactivate', 'depress', 'limit', 'inhibit', 'constrain', 'hinder', and 'downregulate', etc.) Literome inverts the GeneWays pipeline and begins by parsing articles into dependent clauses[44], then extracts biological entities, including genes and proteins. From co-presence within parsed phrases, Literome identifies directed relationships and then filters these by their correspondence to existing gene relationships from the annotated GENIA dataset[45]. "Gene" in this context is used as a shorthand for "gene or gene product", especially in reference to the action source, and henceforth we adopt this abbreviation.

For both datasets, we limited examination of claims extracted from article abstracts to increase the likelihood that they were not merely reiterated, but empirically demonstrated in the associated article. The GeneWays and Literome approaches and associated gene-gene interaction datasets were derived from overlapping collections of gene-related articles present in MEDLINE (GeneWays 197K, Literome 220K). GeneWays and Literome precision is evaluated as 95% and 25%, their percentage of positive interactions 96% and 77%, as shown in Fig. 1, panels a and b. In summary, GeneWays is more accurate, and Literome less constrained: GeneWays and Literome yield directed genetic graphs with 5141 genes involved in 23405 interactions, 10703 genes engaged in 144172 interactions, respectively, yielding an overlap of 4516 genes, but only 6516 overlapping interactions. We perform all of our analyses on both, independently derived datasets.

Next, we derive specific measures associated with how a claim fits into preexisting knowledge about genetic interactions and its breadth of support from article data and meta-data, as illustrated in Fig. 3 and detailed in the supplement. We measure each genetic regulatory interaction's position within preexisting knowledge by (1) the centrality of its gene source and target, as well as (2) the partition size of which it is a member, derived from dense connection among all published genetic claims. We measure the breadth of each interaction's support by the (1) the density of articles published on it each year, (2) number of years over which it has been investigated, (3) number of collaborative communities investigating it, (4) absolute size of the particular community engaged in making each interaction claim, and (5) size of that community relative to others that investigate it. We also derive measures capturing the reputation of the institution hosting the underlying research and the journal publishing the claim, and a number of related variables.

We merged these findings with the massively high-throughput NIH Library of Integrated Network-Based Cellular Signatures (LINCS) L1000 experiment that perturbs 77 distinct cell lines with several different perturbation types, such as cDNA for overexpression of wild-type gene for 5.8K genes[46]. Gene overexpression is a technique that utilizes expression vectors to force high levels of a gene's coding sequence. The resultant effect is high steady state mRNA levels and high steady state protein levels. This enables a vast gene-gene causal experiment: if one gene product is increased and it consistently leads to the increase or reduction in another gene across cell lines, perturbagens, dosage and time, which suggests that the over-expressed gene has a consistent and likely causal association with the other. We compute the mean z-score across experiments for genetic regulatory interactions and transform to (0,1) with the normal cumulative density function, denoting this the average experimental strength of the interaction.

**Bayesian Signals from Science**

We predict scientific certainty and the class of an interaction (neutral, positive or negative) through a Bayesian calculus built atop established statistical and machine learning methods that incorporate the features capturing position within pre-existing knowledge and depth of prior support described above. Some features occur at the level of the genetic regulatory interaction (e.g., position of source and target genes within the network of prior knowledge or the number of research communities publishing on the interaction). Others occur at the level of the published claim (e.g., size of the research community publishing a particular paper about the interaction). All of these features vary with time. We use these features to predict the robust, aggregate results of LINCS L1000 experiments pertaining to the same source and target gene across distinct trials, dosages, tissues and durations.

Cross-tabulation between the value of published claims and the average LINCS L1000 experimental results highlights a contrast. Claims in the biomedical literature tend to be either positive or negative, with a strong positive bias. The distribution of experimental interactions does not share this bias, normally distributed and varying smoothly from negative to positive, peaked and centered at 0.5, indicating a "neutral", inconsistent or nonexistent interaction between those genes. The contrast between published finding and experimental data suggests the extent of the file drawer problem in science where scientists euphemistically "file", but do not publish negative or inconclusive results[47,48]. We note that the correlation between published and experimental results increases markedly as we consider more popular interactions (see Extended Data Fig. 1). These observations led us to partition experimental interactions into 3 categories: neutral, positive and negative. First, we build models to predict the neutrality of an interaction (see SI). Second, we built models to predict whether positive and negative published claims correctly align with positive and negative experimental interactions. We separately built logistic regression and random forest models to estimate the influence of each feature on each outcome. Logistic regressions provide us with interpretable directional estimates, but assume conditional independence between features. Random forests provide us with better estimates of each feature's importance, but reduce interpretability by allowing nonlinear feature interactions. The most important features regarding how a claim fits into the fabric of prior knowledge and its breadth of prior support are defined in Fig.4; for others see Supplementary Information.

Once each claim's correctness is estimated, we can infer the direction of the underlying genetic interaction using Bayes formula, derived under the assumption of conditional independence of claims and the independence of correctness from interaction positivity (see Methods and SI).

**Predicting Claim Accuracy and Interactions**

We evaluate our genetic predictions from distributions of Receiver Operator Curves (ROCs) and the area under those curves (AUCs) presented in Fig.2. Our model of whether a published claim links to a non-neutral genetic interaction, as assessed across the wide range of LINCS experiments, betrays the difficulty of that task (average AUC= 0.58 for GeneWays on 6.8K interactions, AUC= 0.54 for Literome on 25.4K interactions). The most important feature for identifying non-neutral interactions is notable, however: the degree of the source gene in the network of prior published knowledge, reinforcing our understanding of hierarchical genetic regulation. The more central the source gene, the more likely it controls other genes.

Our model of whether a published claim correctly identifies the direction of a genetic interaction, conditional on the interaction not being neutral, is much more powerful (AUC=0.77 for GeneWays on 580 interactions, AUC=0.74 for Literome on 1090 interactions), strongly influenced by both the position of the claim within prior knowledge and its breadth of support (Fig.3). The feature manifesting most predictive power is the size of the partition of published genetic effects that surround the interaction in question. Its positive influence suggests that the structure and direction of nearby interactions may guide researchers to the correct conclusion[6]. The relevance of the claim to researchers working nearby will also increase competition and anticipated scrutiny.

The next most important class of influences are the historical depth and breadth of support. Greater depth and breadth of investigation, absolute size of the relevant research community, and the number of communities studying the claim are associated with empirical correctness. By contrast, empirical incorrectness is correlated with higher relative community size and our index tracing author, institutional and prior knowledge dependencies. Greater relative size indicates that the majority of scientific activity comes from one or a few communities and when the dependency index is high, authors, institutions and citations densely link published claims. Together, these findings reinforce that a lack of social and theoretical independence between claims should reduce our confidence in them and paint a consistent picture of the importance of balanced, independent investigation for scientific certainty.

Using Bayesian inference, we apply these estimates to infer the direction of any given genetic interaction. Our out of sample predictions demonstrate substantially greater signal than random regarding the robust direction of a genetic interaction (AUC = 0.67 for GeneWays; AUC = 0.63 for Literome). These models are less predictive than our models predicting accurate research claims because of inequality in research attention, collectively focused on a few, popular interactions. While scientific certainty about any particular interaction might be satisfied with a moderate number of replications, the inequality of research attention and activity are more likely to furnish the 100th replication of a popular claim than the 2nd of an unpopular one, despite the drop in information this entails for science as a system.

**Policies to Optimize Scientific Certainty**

Our findings suggest scientific policies to increase collective certainty across scientific claims, here evaluated in the context of genetic regulatory interactions. We design one statistical experiment to manipulate the distribution of independent communities examining a research claim and another to manipulate the distribution of claims across interactions, examining their effects on the correctness of all scientific inferences about genetic regulatory interaction. For a research funder like the U.S. National Institutes of Health (NIH) to influence the number of communities studying a specific topic would require them simply to prefer the social, institutional and intellectual independence of each new investigation they fund. For the NIH to broaden the distribution of claims across interactions would require them merely to prefer research on new topics. We demonstrate predictively that such policies, if implemented, could increase correct identification of the direction of genetic regulatory interaction as measured by the AUC of our model.

For the first statistical experiment, we divide interactions from the test sample into disjoint groups by the number of communities or author clusters that publish on them. For the subset of positive and negative interactions we (1) split the dataset into training and testing samples, (2) build the model of claim correctness, then using Bayesian inference (3) predict model certainty for groups of interactions having 1, 2, 3, 4 or more communities in the test sample. The top of Fig. 4 shows that a greater number of communities has a profound, positive effect on the distributions of AUCs for interaction positivity. The more communities studying an interaction, the better we can infer correct insight from the resulting corpus of research.

For the second statistical experiment, we artificially shift the distribution of claim numbers by sampling interactions according to the number of times on which each is published. Specifically, we (1) fix time $t$ for all interactions in the sample and consider only claims published before $t$, (2) prepopulate the sample with ~20% of claims in chronological order, and then (3) repopulate the sample with claims (and thus interactions), year by year, until each interaction contains the complete history of observed claims. In Fig.4c we present two synthetic examples of claim number distribution. The claim number distribution can be approximated with a power law, with density function proportional to the number of interactions having a given number of claims about them raised to the power of some exponent, where lower values correspond to flatter distributions. We demonstrate that flatter claim number distributions, where scientific attention is spread more widely across genetic regulatory interactions, correspond to significantly higher AUCs predicting accurate interactions (increases of $0.03 \pm 0.003$ for GeneWays and $0.007 \pm 0.002$ for Literome).

**Amplifying Signal from Science**
The deluge of published scientific information available to 21st Century researchers, funders, inventors and developers has overwhelmed their capacity to account for all of the signals available from science in their efforts to innovate. Challenges associated with information overload are exacerbated in research about entire scientific systems, such as the regulatory interactions between all human genes as we study here. In such settings, knowledge is necessarily uneven and investing against our ignorance will multiply the value of those investments for broad scientific understanding. In this paper, we demonstrate how the emergence of massive digital archives, machine reading and information extraction tools, alongside automated experiments can help us identify and amplify the signal from science by predicting the

robustness of scientific claims. This project represents the first automated, machine-driven pipeline of which we are aware that reads scientific research papers, extracts information about scientific claims, aligns them high-throughput experiments, and provides a Bayesian update of that knowledge. Our approach takes into account a wide range of features including how claims fit within the web of prior understanding and their breadth of support. And this is precisely what individual scientific experts do when they critically evaluate the literature, based on deep, personal understanding of dynamics and reputations within their specific field, here scaled by machine to many overlapping areas of biomedicine beyond the scope of any single scientific reasoner.

Our models predict replication based on different sets of publications, read by different algorithms, but yield deep consistency regarding what predicts replication and what might be altered by science policy makers to improve the state of scientific knowledge. These findings in the context of human genetic regulatory interactions, however, reveal an essential tension in the undertaking of collective scientific investigation. Robust genetic regulatory interactions are predicted by many investigators devoted to studying popular genes, central in the network of scientific claims and embedded within dense areas of investigation. Nevertheless, greater independence between investigators, communities, institutions, and prior knowledge also dramatically increase claim robustness. This leads to a paradox for science policy. When scientists flock together by studying the same phenomenon it increases our collective understanding. When they flock together in their approach and collaborations, it decreases our collective understanding—it increases the illusion we know more than we do. Our policy experiments suggest if science policy and sponsorship take this tension seriously, they could dramatically increase the robustness of our collective understanding and accelerate innovation. When communities, people and approaches focused on a given genetic regulatory interaction are more diverse, their collective findings more likely converge to powerful estimates derived from massive, high-throughput experiments. On the other hand, when scientists spread out across the space of possible claims, our overall certainty about the entire, interacting system of gene regulation increases. This suggests that when an important scientific process or component merits scientific attention, sponsoring diversity in that attention may pay dividends in robust, replicable understanding. On the other hand, if we seek to gain understanding about the system as a whole, we also increase the signal from published science by sponsoring and rewarding more work on under-examined areas.

Our study has several natural limitations. Despite our replication of all findings with two different samples of research papers and claim extraction algorithms, our study nevertheless only explores claims about genetic interactions. Moreover, in order to evaluate those claims at scale, we only consider claims about regulatory interactions between pairs of genes. This excludes many other, meaningful interactions (e.g., methylation, phosphorylation). Both information extraction algorithms, GeneWays and Literome, had limitations described above, but they balanced one another in terms of precision and recall. Finally, the several linkages across datasets that facilitate our mapping of research claims to LINCS L1000 experimental results necessarily excluded interactions from the literature not present in the experiment and vice versa. To conclude, our analysis of observational data makes it impossible for us to make strong causal claims about the impact of reforming scientific institutions according to the suggestive patterns we document. Notwithstanding these limitations, all of which would have decreased the signal we might expect to isolate from scientific literature, we were able to predict the likelihood of conceptual replication

far above random and identify consistent patterns of focus and independence that, if enhanced, could substantially increase the signal from science.

## Methods

### Information Extraction Algorithms

*GeneWays*. This algorithm and associated database of automatically extracted claims [43,49] contains approximately 496K unique claims (aggregating so each claim is unique per publication) and approximately 313K unique interactions (defined as a triplet including the source gene, target gene, and action, and where action is a verb that takes values including 'bind', 'interact', 'induce', 'associate', 'regulate', etc.) expressed in approximately 197K publications from MEDLINE. Approximately 32% of our claims were extracted from abstracts. We found that claims in publication could either result from independent original research or simply reference a finding from a cited publication. The former were much more likely mentioned in the abstract, and so in our research we considered only claims extracted from abstracts. This operation leaves us with ~172K unique publication-claims and ~130K unique interactions from the abstracts of approximately ~109K unique publications.

A typical record in the GeneWays database has the form: **abg** *prevents* **tert**. To simplify the representation of interactions, we identify all such verbs that can be interpreted as positive or negative directional actions. As positive, we encode: "activate", "actuate", "cause", "control", "direct", "enhance", "facilitate", "force", "increase", "induce", "lead", "overproduce", "promote", "provoke", "stimulate", "transactivate", "trigger", "regulate", "produce", and "upregulate". As negative, we encode: "constrain", "degrade", "destroy", "downregulate", "hinder", "inactivate", "inhibit", "interrupt", "limit", "reduce", "repress", "shut", and "suppress". After projecting the interactions to positive or negative, we are left with ~36K unique interactions, ~68.6K unique claims from ~51K unique publications from PubMed.

For each attribute, Geneways contains a flag indicating whether the claim is negative, where ~4% of claims are negative. According to logic, the negation of "a" increases "b" is the union of both "a" decreases "b" and "a" does not affect "b", non-interactions are never recorded and so we assume a positive interaction is the negation of a negative interaction and visa versa. If we encounter claims with respect to the same interaction extracted from the same paper that negate one another, we discard them. We retain claims from publications that present in our version of MEDLINE from 3K journals that we could identify using an available copy of the Web of Science database.
The final iteration has 23K unique interactions, 44K unique claims from 33K unique publications.

### *Literome*

Literome [50] contains 144K unique interactions, 259K unique claims from 220K unique publications extracted from MEDLINE abstracts by means of distant supervision via Markov Logic with an estimated precision of 25%. We only consider claims extracted from the abstracts and note that Literome has a strong bias towards positive interactions (~98%). For the set of final models described in the articles, we exclude claims with respect to gene TP53 (Entrez id 7157) acting on CDKN1A (Entrez id 1026) because the 150 extracted claims on that interaction were all deemed incorrect or ambiguous as evaluated by a biomedical expert.

**Genetic Dataset from LINCS L1000**

We use the Library of Integrated Network-Based Cellular Signatures (LINCS) that was compiled using the Luminex bead technology called L1000 as the ground truth with respect to gene-gene interactions derived within the same context [46]. The experimental technique of LINCS L1000 is based on tracking gene expression, the procedure by which information from genes chemically perturbed in the experiment causes the synthesis of functional gene products, such as proteins, resulting in an altered cellular phenotype. We use the GSE92742 Level 5 version of LINCS L1000. Level 5 dataset contains signatures from aggregated replicates. The experiments are performed on 77 cell lines, using different perturbation types, durations and dosages. Multiple experiments are performed per combination of cell line, perturbation type, duration and dose. The result of an experiment is a z-score, which quantifies the expression of a particular gene under the action of a perturbagen, relative to the baseline experiment.

We aggregate the z-scores of experiments in the following manner: For a given cell line, perturbagen, dosage and duration we compute the mean value; then across cell lines, perturbagens, dosages and durations we take the maximum of the absolute value for a given interaction. The z-score is then transformed using normal cumulative density function (that takes values in (0, 1)). We denote this $\hat{\pi}^{\alpha}$ and call it the experimental regulatory interaction strength.

For GeneWays, 40% of claims and 32% of interactions remain after merging with LINCS L1000 while for Literome, correspondingly 29 and 25%. After merging GeneWays and Literome onto aggregated LINCS L1000 data we obtain 15.5K and 50.5K claims and 6.8K and 25.4K interactions, respectively. The overlap between GeneWays and Literome claims merged with LINCS L1000 is 2K interactions (31% of all interactions iof GeneWays, 8% of all interactions of Literome) or 827 claims with correlation of the claim variable is .38 (representing 13% of GeneWays claims and 4% of Literome claims). The number of overlapping claims is greater than the number of overlapping interactions due to the majority of interactions being discussed in disjoint sets of publications between GeneWays and Literome. If we restrict the merged GeneWays-LINCS and Literome-LINCS datasets to the strongest positive and negative experimental regulatory interactions (intervals (0, 0.1] and [0.9. 1.0) on the interaction strength cdf) the overlap between GeneWays and Literome is 81 claims (34 interactions) with a correlation on the claim variable of .57. We conclude that GeneWays and Literome datasets are significantly different, but in moderate agreement where they overlap, suggesting that they are largely independent sources of genetic regulatory interaction claims. We note that the distribution of number of claims per interaction follows Zipf's Law.

The correlation between regulatory interaction strength from LINCS L1000 and mean claim value from the literature is negligible, but increases as we introduce a threshold for the number of publications in which the claim appears (see Extended Data Table 1 and Fig. 1).

We define communities associated with each genetic regulatory interaction for claims made within a variety of fixed time intervals: the past 1, 2, 3 and all years leading up to a given year. Each claim is made within a unique publication, and each paper is produced in an institutional, social and knowledge context, reflected by the multiple affiliations, authors, and references mentioned in that paper. Denote the set of affiliations (or authors or references) $V$, publications $U$, such that edges $(u, v)$ between members of these two sets form a bipartite graph, which we reduced to a weighted graph defined on the set of publications $U$, with weights proportional to the number of common affiliations. In every such local weighted graph of publications defined over a given time period (i.e., 1, 2, 3, and all prior years), we identify communities using the information theoretically inspired Infomap algorithm[51] and assign the number of communities, community size and community share for a given claim as derived features. Features deemed unimportant by our analysis (such as journal quality) are described in the SI and listed in Extended Data Table 2.

In order to classify (1) the *neutrality* of a genetic regulatory interaction, (2) the *positivity* (or *negativity*) of a regulatory interaction, and (3) the correctness of a claim, we used random forest and logistic regression models to enable both prediction and interpretation. While random forest allows us to reach near-maximal predictive performance, logistic regression enables the linear interpretation of features, rendering some effects positive and others negative. We choose models of optimal complexity and estimate metrics over the ensemble using procedures described in SI. In Fig. 3, features are presented pictorially with the highest Gini Importance or Mean Decrease in Impurity in the random forest model. Detailed methodology of feature importance calculation is located in the SI. Fig. 3 displays the Gini Importance or Mean Decrease in Impurity for each variable in the random forest model and associated coefficients from the logistic regression, plotted in decreasing importance for the Geneways random forest.

We used Python and scikit-learn for testing models, large scale computations were made possible thanks to CloudKotta infrastructure[52].

# Acknowledgements

# Author contributions

Alexander Belikov proposed and implemented the methodology, validated the model, analyzed the data and drafted the paper. James Evans was responsible for conception and funding of the project, contributed to the design of the methodology and drafted the paper. Andrey Rzhetsky provided feedback on the experimental work and data interpretation, and participated in drafting the paper. All authors read and approved the final manuscript.

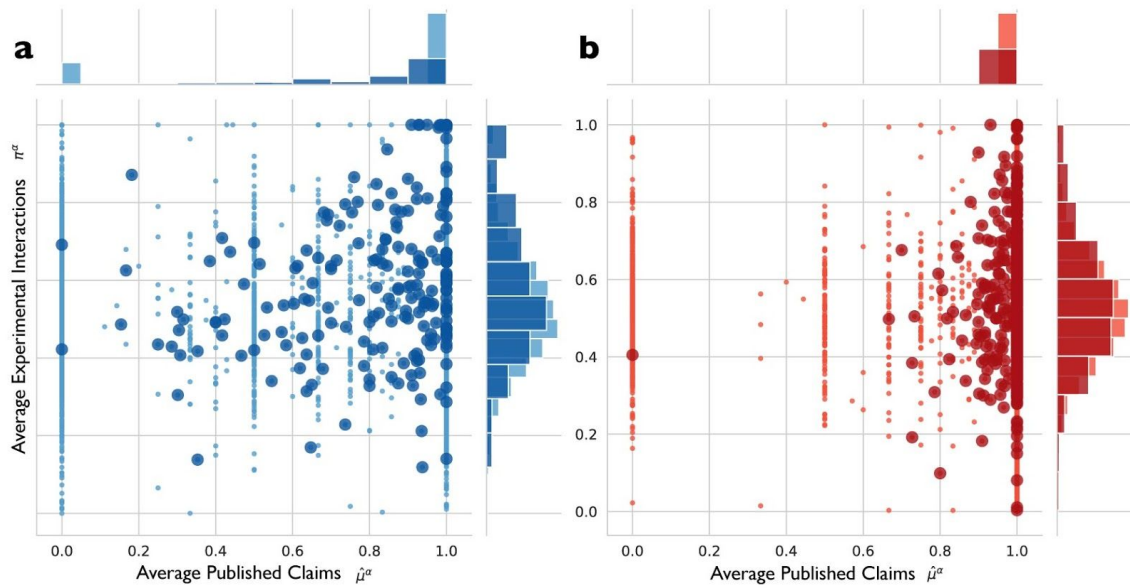

**Fig.1. Joint plot of the mean value of the claim (x-axis) and mean experimental interaction strength (y-axis). a**, GeneWays (blue). **b**, Literome (red). More intense hues of the blue and red (and also greater marker size) correspond to the interactions with 10 or more claims per interaction; for less intense hues (and also smaller marker size) the cutoff is absent, representing the complete distribution.

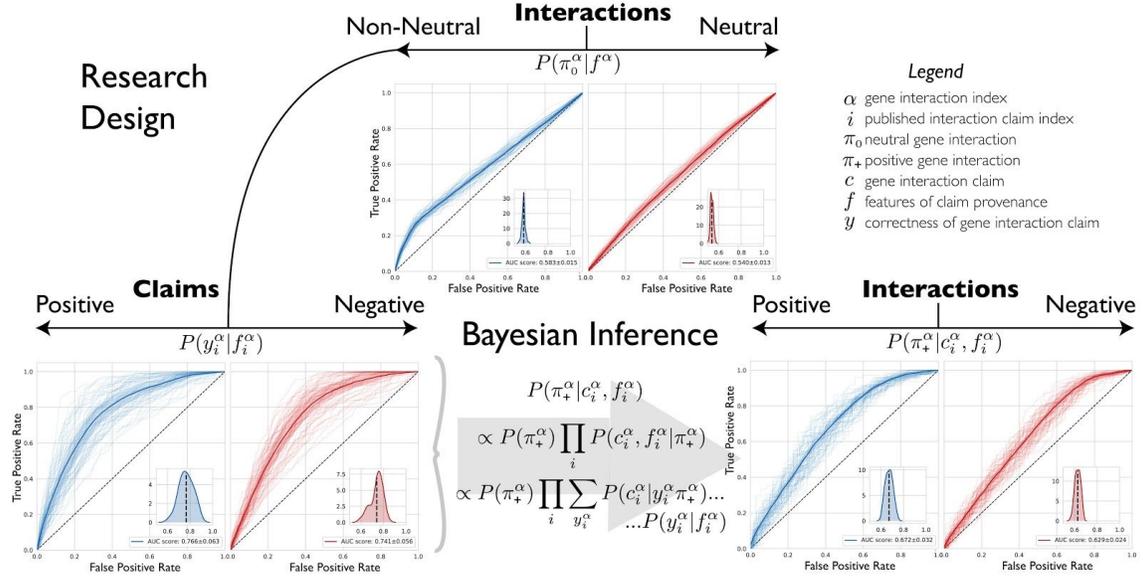

**Fig.2. Research design and prediction results.** We first predicted the neutrality or non-neutrality of each gene-gene regulatory interaction. Then, if the interaction was deemed non-neutral, we predicted whether each claim (of positivity or negativity) from literature was correct. Finally, using Bayesian inference, as illustrated here and detailed in the methods section, we estimated the positive or negativity of genetic regulatory interactions. The ROC curves for the prediction models are blue for GeneWays and red for Literome. Mean ROC curves are in bold surrounded by a 95% c.i., with fainter individual lines corresponding to ROC curves for 60 models; 20 for each of 3 randomly drawn sets of non-overlapping interactions. All models were built from interactions not present in the test set on which ROC curves were evaluated.

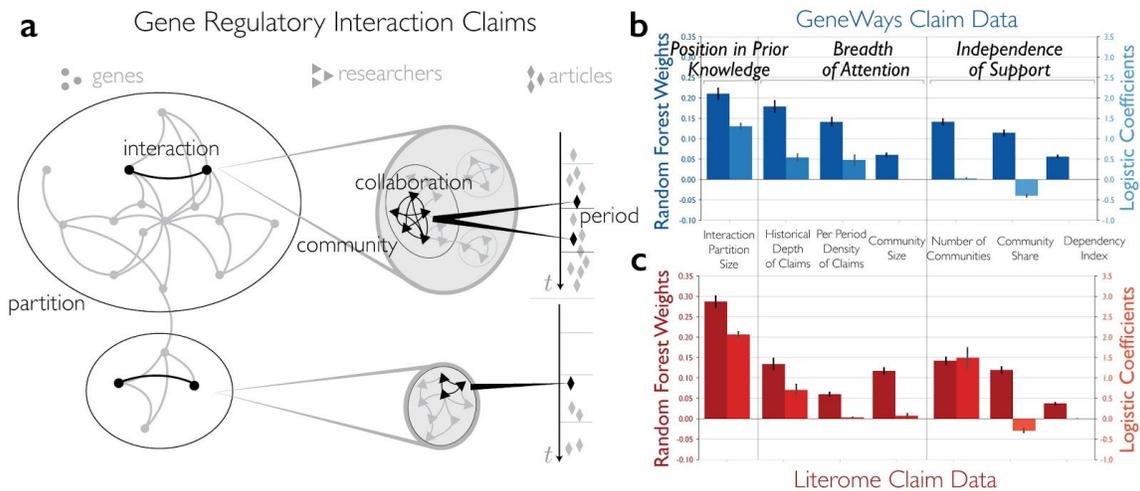

**Fig.3. Feature visualization and estimates from claim-level prediction models. a**, illustrated relationship between genes engaged in regulatory interactions, the communities that research them, and the articles in which this research is published. Interactions cluster into partitions, researchers cluster into communities, and author teams publish articles within fixed periods. Together these structures are used to assess the position of a claim within pre-existing knowledge, the breadth of attention to a claim, and the independence of support for that claim. **b,c** Gini Importance or Mean Decrease in Impurity for features in the random forest models (left vertical scale, bold colors), and coefficients from the logistic regression models (right vertical scale, fainter colors) for GeneWays (**b**) and Literome (**c**). Vertical bars represent 95% c.i. for the mean value of the estimate. See Supplement for details about how specific operationalizations of each of these variables were selected as model features.

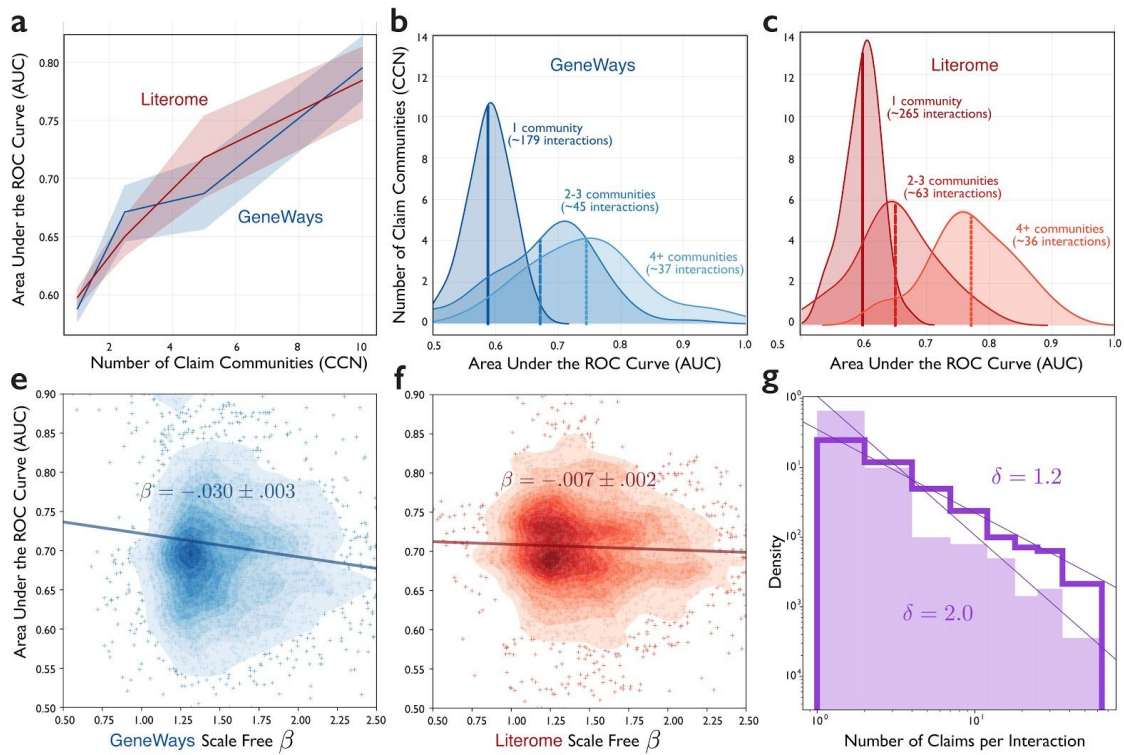

**Fig.4. Science policy experiments revealing the relationship between community independence, collective attention, and certainty about genetic regulatory interactions. a**, relationship between the number of communities studying a particular genetic regulatory interaction and the average AUC of out-of-sample predictions for positive interactions. **b-c**, distributions of the average AUC curves for GeneWays and Literome for interactions with 1, 2-3 and greater than 4 communities. **e-f**, relationship between the shape of the distribution of number of claims per interaction on the AUC of out-of-sample predictions for positive interactions. $\beta$ represents the slope of the claim number per interaction distribution for GeneWays (**e**) and Literome (**f**). **g**, two synthetic examples of claim number distributions, where these distributions can be approximated with power laws, proportional to the number of claims about an interaction raised to the power of some exponent $\delta$, such that lower values correspond to flatter distributions. In flatter claim number distributions, scientific attention spreads more widely across genetic regulatory interactions, which corresponds to significantly higher certainties (AUCs) of accurate interactions.

1. Price, D. J. D. S. & De Solla Price, D. J. Little Science, Big Science. (1963) doi:10.7312/pric91844.

2. Bornmann, L. & Mutz, R. Growth rates of modern science: A bibliometric analysis based on the number of publications and cited references. *Journal of the Association for Information Science and Technology* **66**, 2215–2222 (2015).

3. Hey, T. & Trefethen, A. The data deluge: An e-science perspective. *Grid computing: Making the global infrastructure a reality* 809–824 (2003).

4. Bell, G., Hey, T. & Szalay, A. Computer science. Beyond the data deluge. *Science* **323**, 1297–1298 (2009).

5. Ioannidis, J. P. A. Why most published research findings are false. *PLoS Med.* **2**, e124 (2005).

6. Rzhetsky, A., Iossifov, I., Loh, J. M. & White, K. P. Microparadigms: chains of collective reasoning in publications about molecular interactions. *Proc. Natl. Acad. Sci. U. S. A.* **103**, 4940–4945 (2006).

7. Prinz, F., Schlange, T. & Asadullah, K. Believe it or not: how much can we rely on published data on potential drug targets? *Nat. Rev. Drug Discov.* **10**, 712–712 (2011).

8. Begley, C. G., Glenn Begley, C. & Ellis, L. M. Raise standards for preclinical cancer research. *Nature* vol. 483 531–533 (2012).

9. Danchev, V., Rzhetsky, A. & Evans, J. A. Centralized communities more likely generate non-replicable results. *Elife* (2019).

10. Nissen, S. B., Magidson, T., Gross, K. & Bergstrom, C. T. Publication bias and the canonization of false facts. *Elife* **5**, (2016).

11. Daston, L. J. & Galison, P. *Objectivity*. (Zone Books, 2007).

12. Foreman, P. Weimar Culture, Causality and Quan-tum Theory 1918-1927. *Hist. Stud. Phys. Biol. Sci.* **3**, 2–225 (1971).

13. Surowiecki, J. The wisdom of crowds: Why the many are smarter than the few and how collective

wisdom shapes business. *Economies, Societies and Nations* **296**, (2004).

14. Galton, F. Vox populi (the wisdom of crowds). *Nature* **75**, 450–451 (1907).

15. Hong, L. & Page, S. E. Groups of diverse problem solvers can outperform groups of high-ability problem solvers. *Proc. Natl. Acad. Sci. U. S. A.* **101**, 16385–16389 (2004).

16. Becker, J., Brackbill, D. & Centola, D. Network dynamics of social influence in the wisdom of crowds. *Proc. Natl. Acad. Sci. U. S. A.* **114**, E5070–E5076 (2017).

17. Lorenz, J., Rauhut, H., Schweitzer, F. & Helbing, D. How social influence can undermine the wisdom of crowd effect. *Proceedings of the national academy of sciences* **108**, 9020–9025 (2011).

18. Hicks, D. M. & Katz, J. S. Where is science going? *Sci. Technol. Human Values* **21**, 379–406 (1996).

19. Guimerà, R., Uzzi, B., Spiro, J. & Amaral, L. A. N. Team assembly mechanisms determine collaboration network structure and team performance. *Science* **308**, 697–702 (2005).

20. Hand, E. 'Big science' spurs collaborative trend. *Nature* vol. 463 282–282 (2010).

21. Wuchty, S., Jones, B. F. & Uzzi, B. The increasing dominance of teams in production of knowledge. *Science* **316**, 1036–1039 (2007).

22. Wu, L., Wang, D. & Evans, J. A. Large teams develop and small teams disrupt science and technology. *Nature* (2019) doi:10.1038/s41586-019-0941-9.

23. Jones, B. F., Wuchty, S. & Uzzi, B. Multi-university research teams: shifting impact, geography, and stratification in science. *Science* **322**, 1259–1262 (2008).

24. Merton, R. K. The Matthew Effect in Science: The reward and communication systems of science are considered. *Science* **159**, 56–63 (1968).

25. Azoulay, P., Stuart, T. & Wang, Y. Matthew: Effect or Fable? *Manage. Sci.* **60**, 92–109 (2014).

26. Evans, J. A. Electronic publication and the narrowing of science and scholarship. *Science* **321**, 395–399 (2008).

27. Simkin, M. V. & Roychowdhury, V. P. Do Copied Citations Create Renowned Papers? *Annals of*

*Improbable Research* vol. 11 24–27 (2005).

28. Mullard, A. Reliability of 'new drug target' claims called into question. *Nat. Rev. Drug Discov.* **10**, 643–644 (2011).

29. Freedman, L. P. & Gibson, M. C. The impact of preclinical irreproducibility on drug development. *Clin. Pharmacol. Ther.* **97**, 16–18 (2015).

30. Ioannidis, J. P., Ntzani, E. E., Trikalinos, T. A. & Contopoulos-Ioannidis, D. G. Replication validity of genetic association studies. *Nat. Genet.* **29**, 306–309 (2001).

31. Hirschhorn, J. N., Lohmueller, K., Byrne, E. & Hirschhorn, K. A comprehensive review of genetic association studies. *Genet. Med.* **4**, 45–61 (2002).

32. Lohmueller, K. E., Pearce, C. L., Pike, M., Lander, E. S. & Hirschhorn, J. N. Meta-analysis of genetic association studies supports a contribution of common variants to susceptibility to common disease. *Nat. Genet.* **33**, 177–182 (2003).

33. Open Science Collaboration. Estimating the reproducibility of psychological science. *Science* **349**, (2015).

34. Van Bavel, J. J., Mende-Siedlecki, P., Brady, W. J. & Reinero, D. A. Contextual sensitivity in scientific reproducibility. *Proc. Natl. Acad. Sci. U. S. A.* **113**, 6454–6459 (2016).

35. Zollman, K. J. S. The Communication Structure of Epistemic Communities. *Philos. Sci.* **74**, 574–587 (2007).

36. Payette, N. Agent-Based Models of Science. in *Models of Science Dynamics: Encounters Between Complexity Theory and Information Sciences* (eds. Scharnhorst, A., Börner, K. & van den Besselaar, P.) 127–157 (Springer Berlin Heidelberg, 2012).

37. Baker, M. Biotech giant publishes failures to confirm high-profile science. *Nature* **530**, 141 (2016).

38. Borenstein, M., Hedges, L. V., Higgins, J. P. T. & Rothstein, H. R. *Introduction to Meta-Analysis*. (John Wiley & Sons, 2011).

39. Nussbaum, D. The role of conceptual replication. *Psychologist* **25**, 350 (2012).

40. Sunstein, C. R. *Republic.com*. (Princeton University Press, 2001).

41. Harris, M. A. *et al.* The Gene Ontology (GO) database and informatics resource. *Nucleic Acids Res.* **32**, D258–61 (2004).

42. Friedman, C., Kra, P. & Rzhetsky, A. Two biomedical sublanguages: a description based on the theories of Zellig Harris. *J. Biomed. Inform.* **35**, 222–235 (2002).

43. Rzhetsky, A. *et al.* GeneWays: a system for extracting, analyzing, visualizing, and integrating molecular pathway data. *J. Biomed. Inform.* **37**, 43–53 (2004).

44. Quirk, C. *et al.* MSR SPLAT, a language analysis toolkit. in *Proceedings of the 2012 Conference of the North American Chapter of the Association for Computational Linguistics: Human Language Technologies: Demonstration Session* 21–24 (Association for Computational Linguistics, 2012).

45. Kim, J.-D., Ohta, T., Pyysalo, S., Kano, Y. & Tsujii, J. Overview of BioNLP'09 shared task on event extraction. in *Proceedings of the BioNLP 2009 workshop companion volume for shared task* 1–9 (2009).

46. Subramanian, A. *et al.* A Next Generation Connectivity Map: L1000 Platform and the First 1,000,000 Profiles. *Cell* **171**, 1437–1452.e17 (2017).

47. Rosenthal, R. The file drawer problem and tolerance for null results. *Psychol. Bull.* **86**, 638 (1979).

48. Scargle, J. D. Publication Bias (The 'File-Drawer Problem') in Scientific Inference. *arXiv [physics.data-an]* (1999).

49. Rzhetsky, A. Geneways for Biocomputing. (2006) doi:10.21236/ada459874.

50. Poon, H., Quirk, C., DeZiel, C. & Heckerman, D. Literome: PubMed-scale genomic knowledge base in the cloud. *Bioinformatics* **30**, 2840–2842 (2014).

51. Rosvall, M. & Bergstrom, C. T. Maps of random walks on complex networks reveal community structure. *Proc. Natl. Acad. Sci. U. S. A.* **105**, 1118–1123 (2008).

52. Babuji, Yadu N,  Chard K., Gerow A., Eamon Duede, [Cloud Kotta: Enabling Secure and Scalable Data Analytics in the Cloud,](#) *IEEE International Conference on Big Data,* 302- 310 (2016).



{\Huge Supplementary Information}
\section{Definitions}

\noindent Following are definitions to enable precise articulation of all calculations and procedures.

\noindent$g_s$ - source gene or gene product involved in a genetic regulatory interaction.

\noindent$g_t$ - target gene or gene product involved in a genetic regulatory interaction.

\noindent$\alpha$ - index for each genetic regulatory interaction $(g_s, g_t)$.

\noindent$\hat\pi^\alpha$ - interaction strength estimated from LINCS L1000, averaged over experiments of the same type, maximized over cell type, duration, and dosage.

\noindent$c_i^\alpha$ - claim with respect to interaction $\alpha$ from publication $i$.

\noindent$y_i^\alpha$ - correctness of published claim, defined with respect to interaction $\alpha$ mentioned in publication $i$. We operationalize it here as the difference between the published claim and results from the LINCS L1000 experiment: $y^\alpha_i = 1 - |c^\alpha_i -  \pi_{+}^\alpha|$.

\noindent$\hat \mu^\alpha$ - mean claim value, where the sum runs over all claims with respect to interaction $\alpha$:

\bal
\hat \mu^\alpha = \frac{1}{n}\sum\limits^{n_\alpha}_{i=1} c_i^\alpha
\eal

\noindent$\pi_0^{\alpha}$ - indicator variable for interaction $\alpha$: equal to 1 if $\alpha$ is neutral, 0 if positive or negative. 

\noindent$\pi_+^{\alpha}$ - indicator variable defined on the subset of positive and negative interactions: equal to 1 if $\alpha$ is positive, 0 if negative. 

\section{GeneWays and Literome alignment with LINCS L1000}
\begin{figure}
\includegraphics[width=0.5\textwidth]{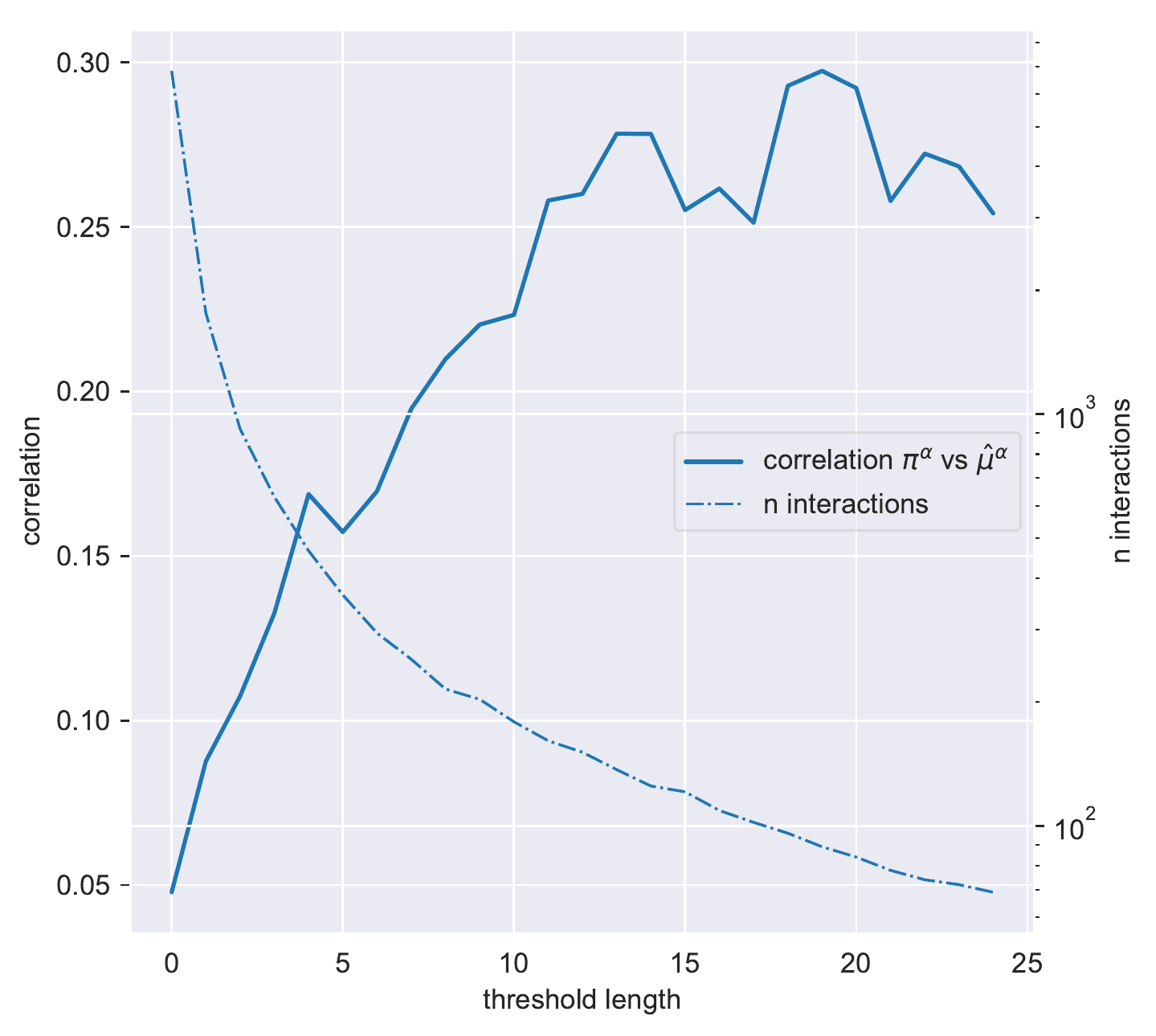}
\includegraphics[width=0.5\textwidth]{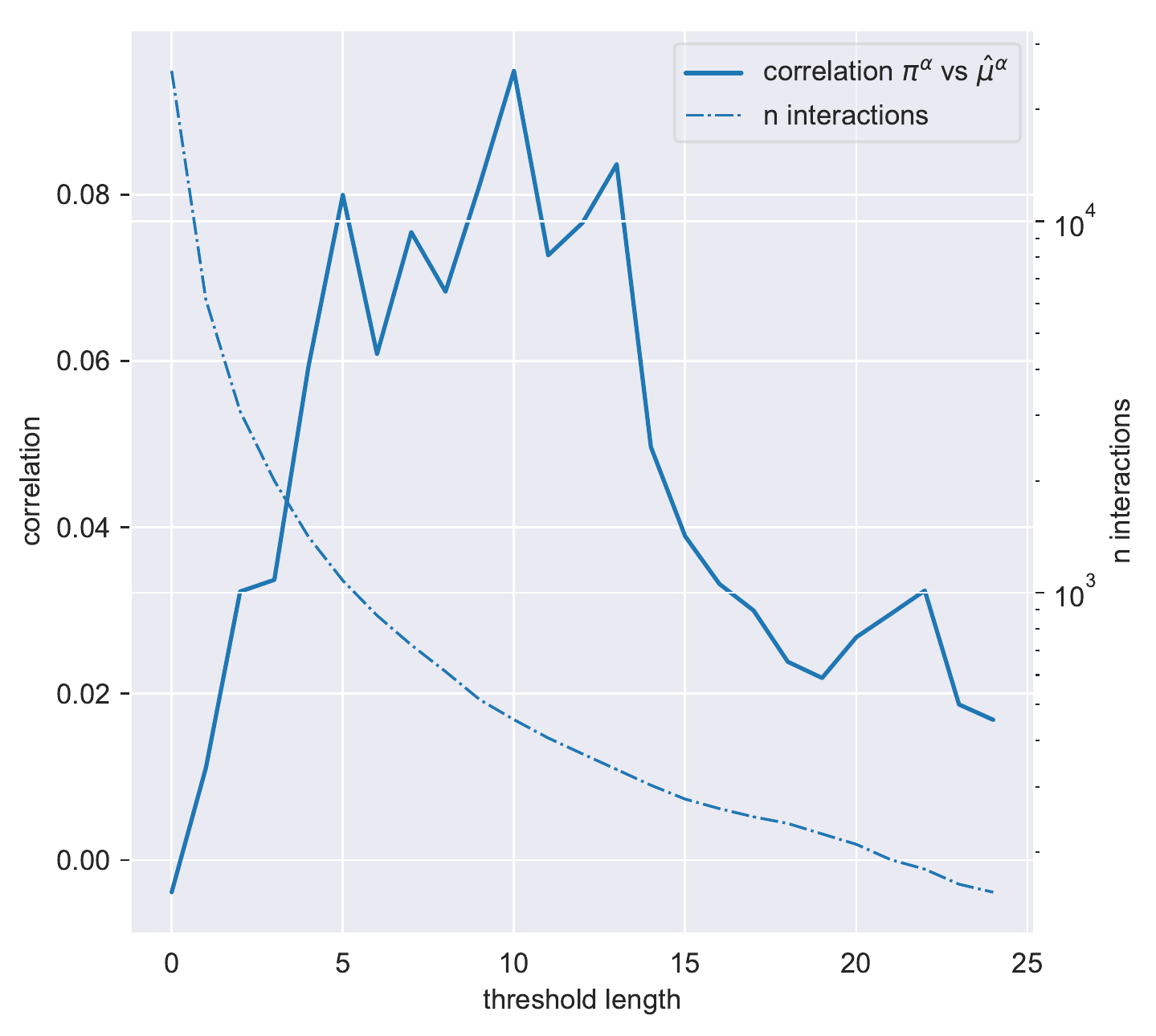}
\caption{Correlation of mean claim value $\mu_\alpha$ and interaction strength $\hat \pi^\alpha$ from LINCS L1000 as a function of threshold on minimum claim sequence length per interaction for GeneWays (left) and Literome (right).}
\label{fig:corr_gw}
\end{figure}

Table \ref{table:corr} and Fig. \ref{fig:corr_gw}.

\begin{table}
    \begin{centering}
    \begin{tabular}{lrrrrr|rrrrr}
        \toprule
        & \multicolumn{5}{c}{GeneWays} & \multicolumn{5}{c}{Literome} \\
        \hline

        $n_c$ &      0 &      2 &      4 &     6 &     8 &      0 &      2 &      4 &      6 &      8 \\
        \hline
        \midrule
        size &   6825 &    921 &    466 &    294 &   215 &  25411 &   3080 &   1413 &    868 &    613 \\
        $\rho$ &  0.048 &  0.107 &  0.169 &   0.17 &  0.21 & -0.004 &  0.032 &  0.059 &  0.061 &  0.068 \\
        mean &  0.749 &   0.78 &   0.79 &  0.804 &   0.8 &  0.978 &   0.98 &   0.98 &  0.981 &   0.98 \\
        \bottomrule\bottomrule
        \hline
    \end{tabular}
\end{centering}
\caption{Sizes of datasets, based on the threshold number of claims required per genetic regulatory interaction ($n_c$); the correlation ($\rho$) between the distribution of agreement over interaction positivity from literature ($\hat \mu^\alpha$) and derived from the LINCS L1000 experimental z-scores ($\pi^\alpha$); and mean interaction positivity ($\hat \mu^\alpha$) from literature.}
\label{table:corr}
\end{table}

\section{Experimental setup}
\subsection{Interaction/claim partition}
We partition gene - gene (or gene product) regulatory interactions from LINCS L1000 into three classes corresponding to negative, neutral and positive interactions. We do so by introducing thresholds $\theta_-$
 and $\theta_+$ that separate correspondingly negative from neutral and neutral from positive regulatory interactions in the space the experimental findings ($\hat\pi^\alpha$). This section explains our method for determination of $\theta_-$ and $\theta_+$.
 
Formally we partition negative, neutral and positive interactions, respectively: $\Ac_{-} = \{\alpha | \hat \pi_\alpha < \theta_-\}$, 
$\Ac_{0} = \{\alpha | \theta_- \le \pi_\alpha < 1 - \theta_+\}$, $\Ac_{+} = \{\alpha | \hat \pi_\alpha \ge 1 - \theta_+\}$, with $0 < \theta_- < 1$ and $0 < \theta_+ < 1 - \theta_-$.

We define an indicator function $\pi^\alpha_0$ on $\Ac$, equal to 1 if interaction $\alpha$ is neutral ($\Ac_{0}$) and 0 otherwise. Defined in this way, $\pi_{0}^\alpha$ is the target variable for the model predicting neutral interactions.

On $\Ac_{-} \bigcup \Ac_{+}$ we define an indicator function $\pi_{+}^\alpha$, which takes value 1 if interaction $\alpha$ is positive ($\Ac_{+}$) and 0 otherwise (negative). Defined in this way, $\pi^\alpha_+$ is the target variable for our model predicting positive interactions. 

For claims attributed to interactions $\Ac_{-} \bigcup \Ac_{+}$ we define claim correctness as the difference between the claim and the experimental results:

\bal
y^\alpha_i = 1 - |c^\alpha_i -  \pi_{+}^\alpha|
\eal

Claim correctness is 1 when the claim is correct (both the claim and the interaction are positive, or when both are negative) and 0 otherwise.

In the following we derive features at two broad levels of analysis: 1) features attributed to each genetic regulatory interaction (i.e., $f^\alpha$), such as the degree or the partition size; and 2) features attributed to claims published in each article about each interaction (i.e., $f^\alpha_i$).

We use interaction level features ($f^\alpha$) in predictive models of the neutrality of a genetic regulatory interaction ($\pi_{0}^\alpha$) and its positivity ($\pi_{+}^\alpha$). We correspondingly use features associated with published claims ($f^\alpha_i$) in predictive models of claim correctness. Because claim correctness is defined for positive and negative interactions, we can use our probabilistic estimates of claim correctness to assist with the prediction of correct interactions ($\pi_{+}^\alpha$).

\subsection{LINCS L1000 thresholds}
In this section we describe our data-driven approach to defining thresholds that partition interactions into neutral ($\Ac_0$), negative ($\Ac_{-}$) and positive ($\Ac_{+}$).

The threshold could be set to a constant value in an {\it ad hoc} manner: $\theta_- = \theta_+ = \varepsilon$, e.g. $\varepsilon=0.15$. Alternatively, it could be set $\theta_-$ and $\theta_+$ by a fixed percentile level:

$$
\theta_- : \varepsilon = \int\limits^{\theta_-}_0 \rho(\pi^\alpha) d\pi^\alpha , \;
\theta_+ :  \varepsilon = \int\limits_{1- \theta_+}^1 \rho(\pi^\alpha) d\pi^\alpha
$$

While both of these definitions of $\varepsilon$ are simple, they depend on the choice of an arbitrary constant.
We propose and implement an approach that takes into account how these interactions are presented in the scientific literature.

We denote the set of all claims $\Cc$.
We note that a partition of $\Ac$ induces a partition of $\Cc$ into $\Cc_+$, $\Cc_-$ and $\Cc_0$, for example $\Cc_+ = \{c^\alpha_i | \alpha \in \Ac_+\}$. We propose to define threshold $\theta_+$ so that it maximizes the distance between $\Cc_+$ and $\Cc_0$, and, correspondingly, $\theta_-$ to maximize the distance between $\Cc_-$ and $\Cc_0$.

We assume that claims for each interaction $\alpha$ are generated from a binomial distribution with parameter $\mu^\alpha$, which is generated from a beta distribution with class-wide parameters $a, b$. Beta distributions for each class can be estimated following the Bayesian update procedure:

$$
g_x(\mu) = Beta\left(a_0 + \sum_{\alpha \in \Cc_x}\sum\limits_{i=1}^{n_\alpha} y^\alpha_i, \quad b_0 + \sum_{\alpha \in \Cc_x} \left(n_\alpha - \sum\limits_{i=1}^{n_\alpha} y^\alpha_i\right)\right)
$$

We then define the distance between two disjoint subsets of $\Cc$ as the Wasserstein distance between two probability measures, in our case - beta posteriors. We note that the Kullback-Leibler (KL) divergence would be less suitable due to Beta function having very localized support for large values of $a$ and $b$.

Distribution $g_{x}(\mu)$ are probability measures defined on a metric space $[0, 1]$ with metric $d$, Euclidean $L^2$ distance in our case. The distance between $\Cc_+$ and $\Cc_0$ is computed as

$$
W(g_+, g_0, \theta_-, \theta_+) = \underset{\gamma \in \Gamma (g_+, g_0)}{\inf}\int d(x,y) d\gamma (x, y),
$$

where $\Gamma (g_+, g_0)$ denotes a collection of all measures with marginals $g_+$ and $g_0$.
Note that $\theta_-$ and $\theta_+$ define $g_+$, $g_0$ and $g_-$.

Increasing thresholds $\theta_+$ and $\theta_-$ corresponds to redistributing claims from $\Ac_0$ to $\Ac_+$ and $\Ac_0$ to $\Ac_-$, respectively. Due to the discrete nature of our datasets, such redistribution occurs discontinuously 
and so we do not expect distances $W(g_+, g_0)$ and $W(g_-, g_0)$ to depend continuously on thresholds. 

We postulate that we want to maximize relative discontinuity $\delta^R$ in the distance rather than the distance $W$ itself:

\bal
\delta^R f(x_0) = \underset{x\to x^+_0}{\lim} \frac{f(x) - f(x_0)}{f(x)},\\
\delta^L f(x_0) = \underset{x\to x^-_0}{\lim} \frac{f(x) - f(x_0)}{f(x)},
\eal

where $\delta^R f(x_0)$ is the relative discontinuity from the right (the limit $x\to x^+_0$ is understood as sequence of $x_n > x_0$), and $\delta^L(x_0)$ - from the left ($x_n < x_0$). Finally we define the optimal values $\theta^*_-$ and $\theta^*_-$ as:

\bal
\theta^*_- = \argmax{\theta_-} \delta^L W (g_-, g_0, \theta_-, \theta_+) \\
\theta^*_+ = \argmax{\theta_+} \delta^R W (g_+, g_0, \theta_-, \theta_+)
\eal

On the one hand, we would like to maximize relative discontinuity. On the other hand, we would like to have a relatively large number of claims in $\Cc_+$ and $\Cc_-$ in order to use statistical methods. In Fig. \ref{fig:data_thr} we show plots of $W (g_-, g_0, \theta_-)$ for a fixed $\theta_+$ and $W (g_+, g_0, \theta_+)$ for a fixed $\theta_-$ for both  the GeneWays and Literome datasets.

In Fig. \ref{fig:threshold_3d} we show plots of distances $W (g_-, g_0, \theta_-, \theta_+)$ and $W (g_+, g_0, \theta_-, \theta_+)$ as functions of $\theta_-$ and $\theta_+$ for GeneWays. As expected, the dependence on $\theta_+$ for $W (g_-, g_0, \theta_-, \theta_+)$ and $\theta_-$ for $W (g_+, g_0, \theta_-, \theta_+)$ are weak. The corresponding plots for Literome are similar, although less pronounced. Armed with this observation we simplify our optimization problem:

\bal
\theta^*_-, \theta^*_+  = \argmax{\theta_-, \theta_+} \delta^L W (g_-, g_0, \theta_-, \theta_+) \delta^R W (g_+, g_0, \theta_-, \theta_+)
\eal

and obtain optimal values of $\theta_-$ and $\theta_+$ to obtain (0.305, 0.218) for GeneWays and (0.256, 0.157) for Literome.

As a result of this procedure, we select positive and negative interactions: 2476 claims about 580 interactions for GeneWays and 2720 claims about 1090 interactions for Literome. Note that the main results reported in the paper are qualitatively the same if we take fixed thresholds for class definition.

\subsection{Models}

We build our models of interaction classification in a hierarchical way. Our first model  answers the question whether interaction $\alpha$ is neutral: $P(\pi^\alpha_0 = 1 | f^\alpha)$.

Conditional on interaction $\alpha$ being non-neutral we can estimate whether it is positive: $P(\pi_+^\alpha = 1 | f^\alpha, \pi^\alpha_0 = 0)$. The full probability that interaction $\alpha$ is positive is 
\bal
P(\pi_+^\alpha = 1 | f^\alpha) = P(\pi_+^\alpha | f^\alpha, \pi^\alpha_0 = 0) P(\pi^\alpha_0 = 0 |  f^\alpha)
\eal

In summary, whenever we have access to scientific claims from the literature and we can model their correctness based on independent evidence, the estimate of interaction positivity $\alpha$ can be augmented via Bayes formula in the following way. We can estimate the positivity of an interaction ($\pi_+^\alpha$) as a function of claims ($c_i^\alpha$) and features ($f^\alpha_i$) by using our estimate of claim correctness ($y^\alpha_i$), assuming conditional independence between claims. 

\bal
P(\pi_+^\alpha| \{c_i^\alpha, f^\alpha_i\})
\propto P(\{c_i^\alpha, f^\alpha_i\} | \pi_+^\alpha) P(\pi_+^\alpha) \propto \prod\limits_i P(c_i^\alpha, f_i^\alpha| \pi_+^\alpha) P(\pi_+^\alpha)\eal
\bal
\propto \prod\limits_i P(\pi_+^\alpha) \sum\limits_{y^\alpha_i} P(c^\alpha_i| y^\alpha_i \pi_+^\alpha)  P(y^\alpha_i | f^\alpha_i)
\eal

Alternatively we use interaction level features for the model of positive interactions: $P(\pi^\alpha_+ = 1 | f^\alpha)$.

We note that some of the features detailed in the following section do not assume the conditional independence of each claim. We minimize the size of these dependencies and their potential to distort our estimates, as we describe in detail below in the section on sampling, by separating genetic regulatory interactions across training and testing samples, such that no dependencies fit within the training data can artificially inflate our predictions in the testing data.  

\section{Features}

Below we define features used in models of interaction neutrality, positivity and claim correctness. In order to predict interaction neutrality, we derive features from the publicly available knowledge network of published claims. All features are defined using data available before, or in special cases contemporaneous with the prediction in question.  

Claim features are defined at multiple levels: 1) at the level of genetic regulatory interactions indexed by $\alpha$ (e.g., the centrality of the interaction in the network of other published interactions); 2) at the level of the claim `batch`, defined as the set of claims with respect to the same interaction ($\alpha$) within a predefined time interval (at time $t$ and window-size $w$), indexed by $\alpha, t, w$ (e.g., number of claims made with respect to an interaction before a given year); and finally 3) at level of the publication ($i$) indexed by $\alpha, i$, where $i$ indexes all publications pertaining to interaction $\alpha$ (e.g., citation impact of the journal or status rank of the university affiliations held by authors when the published the article).

\subsection{Interaction-level features} 

\subsubsection{Mean claim percentile} Mean claim value ($\hat \mu^\alpha$) is the average value over published Boolean (positive or negative) claims $c^\alpha_i$ pertaining to interaction $\alpha$. We define mean claim percentile (MCP) $p_{\alpha} = P(x < \hat \mu^\alpha)$ and absolute value of the median mean claim percentile (AMMCP) $\delta p_{\alpha} = |p_{\alpha} - 0.5|$. Calculation of the AMMCP is motivated by high skewness in the distribution of mean claim values. We expect mean claim percentile to be predictive in our model of interaction positivity, and absolute median of the mean claim percentile to be predictive in our model of interaction neutrality.

\subsubsection{Gene degree}
We consider the directed network of genes, where edges are ordered pairs of source/target genes or gene products $(g_s, g_t)$, corresponding to the regulatory action of $g_s$ on $g_t$ (interaction $\alpha$) mentioned in a publication. The incoming degree of a gene is the number of all source gene ($g_s$) for which the focal gene is a target ($g_t$). By analogy, the outgoing degree of a gene is the number of all target genes ($g_t$) for which the focal gene is a source ($g_s$). 

\subsubsection{Interaction partition}
We combine GeneWays and Literome datasets and assign weights to the edges according to the number of claims made before time $t$. By convention, we normalize the GeneWays and Literome data so that a publication contains only one mention of any particular interaction. We then use the InfoMap and Multilabel community detection algorithms \cite{Rosvall2009} to partition the interactions into clusters using the \textit{igraph} python package\cite{igraph}. We identify dynamic community structure as a function of time with no loss of memory. For example, the interaction partition for interaction $\alpha$ in 1970 is defined for all interactions published prior to and including 1970.

We define the effective size of each interaction partition as the geometric mean $S_{\alpha} = \sqrt{S_{g_s} S_{g_t}}$, of $S_{g_s}$ and $S_{g_t}$, the sizes of communities containing $g_s$ and $g_t$ correspondingly and call this quantity Interaction Partition Size (IPS). An interaction is an edge between genes or gene products, so the two genes can either belong to the same or different partitioned communities. We call such a flag, taking value \textit{True} in former case (same partition) and \textit{False} otherwise, Interaction Partition Position (IPP).

\subsubsection{Interaction history length} We define the interaction of history length ($\Delta$ years) as the difference in years between the last and the first published claims on this genetic regulatory interaction. 

\subsection{Batch level features}

\subsubsection{Claim Popularity \& Claim Density} Let $\Cc_\alpha$ be the partition of all claims by interaction and $\Cc(t)$ the partition of claims by year of publication. Then the number of claims with respect to interaction $\alpha$ at time $t$ is $\nu_\alpha(t) = \left |\Cc_\alpha \cap \Cc(t) \right |$.

We define Claim Popularity (CP) in time window $w$ as the number of claims published in time interval $(t - w, t)$. For the strict case in which we do not consider claims published in the same year, $CP^\alpha(t, w) = \sum\limits_{t - w < t' < t} \nu_\alpha(t')$. We also consider the non-strict, right continuous case in time interval $(t - w, t]$ where we additionally consider claims published in the same year, 
$CP^\alpha_{rc}(t, w) = \sum\limits_{t - w < t' \le t} \nu_\alpha(t')$.

We define Claim Density (CD) as the number of published claims about interaction $\alpha$ within time window $w$: $\rho(t) = CP^\alpha(t, w)/(t - t_0 + 1)$ and $\rho_{rc}(t) = CP_{rc}^\alpha(t, w)/(t - t_0 + 1)$.

\subsubsection{Claim density uniformity}
We also consider Claim Density Uniformity (FLAT): for a given interaction we consider the number density of claims in time window $(t_0 - w, t)$ and $(t_0 - w, t]$ and compare it to a uniform distribution using the Kolmogorov-Smirnov test and use the KS statistic as a feature. A low KS statistic corresponds to a ``flat'' distribution of claims in time.

\subsubsection{Journal quality} To measure journal quality (JQ) we use the article influence metric \cite{West2010-vn}, which draws on a recursively weighted centrality of each article in terms of article citations, computing using the eigenvector of the article citation matrix. We apply this to our version of the Web of Science database (containing 57M unique publications) to derive the ranking of journals from 1990 through 2014 (only 4\% of GeneWays and 4\% of Literome claims, intersected with LINCS L1000, are dated before 1990). We use the prior 5 year window of journals citing journals to derive the article influence metric \textit{(ai)}, computed for each journal at each time point. This calculation of journal quality drew on capabilities of the Cloud Kotta platform \cite{Babuji2016}.

\subsubsection{Affiliation rank}
In order to evaluate the affiliation rank (AR) we use the top 100 QS World University Rankings of biological sciences from 2011. From the non-zero intersection of both GeneWays and Literome claims with LINCS L1000 approximately 21\% have the affiliation in the top 100 University Ranking of biological sciences, whereas the fraction of claims for which no affiliation could be identified from the data available is only 3\% for GeneWays and 4\% for Literome.

\subsubsection{Citation count}
From the Web of Science we obtain current citation counts for 97\% of the publications of interest from GeneWays and Literome datasets. The mean citation count for GeneWays dataset is approximately 57, while for Literome it is 62. Along with the raw citation count, we also fit the citation history for each publication to a lognormal distribution:

\bal
P(C_t) \propto \frac{1}{t \sigma\sqrt{2\pi}} e^{-\frac{(\ln t - \mu)^2}{2\sigma}}
\eal

and use parameters of that log-normal distribution $\mu$ and $\sigma$ as features. The fit allows us to estimate the projected total citation count $A$, which we also consider as a feature.

\subsubsection{Herfindahl index of affiliations and authors}
We calculate the Herfindahl index of affiliations and authors to capture the inequality of attention and associated dependence between articles published by the same authors at the same institutions. We disambiguate affiliations using the hierarchical distance technique \cite{Belikov2020}, which takes advantage of the trace of the ordered matrix of institution terms (e.g., to distinguish between Washington University and the University of Washington). We then compute the relative weight for each affiliation $f_{j\alpha}(t)$ and interaction $\alpha$ at time $t$ as the sum over contribution for all preceding time.

Let $A_i$ be the set of affiliated institutions listed in publication $i$. We use the simplifying assumption that each publication $i$ carries weight $1$, which is divided equally between all the unique identifiable affiliations of authors. This is useful because the Web of Science did not explicitly link affiliations to authors for 20th Century publication data. We define the contribution of a publication to the weight of an affiliation as:

\bal
w_{ij\alpha}(t) =
    \begin{cases}
    \frac{1}{|A_i(t)|}, \quad j \in A_i(t)\\
    0, j \not\in A_i(t)
    \end{cases}
\eal

At a given time the weight of a affiliated institution $j$ in the history of an interaction $\alpha$ is the sum over the weights from relevant publications $w_{j\alpha}(t) = \sum\limits_{t'<t,\, i : j\in A_i} w_{ij\alpha}(t')$. We define normalized weight (NW) $f_{j\alpha}(t)$ of affiliation $j$:

\bal
f_{j\alpha}(t) = \frac{w_{j\alpha}(t)}{\sum_j w_{j\alpha}(t)},
\eal

and normalized Herfindahl index (NHI) $nhi_\alpha(t)$ as
\bal
hi_\alpha(t) = \sum f^2_{j\alpha}(t); \qquad
nhi_\alpha(t) = \frac{hi_\alpha(t) - 1/K_\alpha(t)}{1 - 1/K_\alpha(t)},
\eal

where $K_\alpha(t)$ is the number of affiliations associated with interaction $\alpha$ up to time $t$:

\bal
K_\alpha(t) = \left|\bigcap\limits_{t' < t,\, i \in B_\alpha} A_i(t')\right|.
\eal

In a comparable way, we derive relative weights and the normalized Herfindahl index for authors as well. These indices capture how much a few institutions and authors are responsible for research attention to a claim. If the Herfindahl indices are low, then attention to the claim is spread across many institutions and authors.    

In the correlation plots (Fig. \ref{fig:correlations:cor:batch}) normalized weights and normalized Hefindahl index are denoted NW\_affs and NHI\_affs and correspondingly NW\_authors and NHI\_authors for affiliated institutions and authors: NW is defined at the claim level, while NHI - at the batch level.

\subsubsection{Dependency index}
We define two new network measures to directly measure their indirect dependency upon one another. These naturally relate to, but expand on the Herfindahl index measures described above. Our dependency indices are measured at the level of an individual claim, and also at the level of a batch of claims. Our claim-level dependency index measures how well each node from one component of the network connects to others of the same type through nodes in the other component. If genetic regulatory interaction claims have high dependency on one another through the same shared authors, affiliated institutions, or shared references, then they are not independent from one another. Our batch-level dependency index measures how one entire component -- one type of nodes -- in a bipartite graph is connected through the other. As with the claim-level dependency-index, but at the collective level of all claims within a batch, more dependency between claims means less independence.   

Consider a bipartite graph $G = (U, V, E)$, where the only possible edges connect nodes from set $U$ to set $V$. In our case, $U$ represents the set of publications, and $V$ -  the set of affiliations, authors or referenced articles. We characterize the connectedness of a bipartite graph $G$ by a single number, not by the relative number of edges $\frac{|E|}{|U||V|}$ as a metric, but rather to encode the shape of the degree distribution of $U$ with a single number.

We propose to use the relative $\lambda$th moment in $f$-percentile as a metric describing how well $U$ is supported on $V$ and call it the batch-level dependency index. Let $d_i$ be the degree of vertex in $V$. There are $|V|$ such vertices. We denote the sum over the $f$ percentile of greatest degrees, such that $d_i > k_f$, where $k_f$ is found from the equation $\frac{\sum_i \mathds{1}\{d_i > k\}}{ |V|} = f$, where $\mathds{1}$ is the indicator function, by a prime ($'$).
And so the $\lambda$th moment in $f$-percentile is 

\bal
\langle d^\lambda \rangle_f = \frac{\sum' d^\lambda_i}{f|V|}
\eal

To normalize the degree moment we define the fractional dependency index as the ratio of $\langle d^\lambda \rangle_f$ to the maximum degree to the power $\lambda$: $d^\lambda_{max} = |U|^\lambda$.

\bal
\sigma_{f,\lambda}(G) = \frac{\langle d^\lambda \rangle_f}{|U|^\lambda}
\eal

For example, consider three research claims published across three separate articles: $c_1$ written by Alice and Bob, $c_2$ by Alice and John and $p_3$ by Alice, John and Mary. In this case authors play the role of the $V$ set, with Alice, John, Mary and Bob having correspondingly degrees 3, 2, 1, 1. If $f$ is set to 0.5 and $\lambda=1$, $\langle d \rangle_{0.5} = \frac{2+3}{2}$. With $|U| = 3$,  $\sigma_{0.5, 1} = 0.83$.

We define the dependency index for $u \in U$ to all other vertices in $U$ through shared connections to vertices in $V$ in the following manner: let $d_v$ be the degree of vertex $v$ in $V$.

\bal
\mbox{aff}_G(u) = \frac{\sum\limits_{v: \exists (u,v)} (d_v - 1)}{d(u)(|U| -1)}
\eal

It follows from our definitions that $0 \le \sigma_{f,\lambda}(G) \le 1$ and $0 \le \mbox{aff}_G(u) \le 1$.

For each interaction $\alpha$ and time $t$ at which the claims on interaction $\alpha$ were made, we consider a subset of publications made during time interval $(t-k, t]$, which forms $U$, then set $V$ consists of the corresponding affiliations, authors or references. We also consider the case of unbounded past-looking windows $(-\infty, t]$.

For the support metric we use $\lambda=2$, $f=0.2$ for references and $f=0.5$ for affiliations and authors, as there are many more distinct references than affiliations or authors.

In the correlation plot (Figs. \ref{fig:correlations:cor:claim}) dependency indices are denoted as CDEP and BDEP respectively. As mentioned above, CDEP is defined at the claim level, while BDEP is defined at the batch level. 

\subsubsection{Claim community number (CCN), community size (CSI) and community share (CSA)}
The number of claim communities captures the separated, largely independent research efforts, defined across authors, institutions, and motivating referenced articles. Community size measures the number of researchers within the connected community surrounding a given claim, and community share the proportion of all claims within the batch are represented by those within that community. See

As in the case of our dependency indices, we consider a bipartite graph where set $U$ consists of publications and set $V$ affiliations, authors or references. We identify communities using the Infomap method in the projection of $G(U,V, E)$ on $U$ with weights of edges defined as 
\bal
w_{ab} = \frac{|Nei(a) \bigcap Nei(b)|}{ |Nei(a) \bigcup Nei(b)|}, \quad Nei(a) = \{ v \in V | (a, v) \in E\},
\eal

where $Nei(a)$ is the set of neighbors of $a$.
In Figs.\ref{fig:correlations:neut} and \ref{fig:correlations:pos} we present the correlations between the interaction level features and interaction neutrality $\pi^\alpha_0$ and interaction positivity $\pi^\alpha_+$ respectively. In Figs.\ref{fig:correlations:cor:batch} \ref{fig:correlations:cor:claim} we present
the correlations of batch and correspondingly claim level features with claim correctness $y^\alpha_i$. 

\section{Experimental setup}

\subsection{Sampling}

Here we detail our sampling techniques and discuss the use and role of predictive models in our analysis. We chose logistic regression and random forest models for our prediction tasks. Logistic regression enables interpretability, while random forests brings us closer to maximal predictive performance.

In order to evaluate out-of-sample predictions and estimate the distributions of metrics of interest (such as area under the curve (AUC) for receiver-operator characteristic (ROC)) for interaction level models, we generate 20 3-fold random samples by genetic regulatory interaction. This amounts to having 60 model-level datapoints, which form the basis of each point or line of performance in Fig. 2 and 4.

For models of claim correctness and interaction positivity (which claim correctness), we need to alter the sampling procedure in order to avoid an artificial inflation of performance. Claims are conditional on interactions, and so if they are sampled randomly, the same interaction may have claims in both test and the training samples, falsely boosting the appearance of performance.
We employ the following technique: we array interactions according to their popularity distribution function, which we model with Zipf's law (a discrete power law distribution). We then randomly sample claims associated with interactions from the training sample and attribute the remainder to the test sample.
In this way, in the evaluation of the positivity model we are not testing it on claims that rely on interactions seen during the training phase.

\subsection{Model selection}

For random forest models, we vary the depth of the trees from which they ensemble, the minimum leaf size and the number of estimators to find a combination that optimizes the AUC of the test set. We find that unlike the AUC on the training set, which increases with model complexity, the AUC of the test remains approximately flat as a function of model parameters. For all experiments we fix the tree depth to 2, number of estimators to 100 and the minimum number of samples in the leaf to 2\% of the sample size. For logistic regression model we use an L1 penalty to introduce sparsity and vary the penalty coefficient until there are exactly 5 non-zero coefficients. For all AUC plots, the error bands denote 95 percent confidence region for the mean AUC  Figs. [\ref{fig:complexity_neutral_auc}-\ref{fig:complexity_claims_auc}].

We conclude that we are generally in regime of high variance and therefore it is desirable to use models of low complexity, and apply such techniques as bagging.

While to answer the question if an interaction is positive and not negative we focus mainly on the Bayesian approach utilizing claim correctness, we also evaluated a model based on purely interaction level features, derived as for the model claim neutrality from the network of sociological claims. The performance of this model is comparable to the one obtained using Bayes law, as can be seen in Fig. \ref{fig:complexity_posneg_auc}. The importances for this model are shown in \ref{fig:imps_posneg} 

\subsection{Evaluation of feature importance}

In evaluation of interaction neutrality and positivity models,
we calculate feature importances\footnote{ We use impurity-based feature importances: the total reduction of the criterion brought by that feature, also known as the Gini importance.} of the random forest model and the coefficients of the logistic regression as simple averages over samples

For the model of claim correctness, we aggregate features into feature families as presented in Table \ref{table:families2}.
For each sample and family we sum the importances for the random forest models and the coefficients for the logistic regressions and then compute the mean over samples.

We use the same procedure for random forest and logistic regression to describe the robust importance of feature families. In the case of logistic regression, it also sheds light on the sign of the effect.

\begin{table}
    \small
    \begin{centering}
    \begin{tabular}{ccp{8cm}}

        \toprule
        feature family & number of features & description \\
        \hline\hline
        
        MCP & 1 & mean claim percentile \\ 
        AMMCP & 1 & absolute median mean claim percentile \\
        affiliation count & 1 & number of affiliations per publication \\
        author count & 1 & number of authors per publication \\
        CDEP & 12 & claim-level dependency indices for affiliations, authors and references times 4 windows sizes \\
        JQ & 1 & journal quality \\
        AR & 1 & affiliation ranking \\
        popularity (CP), density (CD) & 16 & popularity and claim densities, defined by strict and non strict right inequality, times 4 window sizes \\
        citations & 8 & citation metrics: number of citations in the first 3 years, 3 parameters of the lognormal fit and their logarithms, flat whether the lognormal fit was successful \\
        degrees & 4 & source degree in/out, target degree in/out \\ 
        $\Delta$ year & 1 & time difference between the first and the last available publications on interaction in years \\
        FLAT & 8 & uniformity of popularity, defined by strict and non strict right inequality, times 4 window sizes \\
        \bottomrule\bottomrule
        \hline
    \end{tabular}
\end{centering}
\label{table:families}
\end{table}

\begin{table}
    \small
    \begin{centering}
    \begin{tabular}{ccp{8cm}}
        \toprule
        feature family & number of features & description \\
        \hline\hline
        IPS & 18 & interaction partition size \footnote[1]{We consider the following set networks/community detection algorithms defined (infomap, dir, unweighted), 
        (infomap, dir, weighted),
        (infomap, undir, unweighted),
        (infomap, undir, weighted),
        (ml, undir, unweighted),
        (ml, undir, weighted)} \\
        IPP & 6 &  interaction partition position \\
        NW & 2 & normalized weight for authors and affiliations \\
        NHI & 2 & normalized Herfindal index for authors and affiliations \\
        CCN & 12 & claim community number, 4 windows, size for authors, affiliations references \\
        CSI & 12 & community size, 4 windows, size for authors, affiliations references \\
        CSA & 12 & community share, 4 windows, size for authors, affiliations references \\
        CDEP & 12 & claim-level dependency index, 4 windows, size for authors, affiliations references \\
        BDEP & 12 & batch-level dependency index, 4 windows, size for authors, affiliations references \\
        time & 2 & year off and year off$^2$, time in years between the current publication and the first publication on a given interaction \\
        \bottomrule\bottomrule
        \hline
    \end{tabular}
\end{centering}
\caption{Feature families.}
\label{table:families2}
\end{table}

\section{Policies}

Based on our analyses, the striking importance of community number in predicting robust findings (Fig. 3) and the uneven distribution of research attention (Fig. 1), we consider two policies that could increase the overall identification of the sign of interaction as measured by the AUC of the model evaluated on our test set.

In the first virtual experiment, we divide interactions in the test set by the number of batch communities estimated according to authorship. In the second experiment, we evaluate the AUC on hypothetical, historical sub-samples, for which we artificially vary the number of claims to alter to the claim distribution across all genetic regulatory interactions. 

Examples of such subsamples are presented in Fig. \ref{fig:len_distr} and 
\ref{fig:len_distr_ig}.

For the second experiment we estimate AUCs and information gain (IG), defined 
$$
IG = \mbox{ent}(p^{(0)})- \frac{1}{k} \sum\limits_{\alpha=1}^{\alpha=k} \mbox{ent}(p^\alpha)
$$. 
Here $p^\alpha$ is the estimated distribution of interaction $\alpha$ and $ent(p^\alpha) = -\sum p^\alpha_i \log p^\alpha_i$ and $p_0$ is a non-informative Bernoulli with parameter $\nu = 0.5$
We note that flatter distributions corresponds to higher AUCs and greater average information gains in a statistically significant way, cf. Figs. \ref{fig:len_distr} and \ref{fig:len_distr_ig}.

\clearpage

\clearpage
\begin{figure}
  \begin{minipage}[b]{0.5\linewidth}
    \centering
    \includegraphics[width=.9\linewidth]{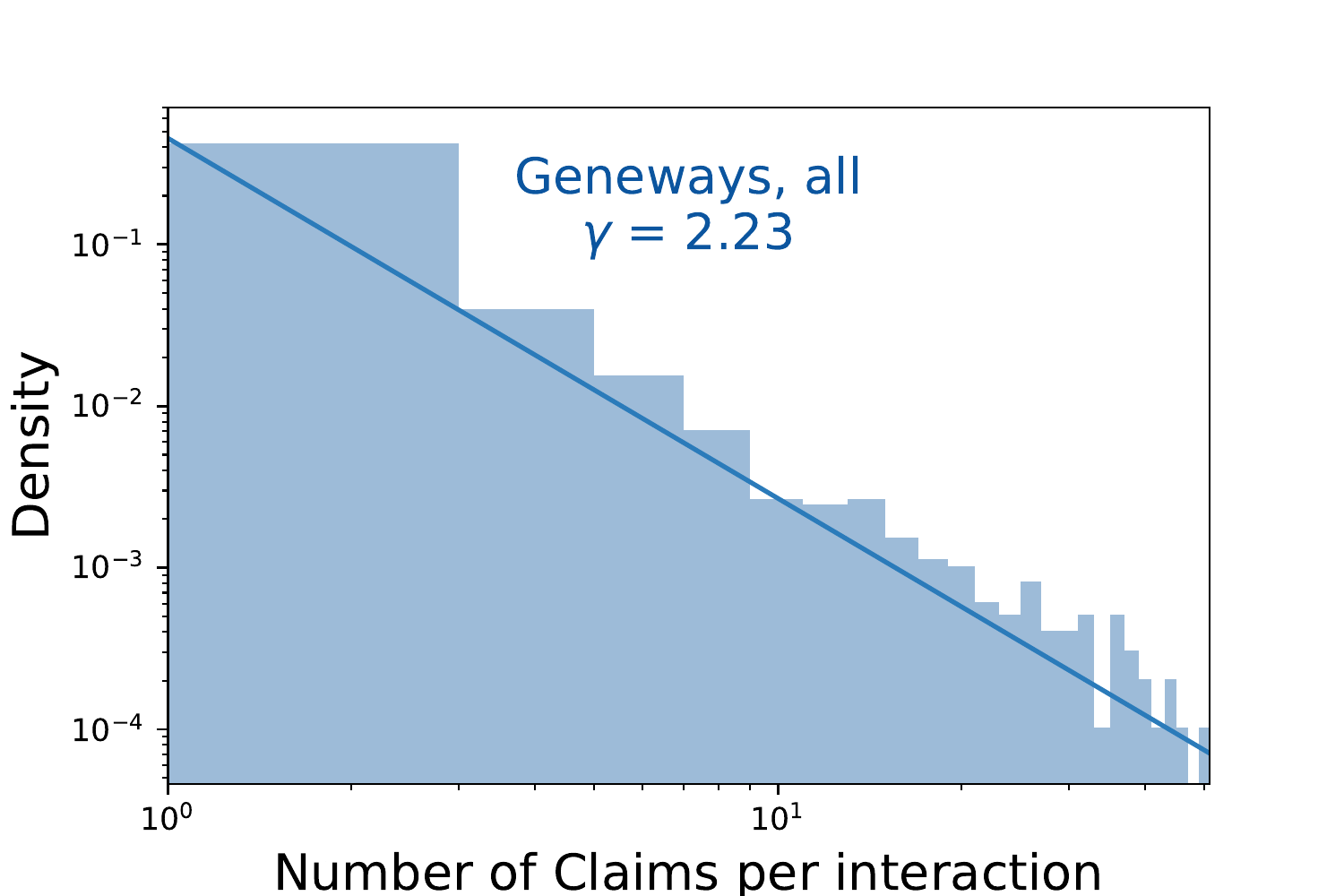} 
    \vspace{4ex}
  \end{minipage}
  \begin{minipage}[b]{0.5\linewidth}
    \centering
    \includegraphics[width=.9\linewidth]{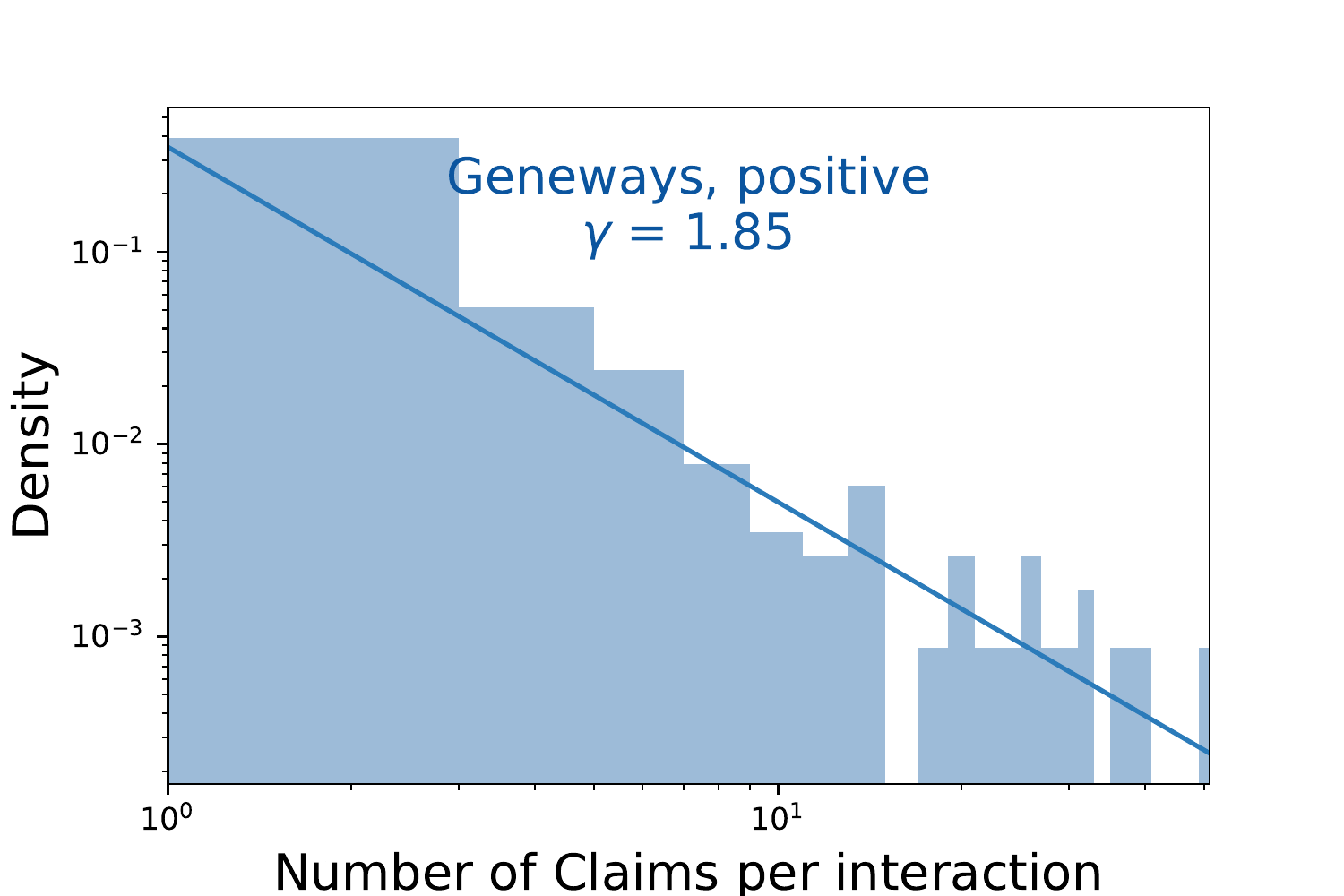} 
    \vspace{4ex}
  \end{minipage} 
  \begin{minipage}[b]{0.5\linewidth}
    \centering
    \includegraphics[width=.9\linewidth]{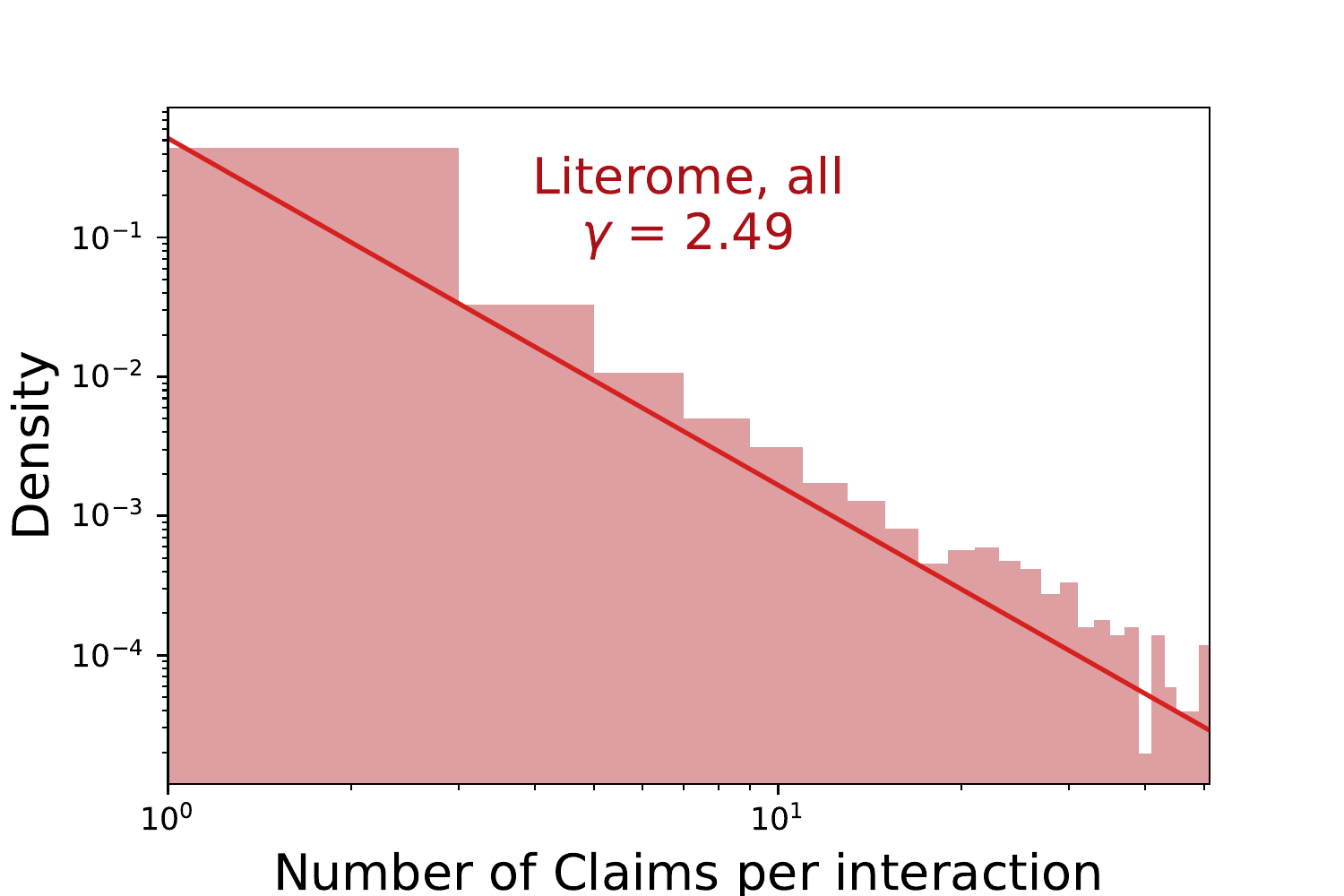} 
    \vspace{4ex}
  \end{minipage}%
  \begin{minipage}[b]{0.5\linewidth}
    \centering
    \includegraphics[width=.9\linewidth]{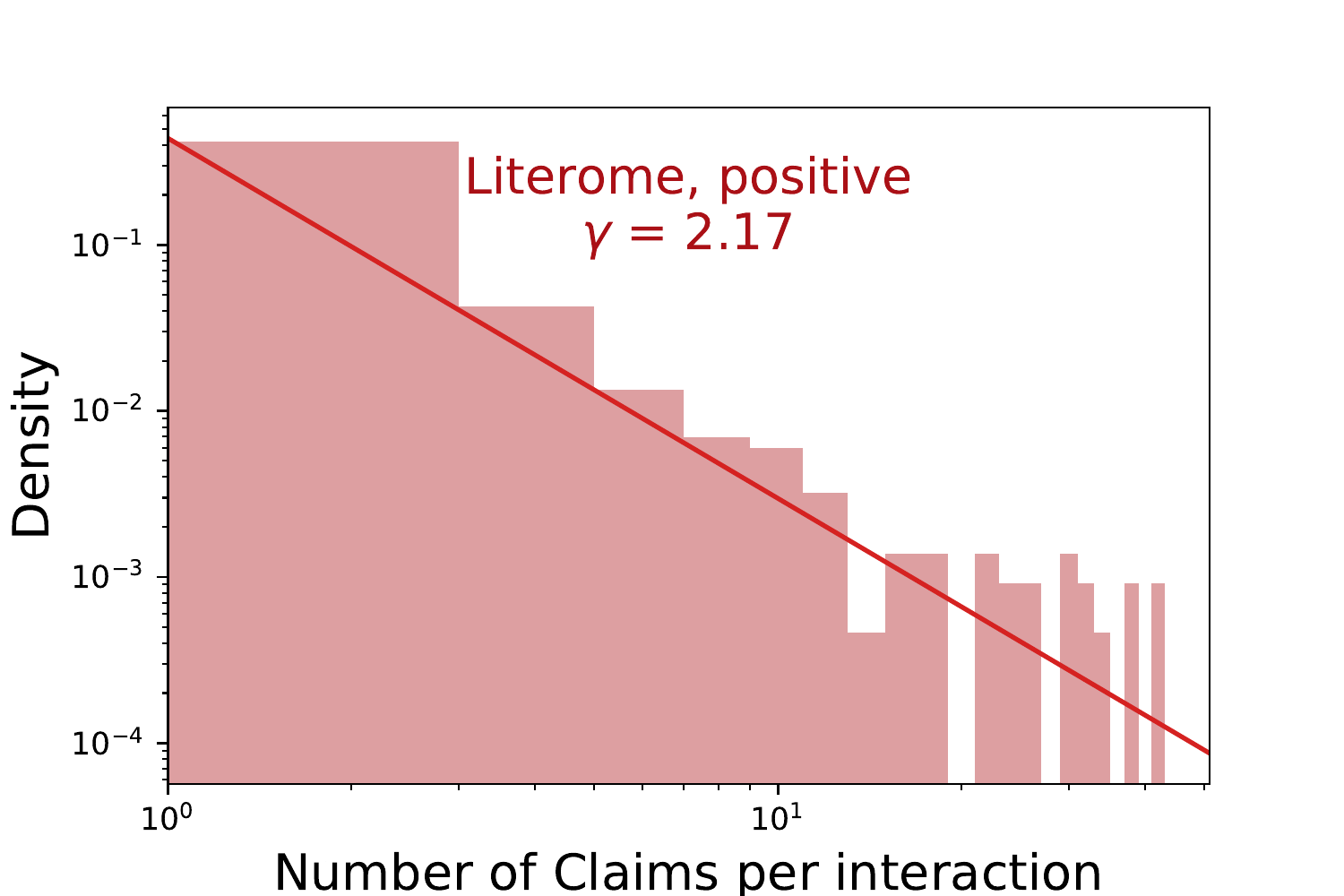} 
    \vspace{4ex}
  \end{minipage} 
    \caption{Claim number density for GeneWays (top panel) and Literome (bottom panel) all interaction (left) and selected positive/negative interactions.} 
  \label{fig:ndensity} 
\end{figure}

\clearpage
\begin{figure}
    \centering
    \minipage{0.5\textwidth}
        \centering
        \includegraphics[width=0.9\textwidth]{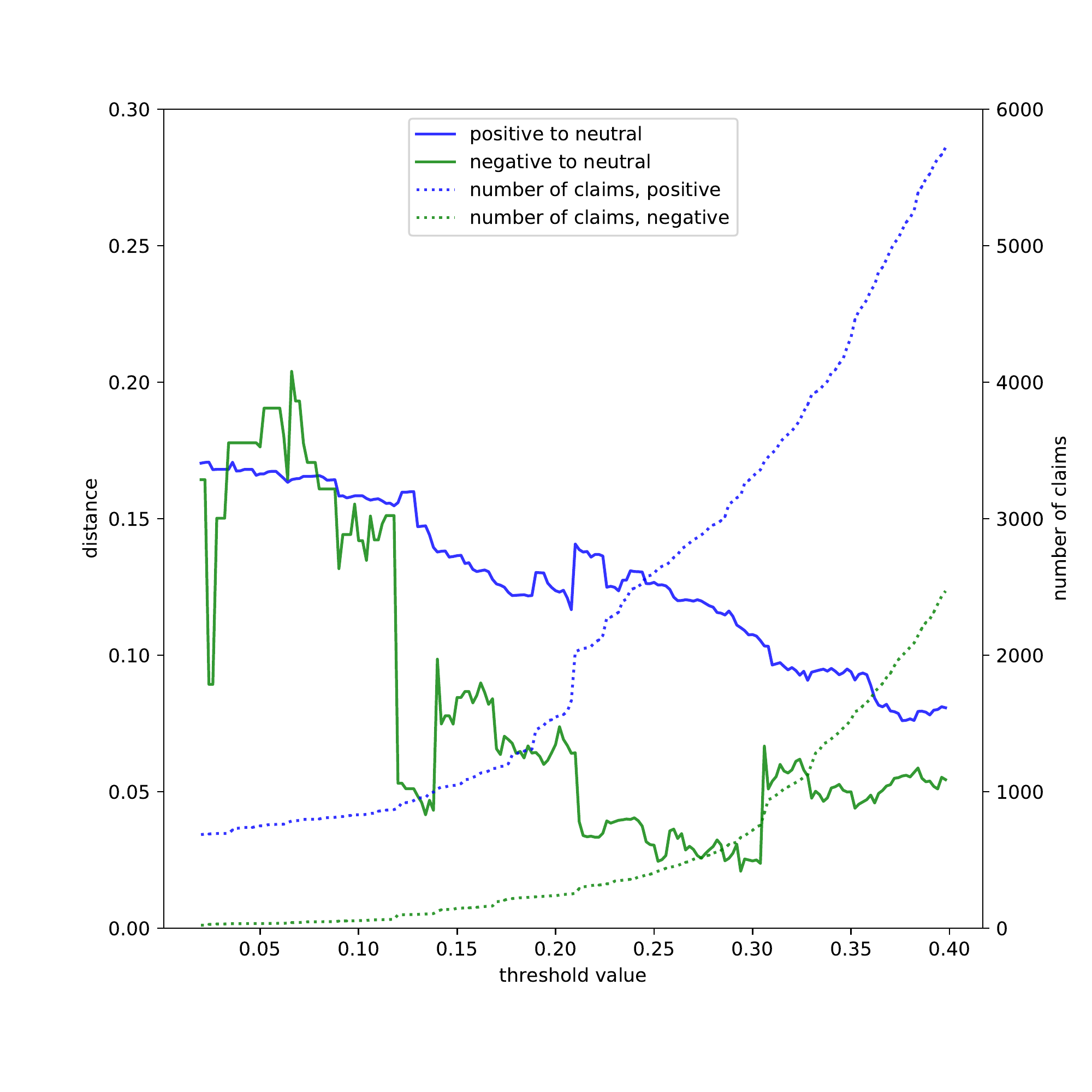}
    \endminipage\hfill
    \minipage{0.5\textwidth}
        \centering
        \includegraphics[width=0.9\textwidth]{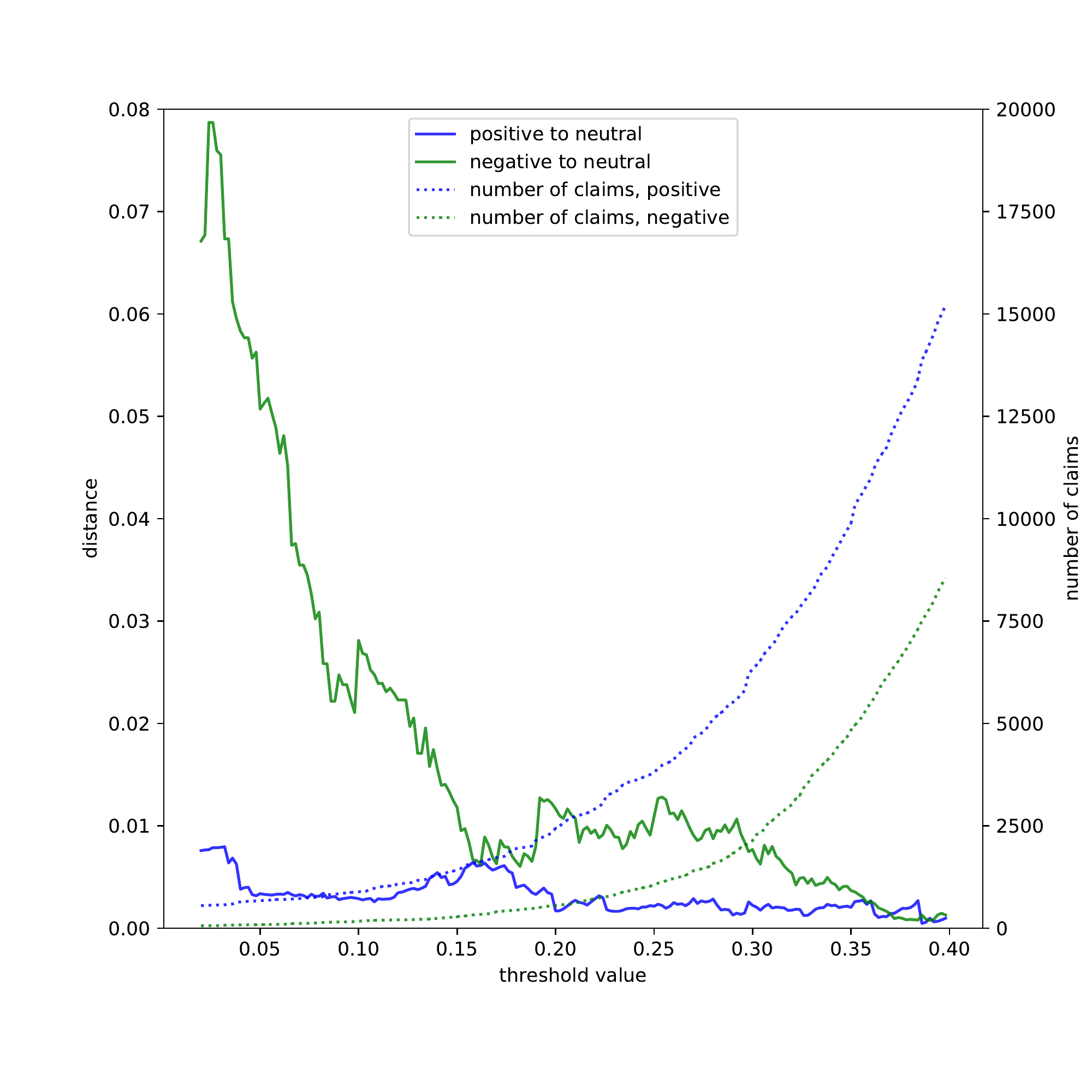}
    \endminipage
    \caption{Distance between the classes of neutral $\Cc_0$ and negative $\Cc_-$ interactions $W (g_-, g_0, \theta_-)$ (solid green line), number of of claims on the negative class $\Cc_-$ $n_-(\theta_-)$ (dotted green line), as a function of $\theta_-$; distance between the classes of neutral $\Cc_0$ and positive $\Cc_+$ interactions $W (g_+, g_0, \theta_+)$, solid blue line, $n_+(\theta_+)$ (dotted blue line) as a function $\theta_+$ for GeneWays (left) and Literome (right).}
    \label{fig:data_thr}
\end{figure}

\clearpage
\begin{figure}
    \centering
    \minipage{0.5\textwidth}
        \centering
        \includegraphics[width=1.0\textwidth]{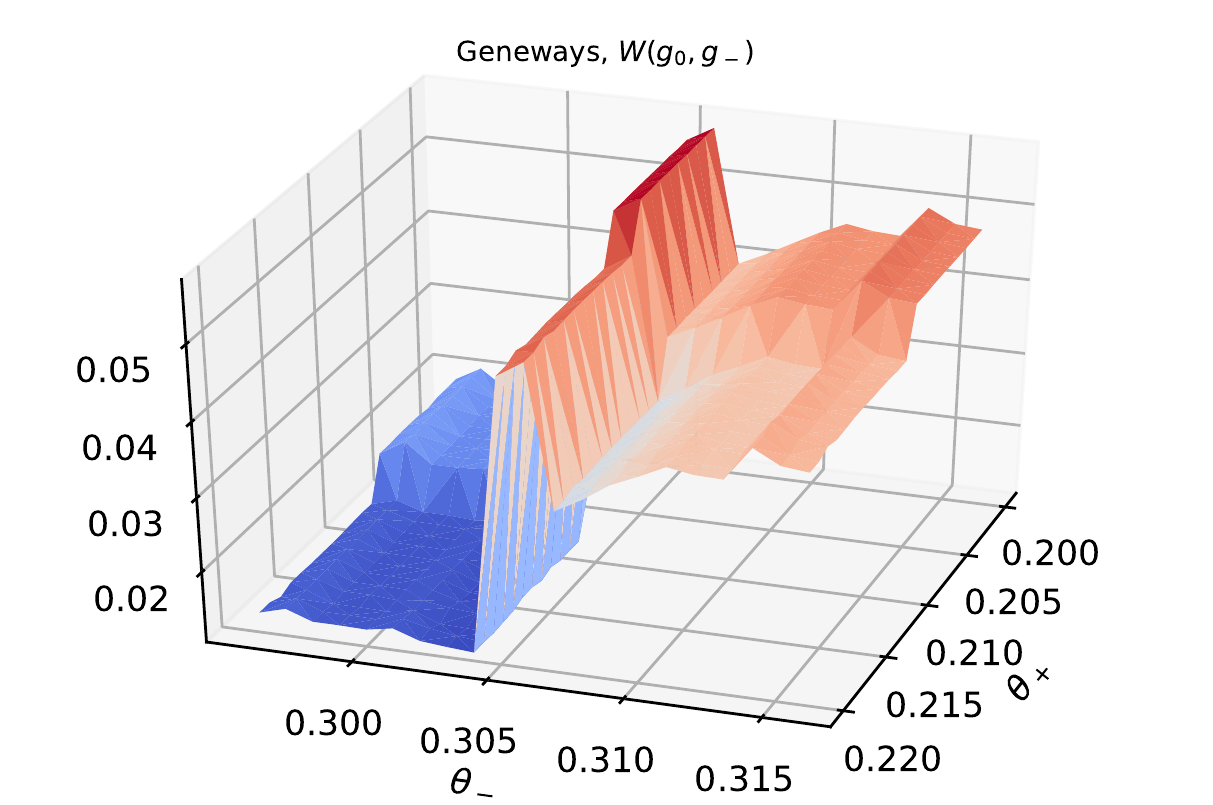}

    \endminipage \hfill
    \minipage{0.5\textwidth}
        \centering
        \includegraphics[width=1.0\textwidth]{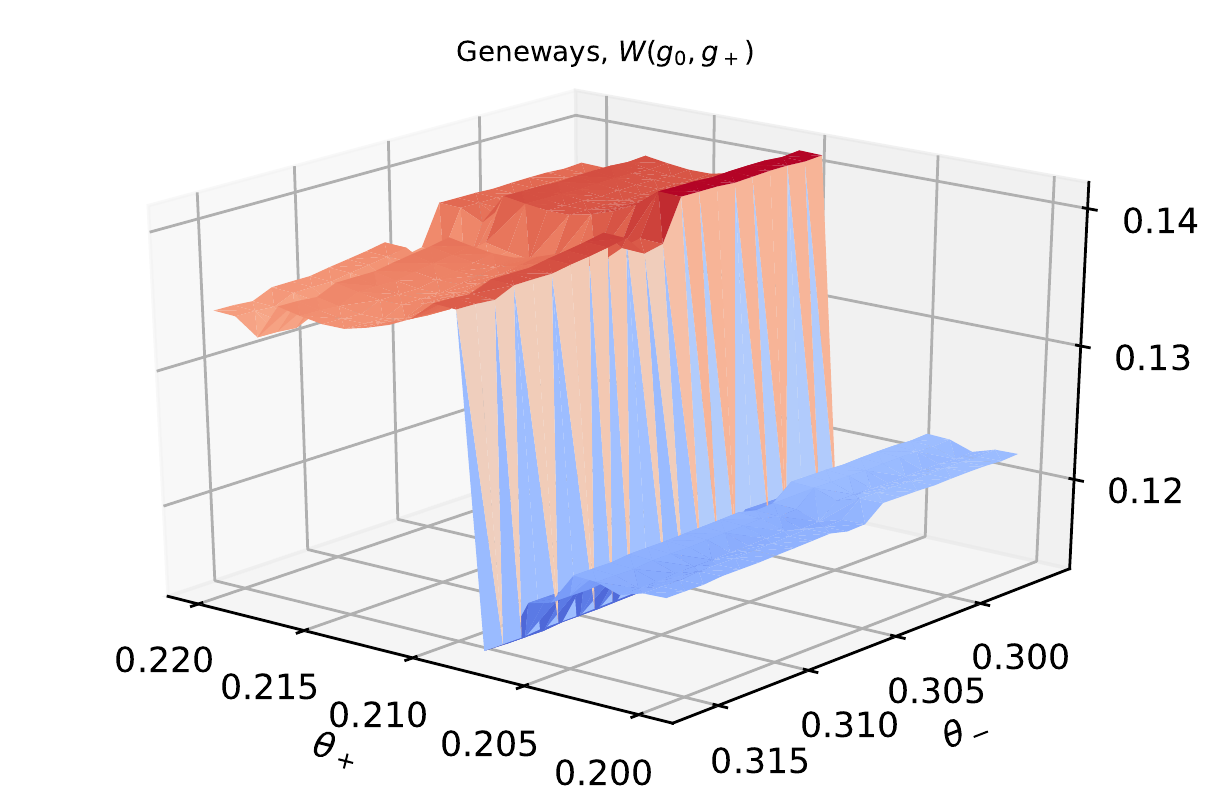}
    \endminipage \hfill
        \caption{GeneWays. Left: distance between neutral $\Cc_0$ and negative $\Cc_-$ interactions $W(g_0, g_-, \theta_- \theta_+)$; right: distance between neutral $\Cc_0$ and positive $\Cc_+$ interactions 
        $W(g_0, g_+, \theta_-, \theta_+)$.}
    \label{fig:threshold_3d}
\end{figure}

\clearpage
\begin{figure}
\includegraphics[width=1.\textwidth]{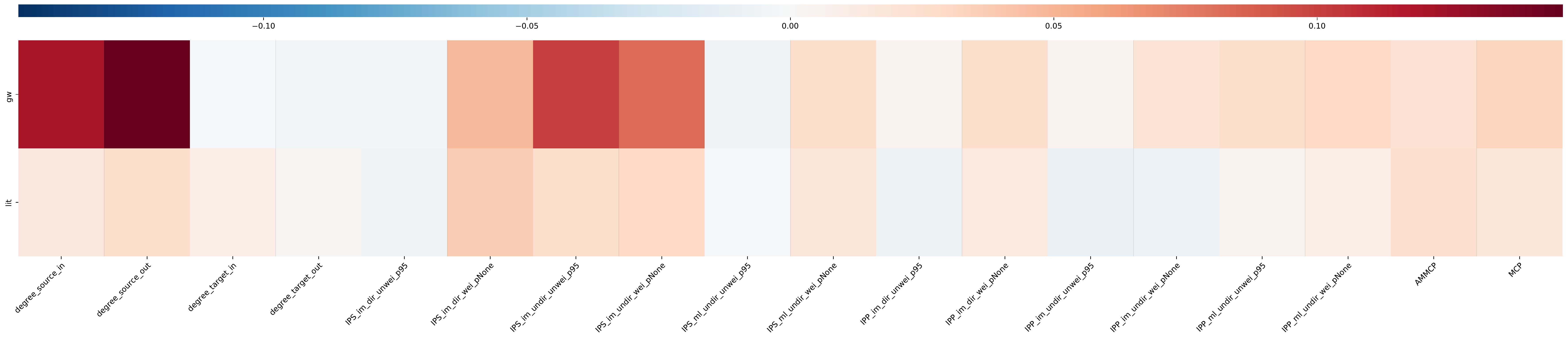}
\caption{Pearson correlation heat map vector between $\pi^\alpha_0$ and interaction level features for GeneWays (top panel) and Literome (bottom panel).}
\label{fig:correlations:neut}
\includegraphics[width=1.\textwidth]{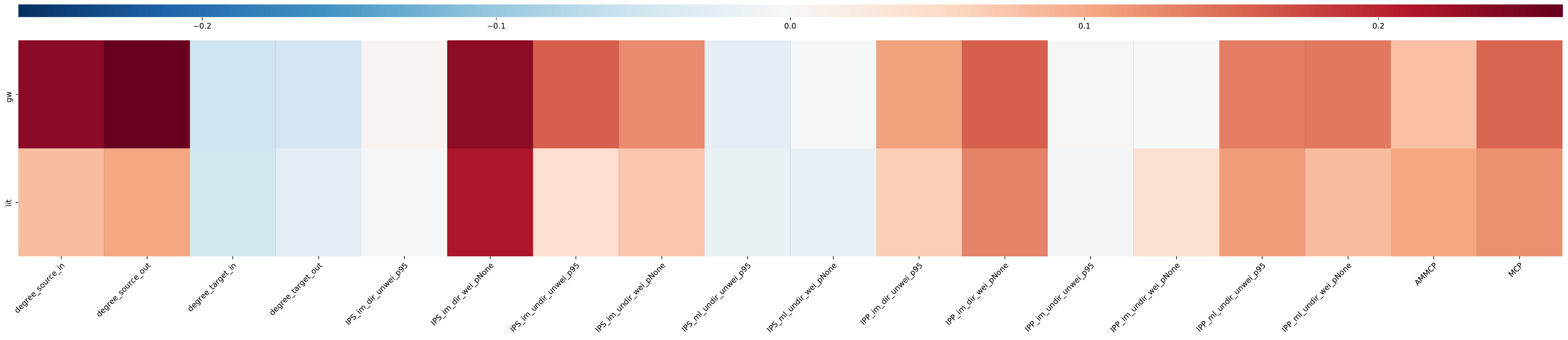}
\caption{Pearson correlation heat map vector between $\pi^\alpha_+$ and interaction level features for GeneWays (top panel) and Literome (bottom panel).}
\label{fig:correlations:pos}
\end{figure}

\begin{figure}
\includegraphics[width=1.\textwidth]{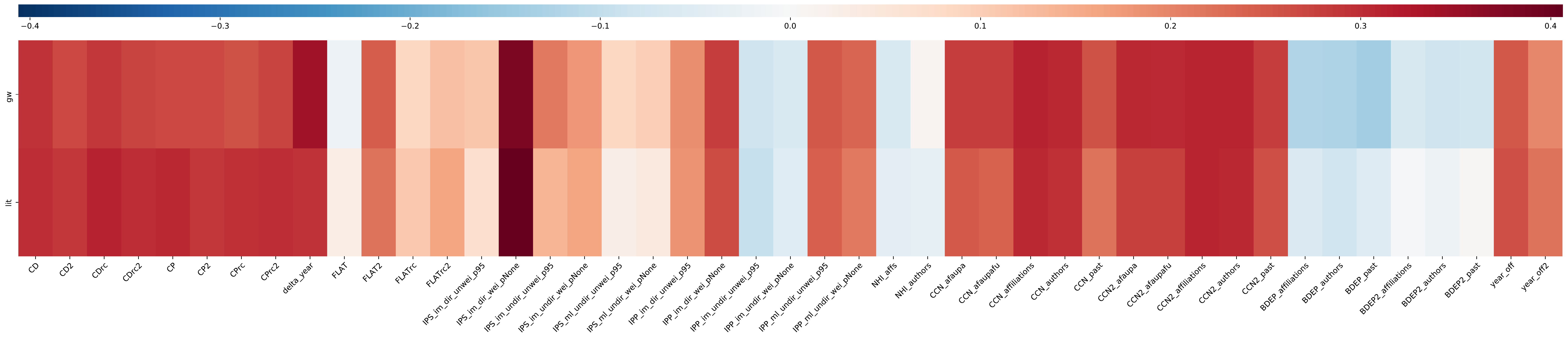}
\caption{Pearson correlation heat map vector between claim correctness $y^\alpha_i$ and batch level features for GeneWays (top panel) and Literome (bottom panel).}
\label{fig:correlations:cor:batch}
\includegraphics[width=1.\textwidth]{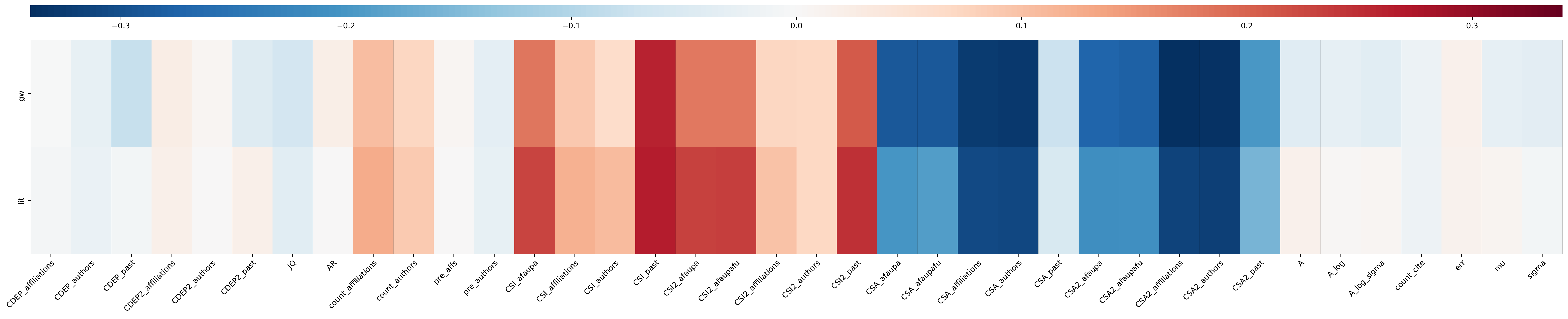}
\caption{Pearson correlation heat map vector between claim correctness $y^\alpha_i$ and claim level features for GeneWays (top panel) and Literome (bottom panel).}
\label{fig:correlations:cor:claim}
\end{figure}

\clearpage
\begin{figure}
\includegraphics[width=1.0\textwidth]{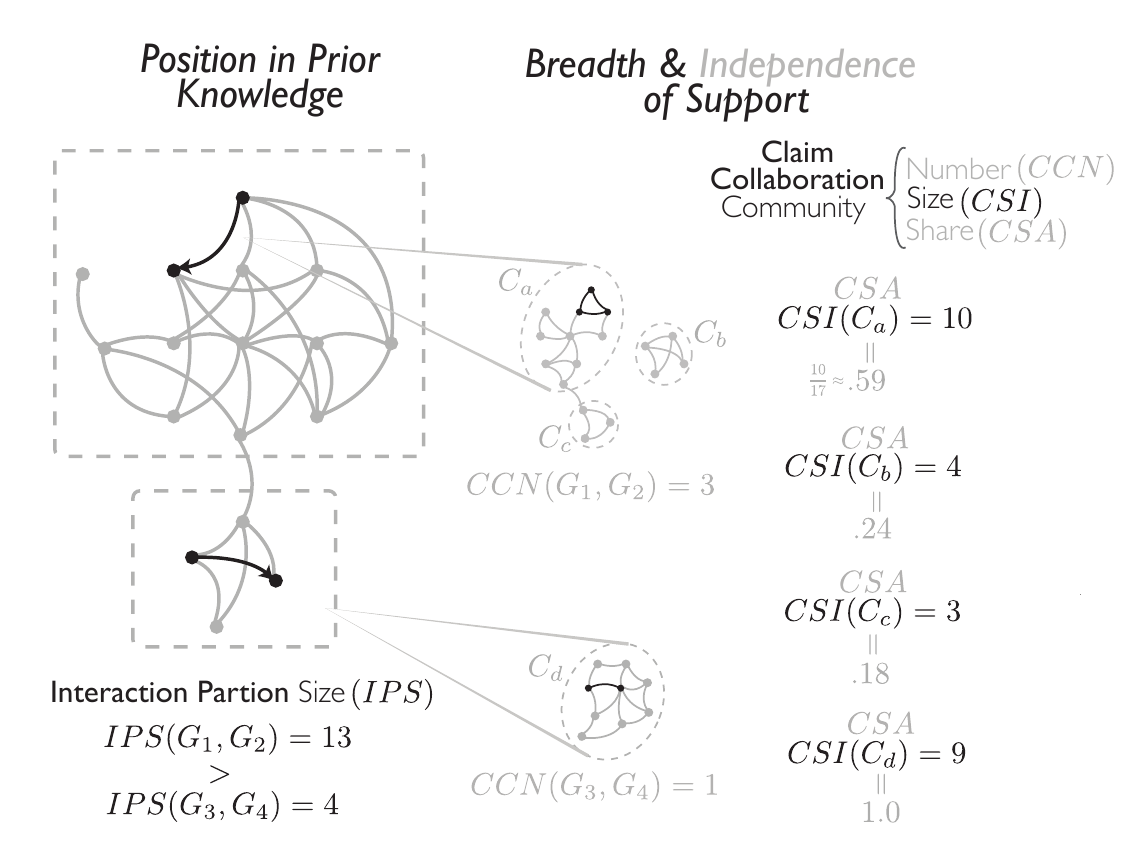}
\caption{Pictoral example of selected interaction and claim variables}
\label{fig:pictoral}
\end{figure}

\clearpage
\begin{figure}
\centering
    \mbox{
        \includegraphics[width=0.33\linewidth,clip=]{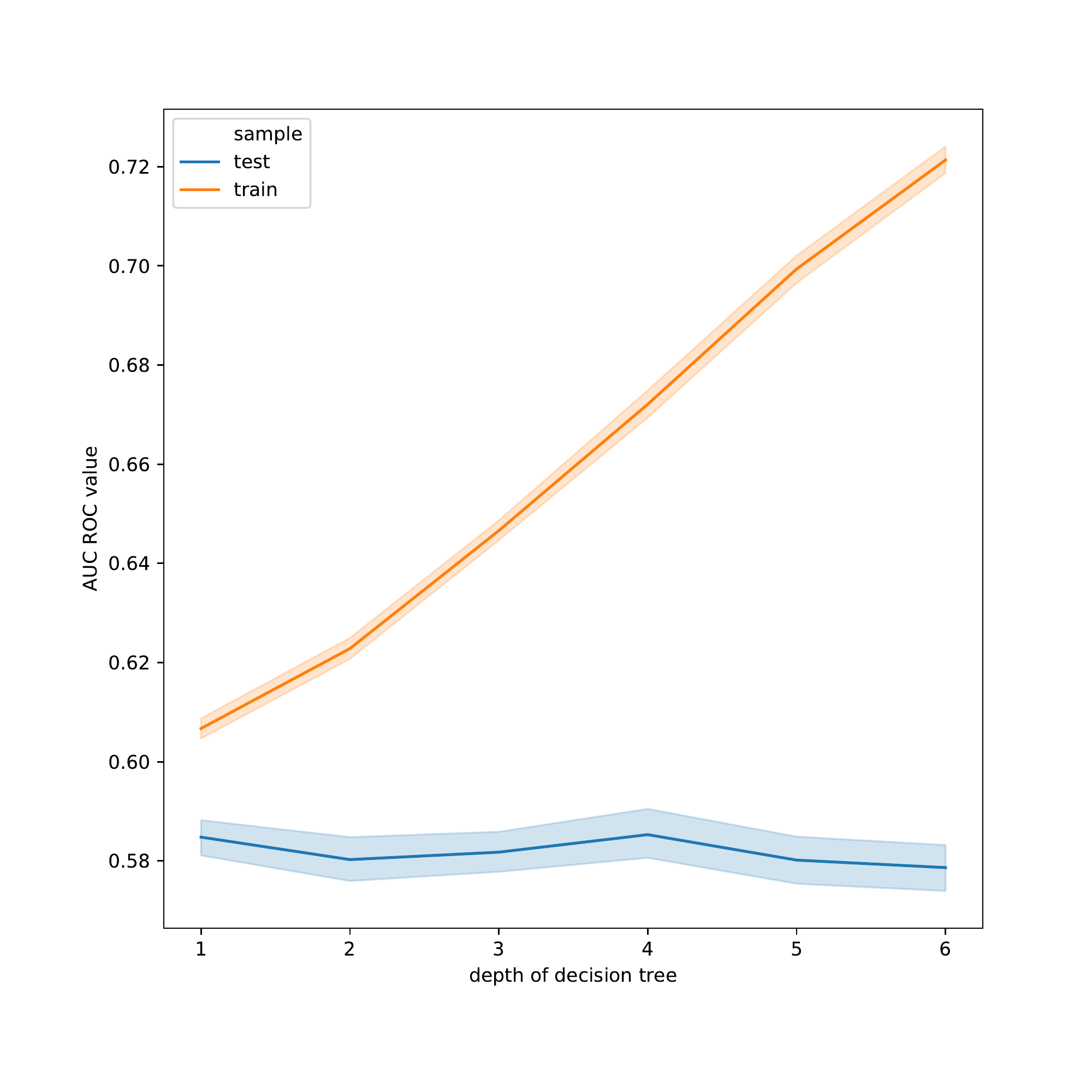}
        \includegraphics[width=0.33\linewidth,clip=]{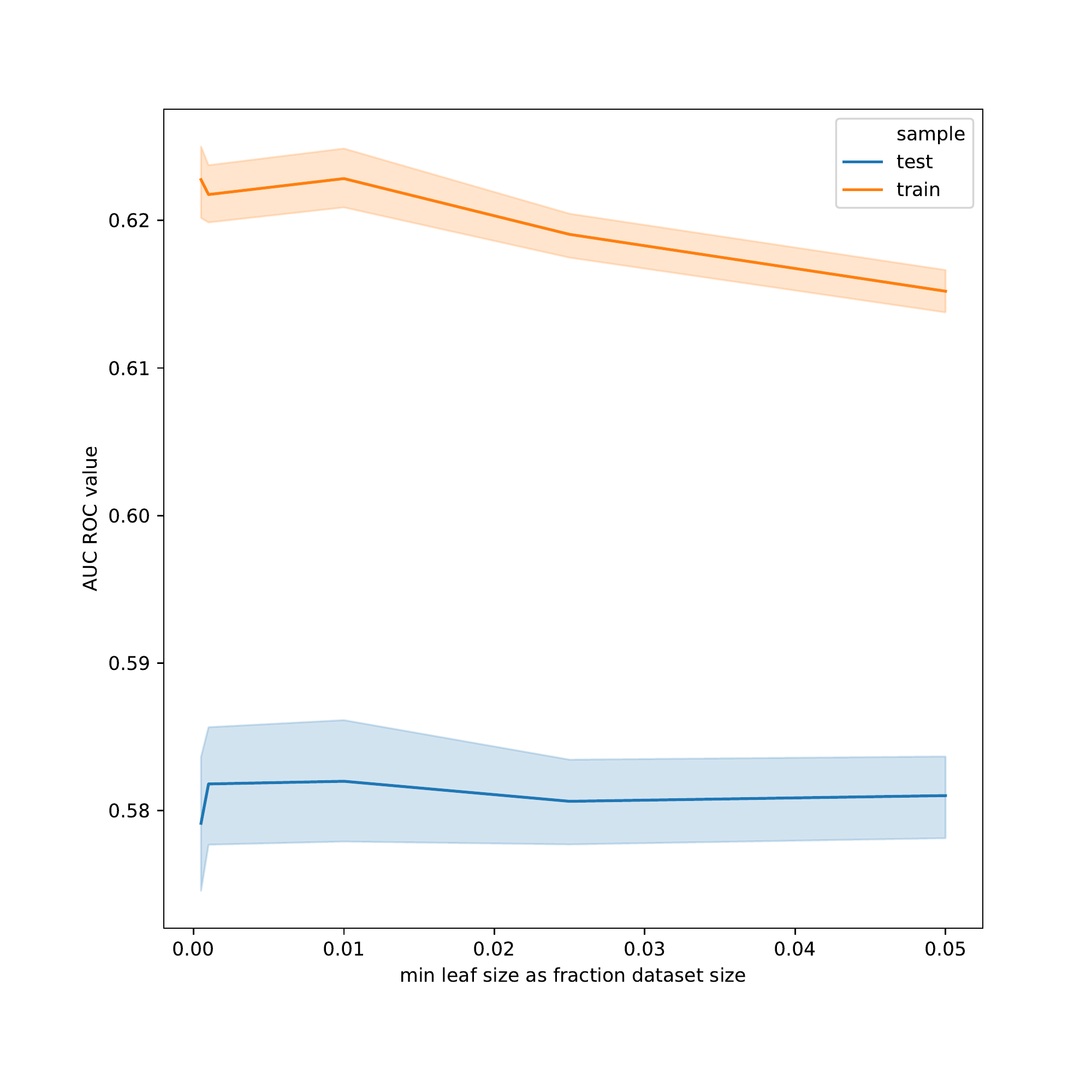}
        \includegraphics[width=0.33\linewidth,clip=]{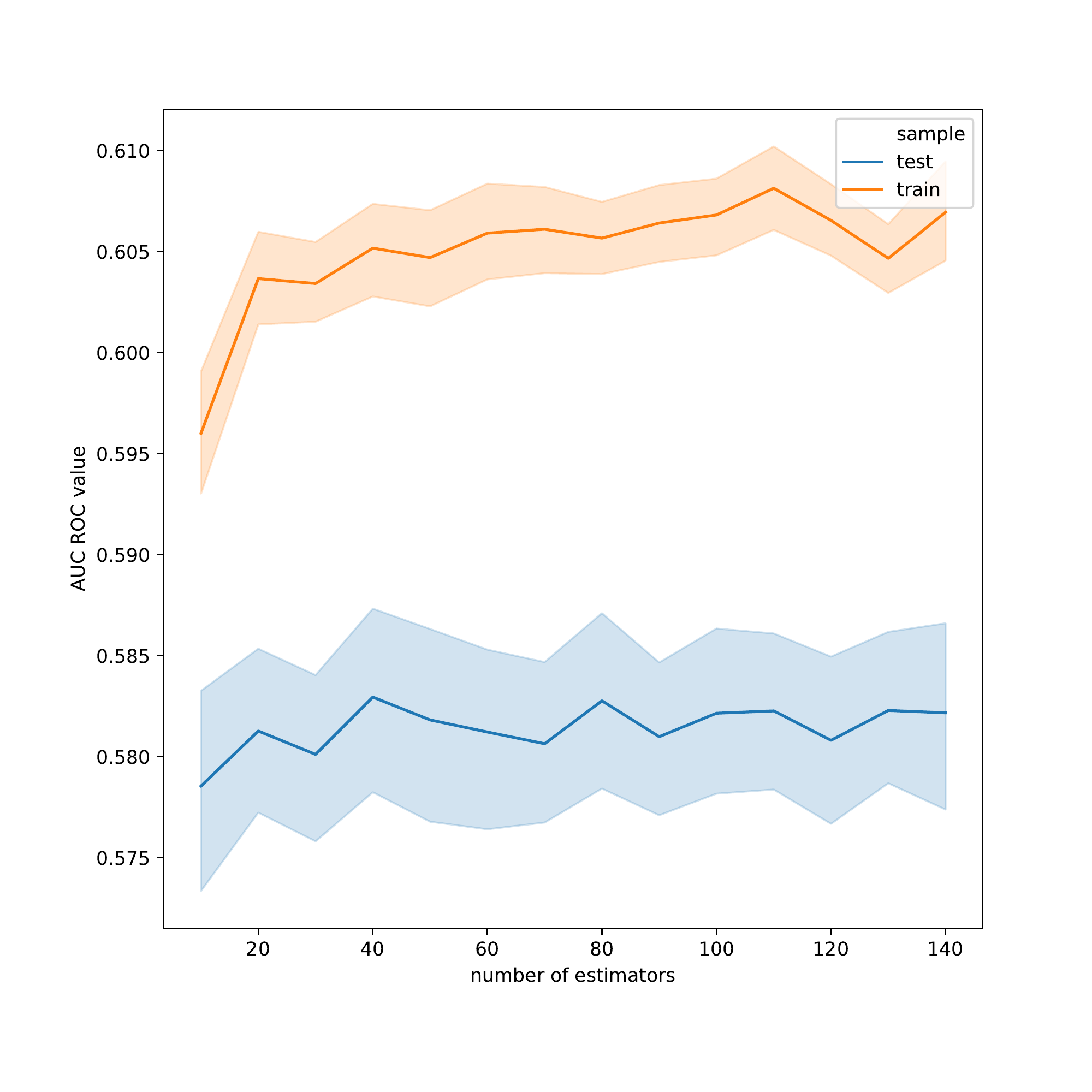}}
        
    \mbox{
        \includegraphics[width=0.33\linewidth,clip=]{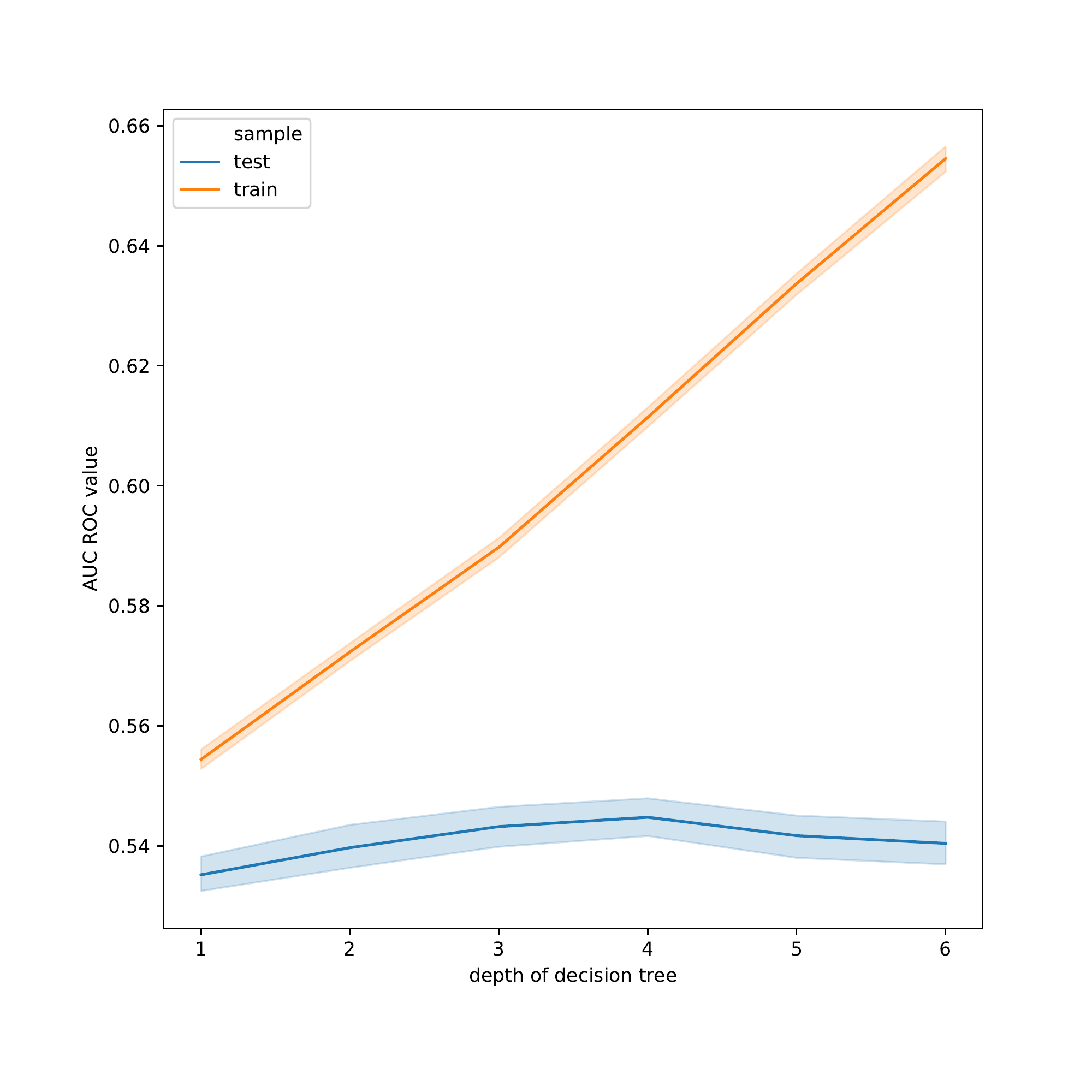}
        \includegraphics[width=0.33\linewidth,clip=]{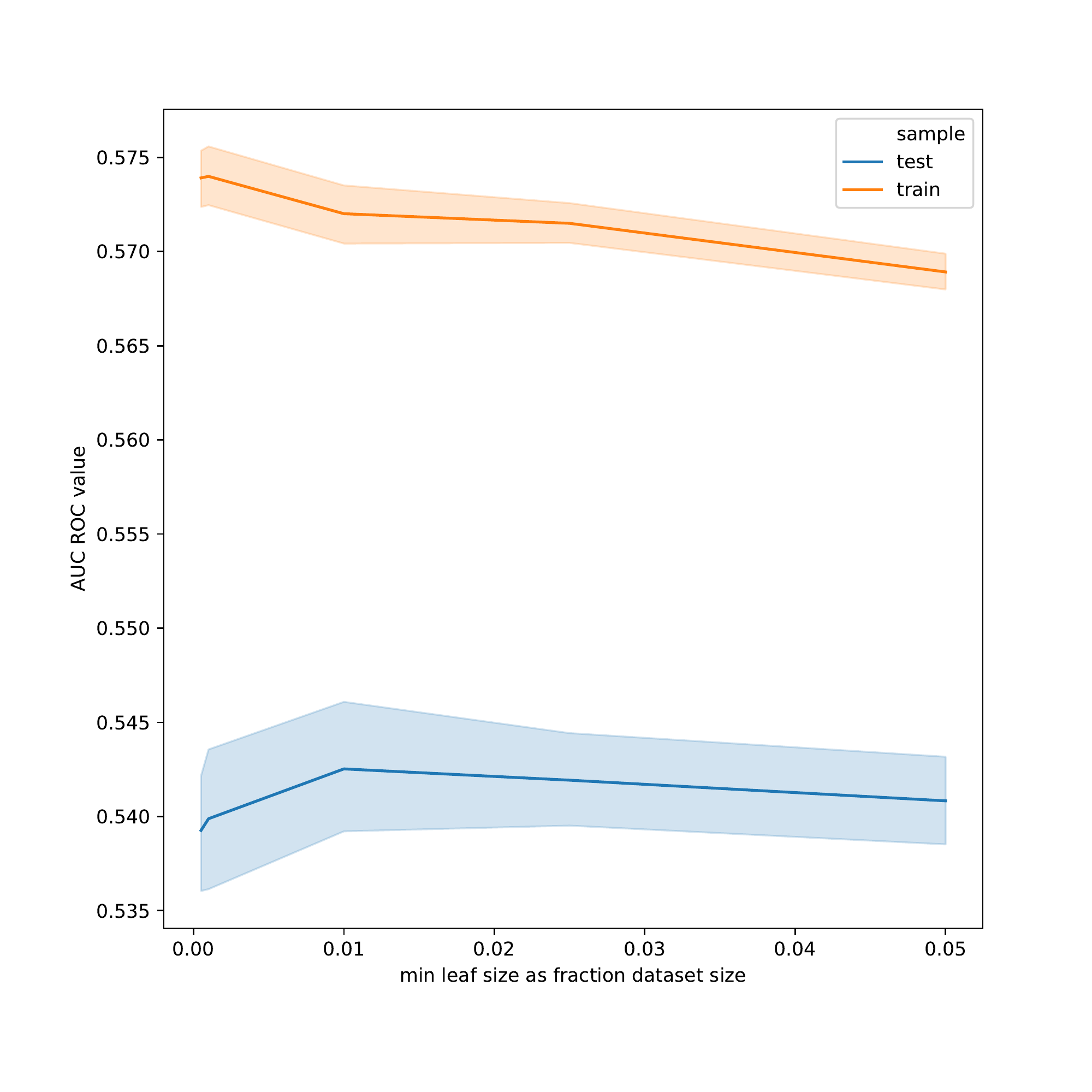}
        \includegraphics[width=0.33\linewidth,clip=]{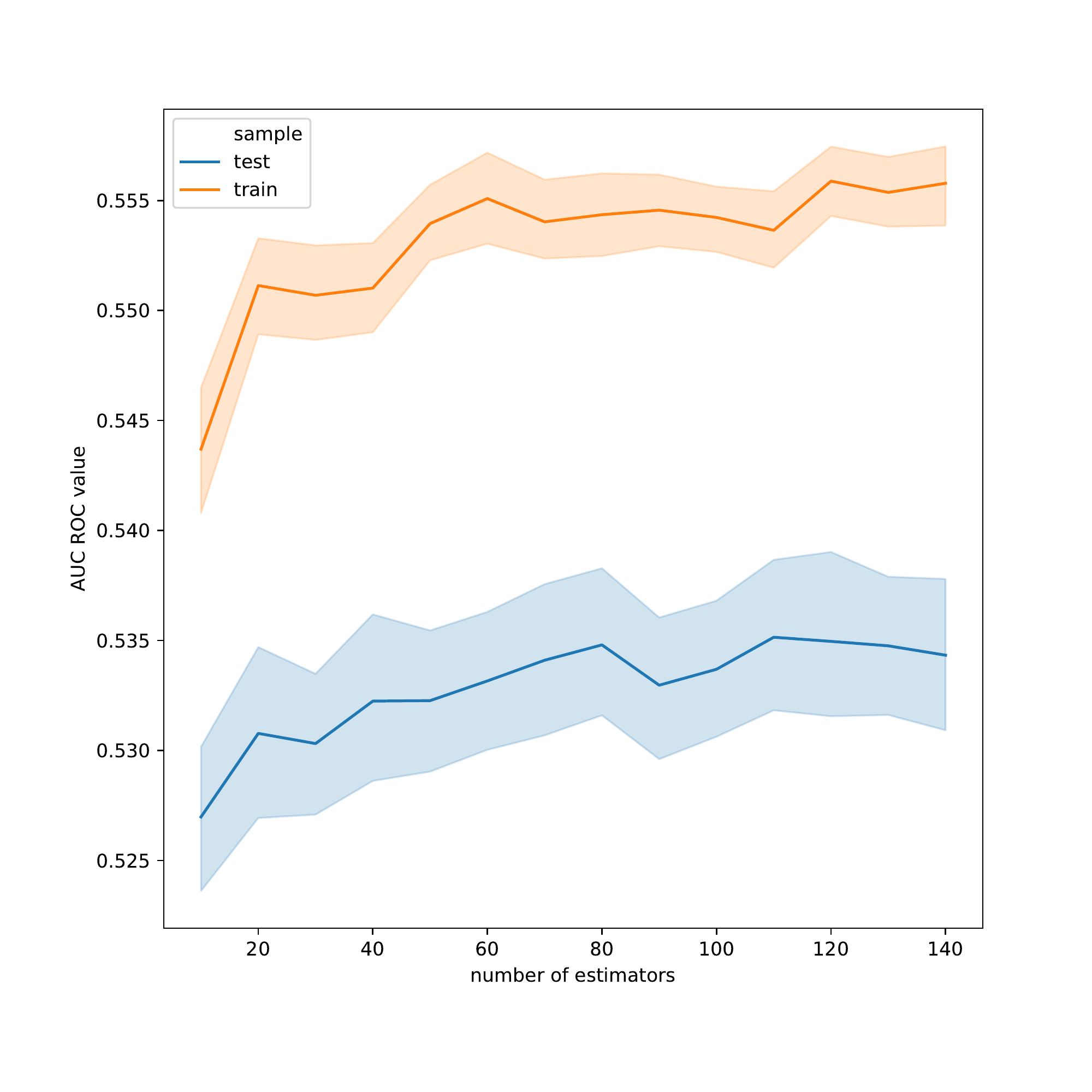}}
\caption{Neutral interactions model selection. GeneWays (top row), Literome (bottom row).
    Left: the distribution of ROC AUC as a function of depth of random forest.
    Center: the distribution of ROC AUC as a function of minimum number of samples in a decision tree leaf.
    Right: the distribution of ROC AUC as a function of the number of trees in a random forest.
    }
\label{fig:complexity_neutral_auc}
\end{figure}

\clearpage
\begin{figure}
\centering
    \mbox{
        \includegraphics[width=0.33\linewidth,clip=]{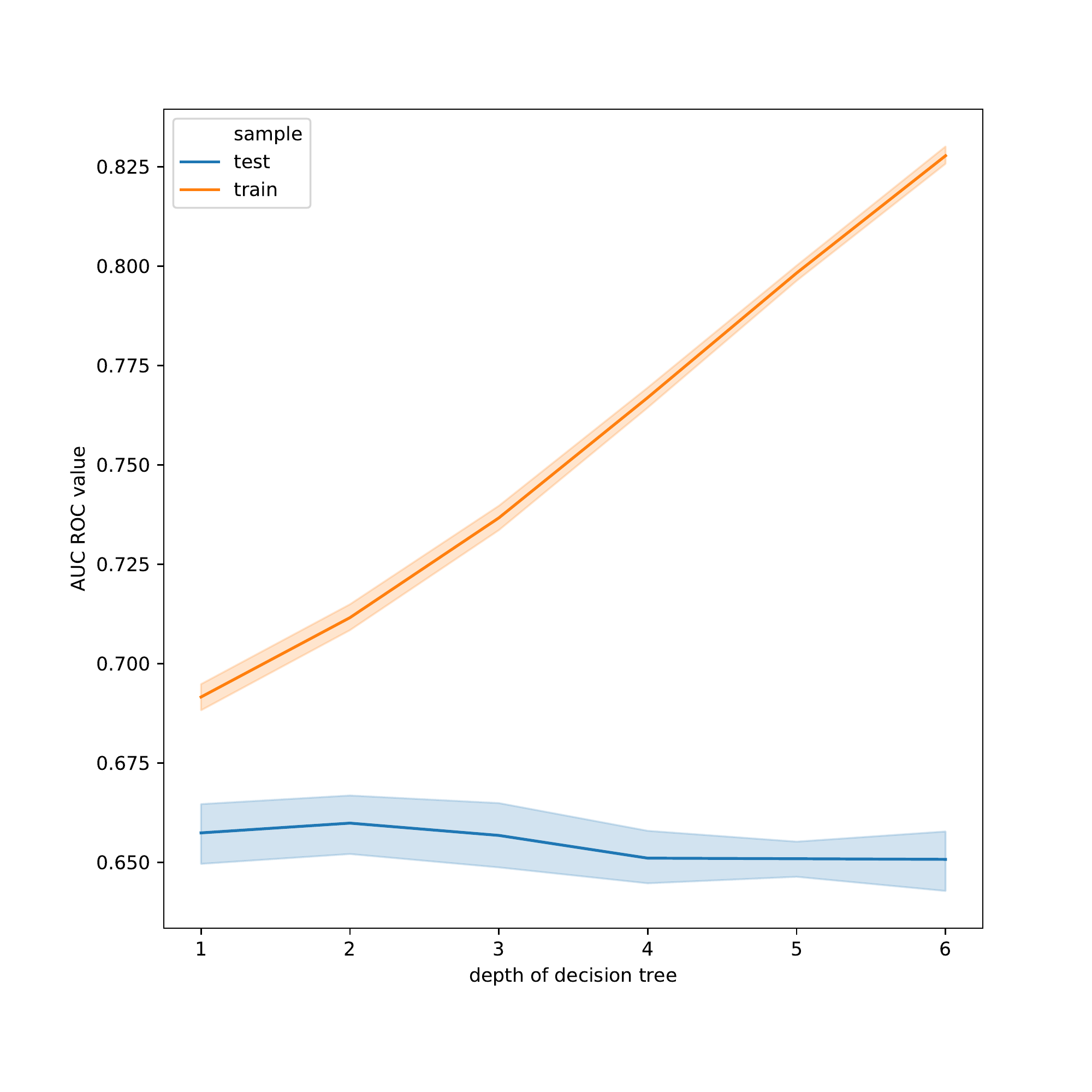}
        \includegraphics[width=0.33\linewidth,clip=]{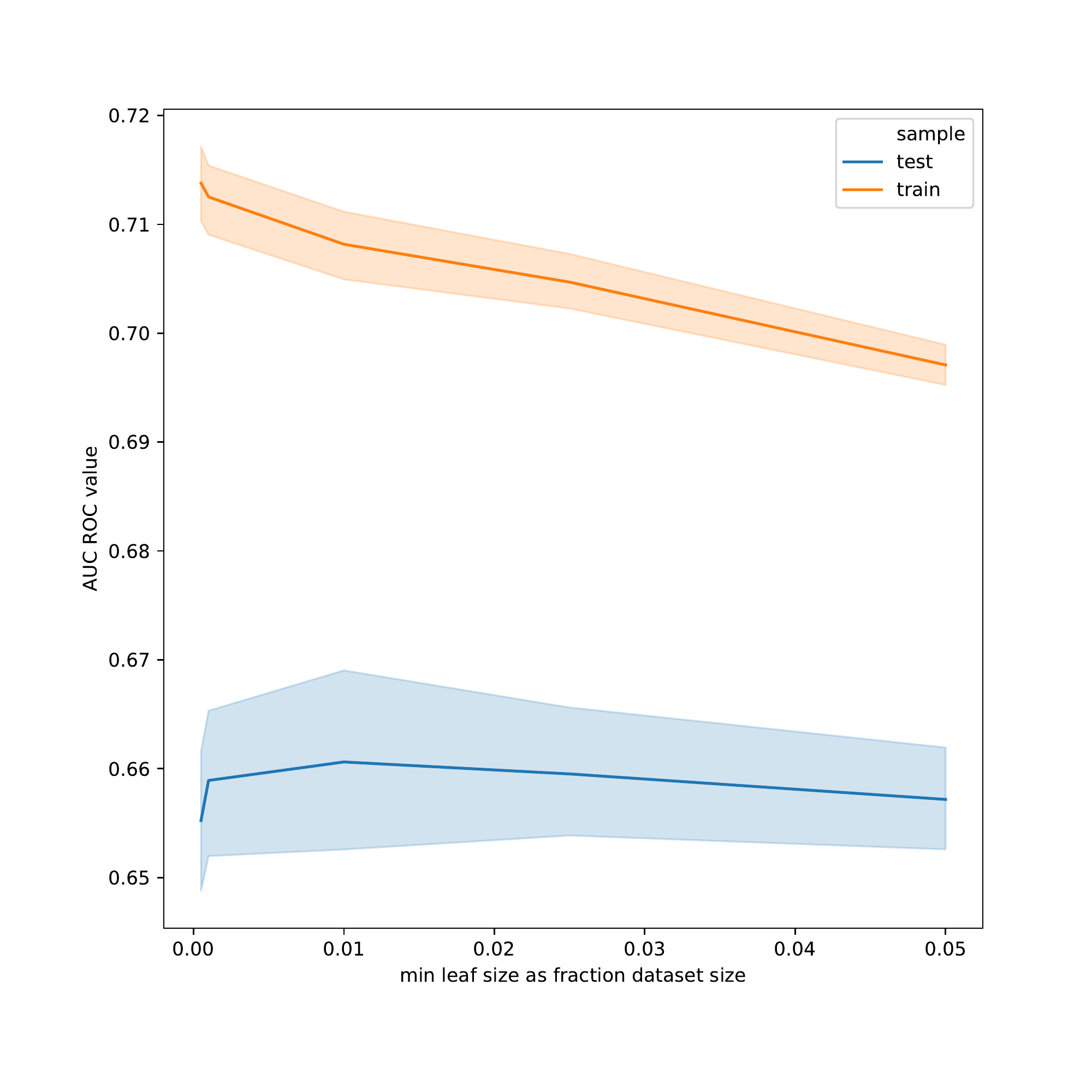}
        \includegraphics[width=0.33\linewidth,clip=]{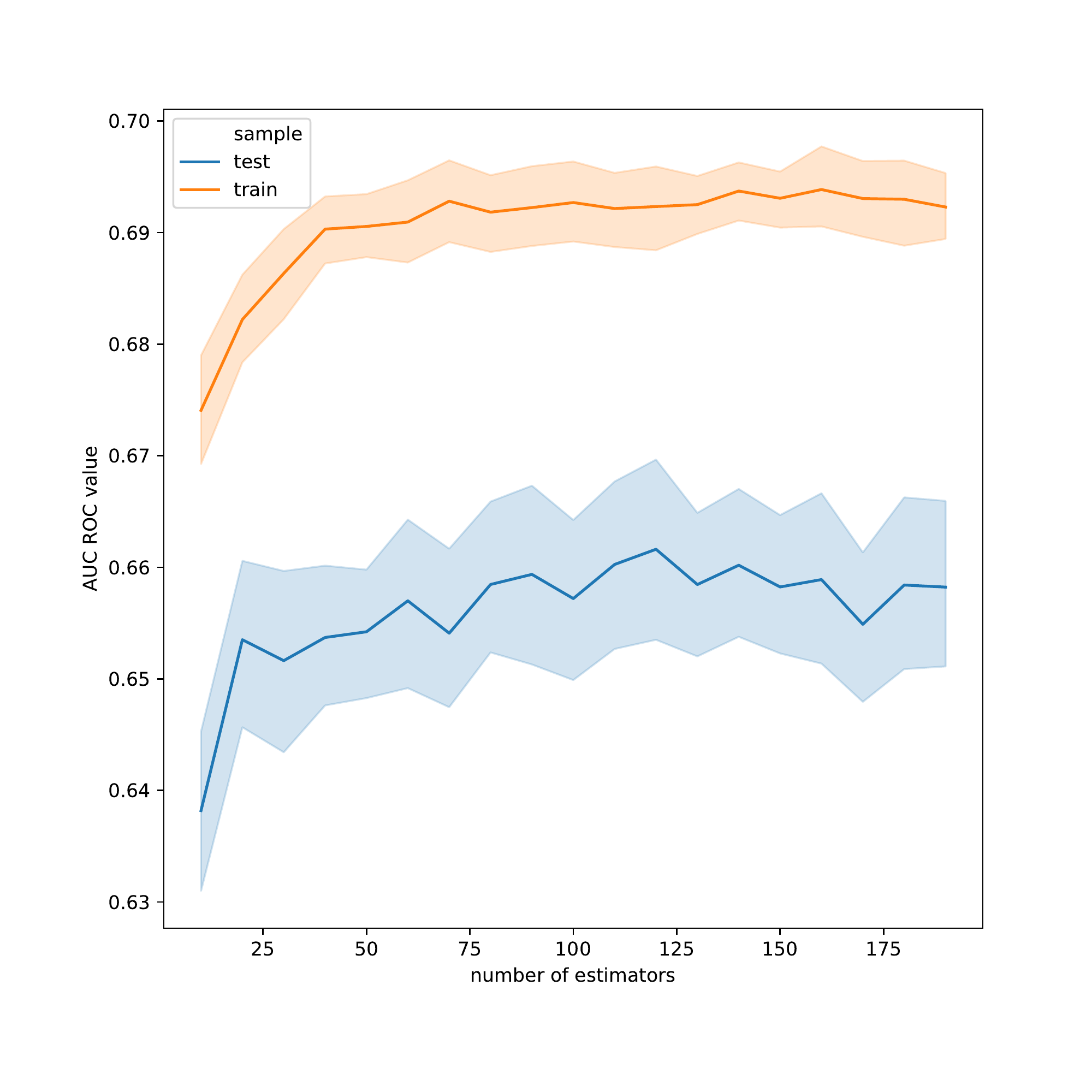}}
        
    \mbox{
        \includegraphics[width=0.33\linewidth,clip=]{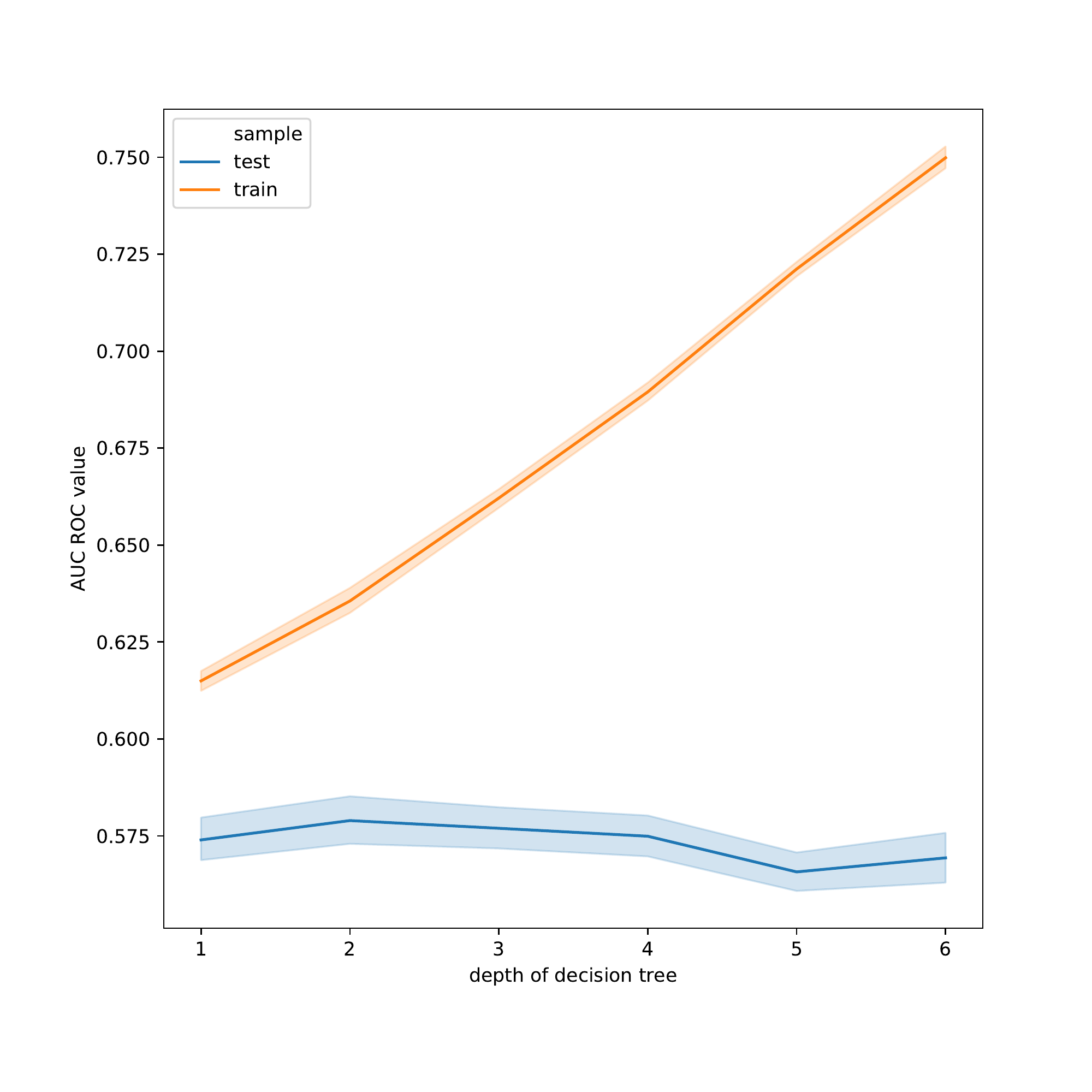}
        \includegraphics[width=0.33\linewidth,clip=]{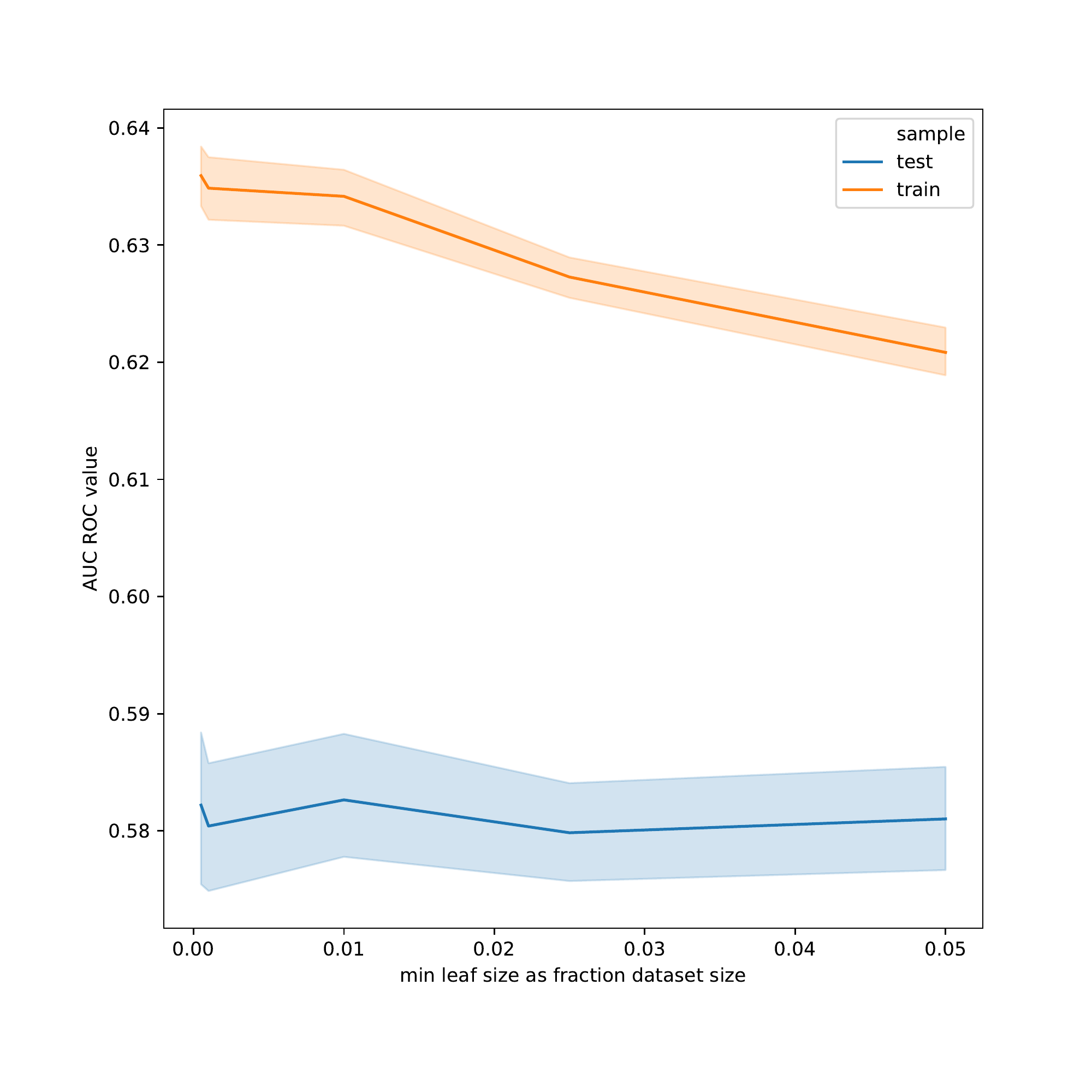}
        \includegraphics[width=0.33\linewidth,clip=]{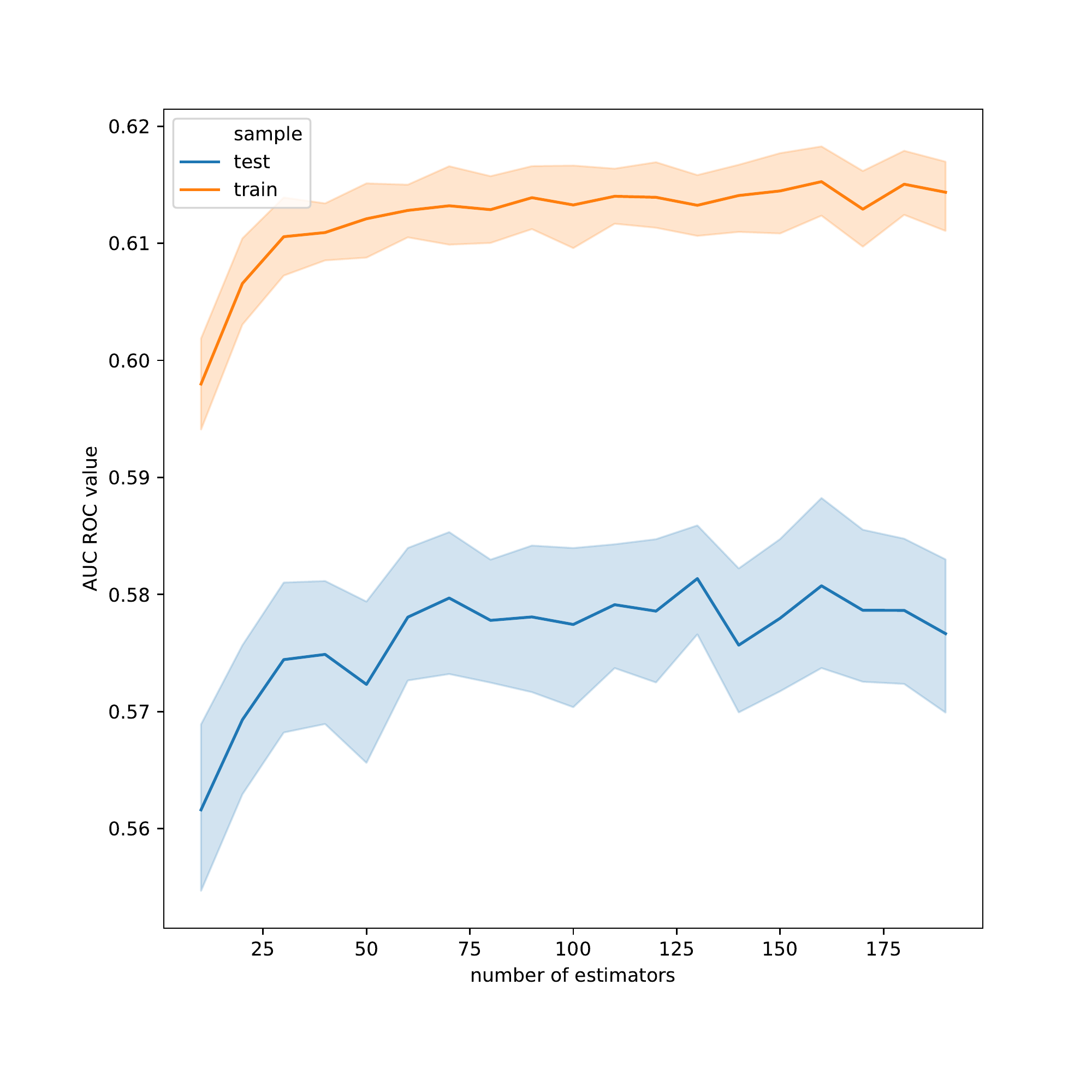}}
\caption{Positive interactions model selection. GeneWays (top row), Literome (bottom row).
    Left: the distribution of ROC AUC as a function of depth of random forest.
    Center: the distribution of ROC AUC as a function of minimum number of samples in a decision tree leaf.
    Right: the distribution of ROC AUC as a function of the number of trees in a random forest.
    }
\label{fig:complexity_posneg_auc}
\end{figure}

\clearpage
\begin{figure}
\centering
    \mbox{
        \includegraphics[width=0.33\linewidth,clip=]{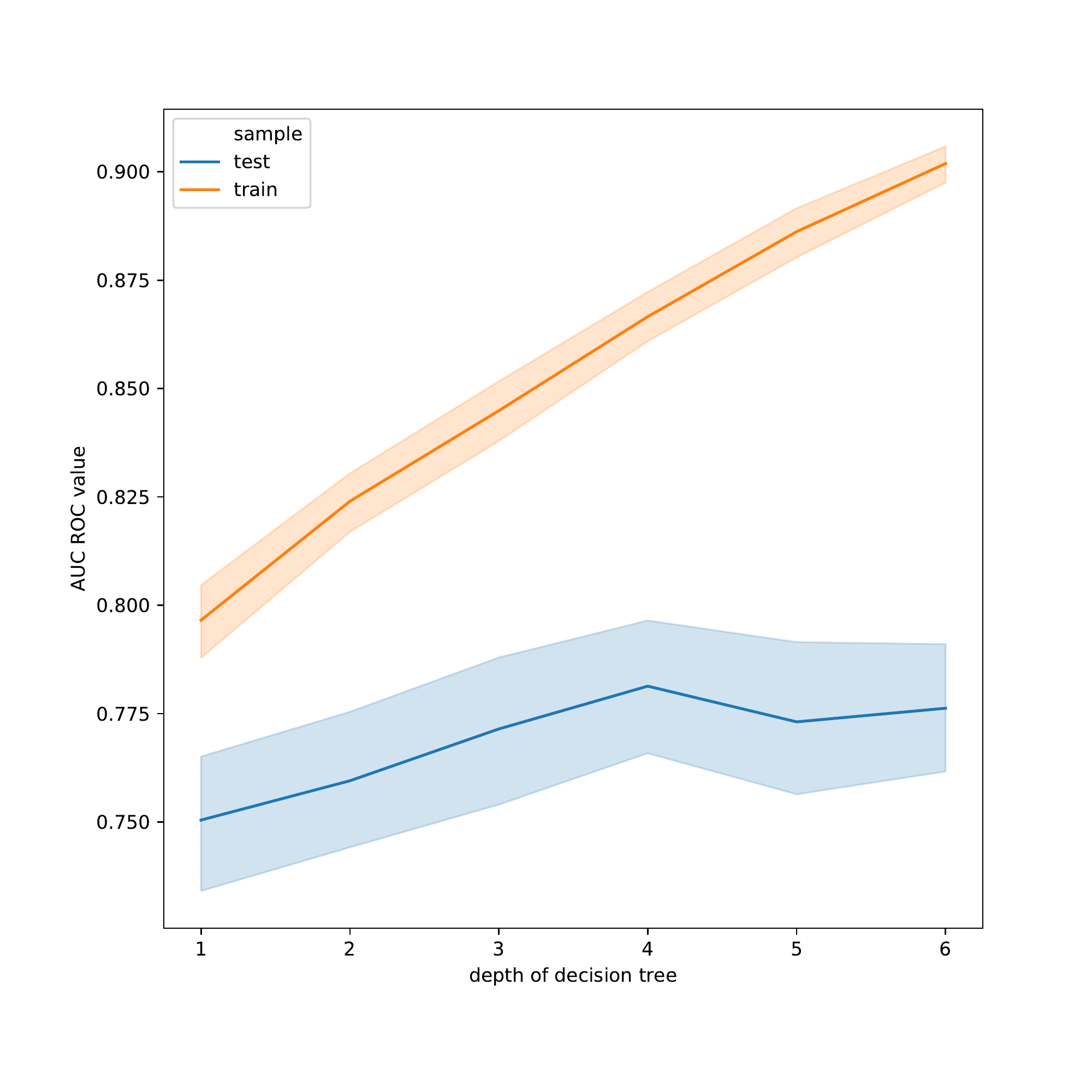}
        \includegraphics[width=0.33\linewidth,clip=]{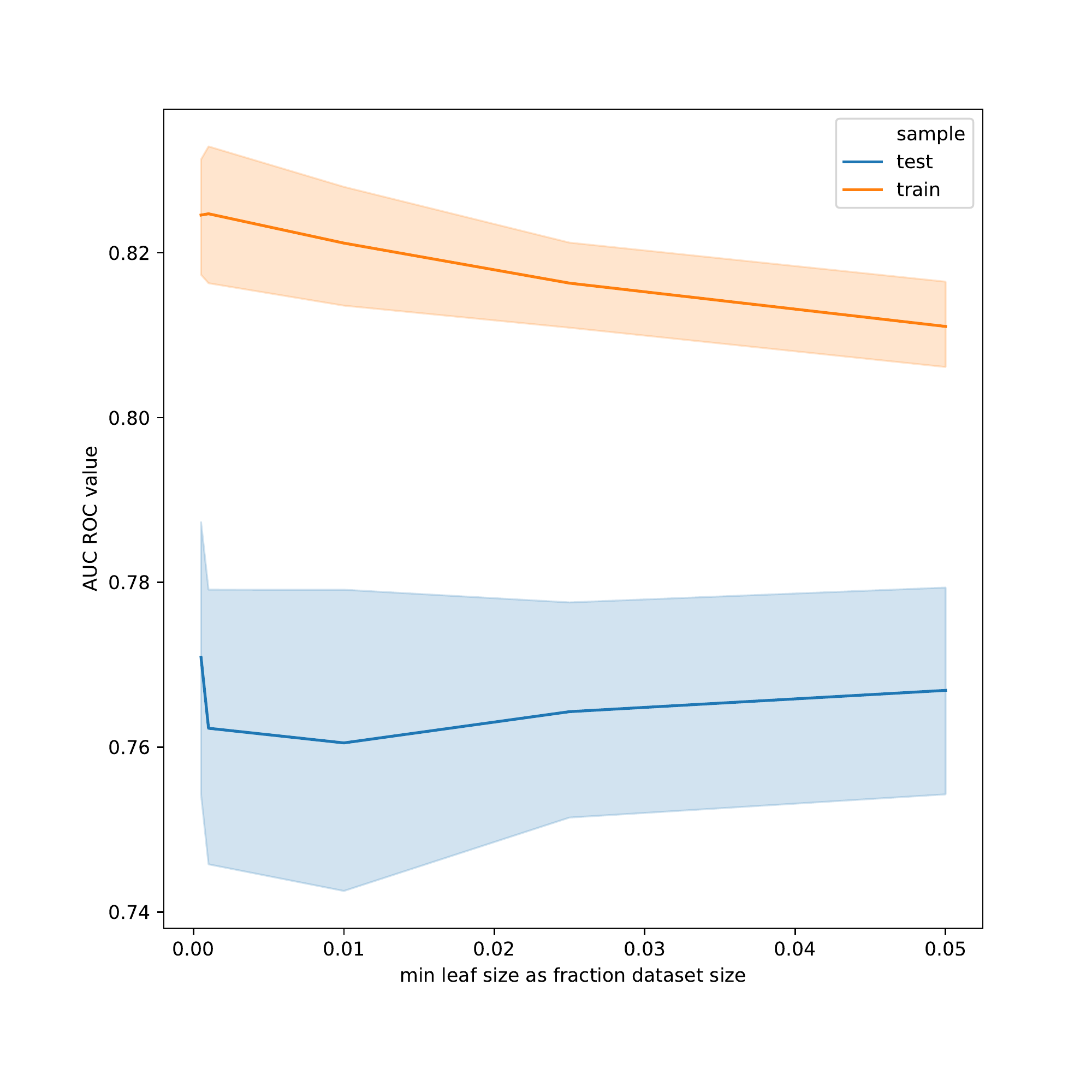}
        \includegraphics[width=0.33\linewidth,clip=]{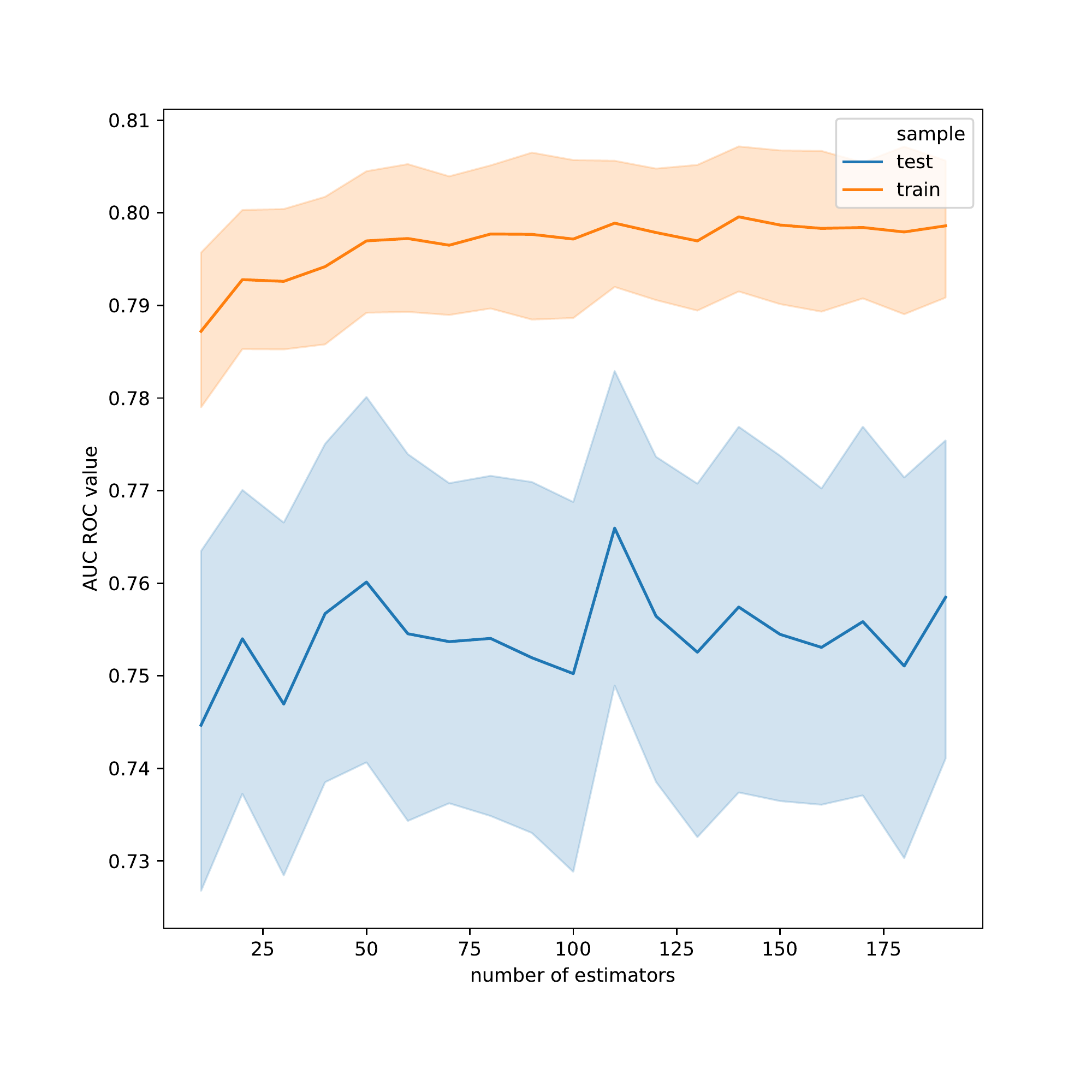}}
    \mbox{
        \includegraphics[width=0.33\linewidth,clip=]{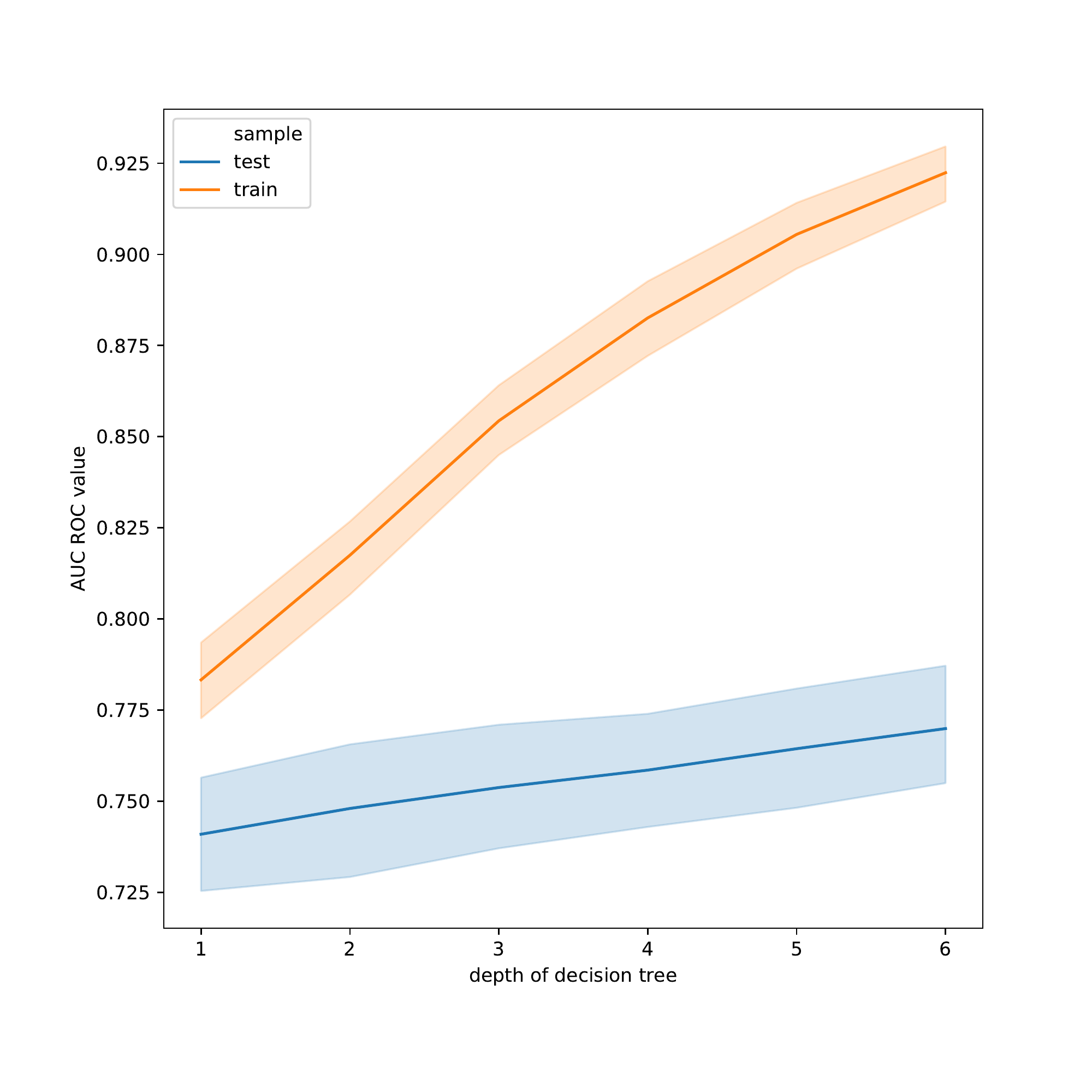}
        \includegraphics[width=0.33\linewidth,clip=]{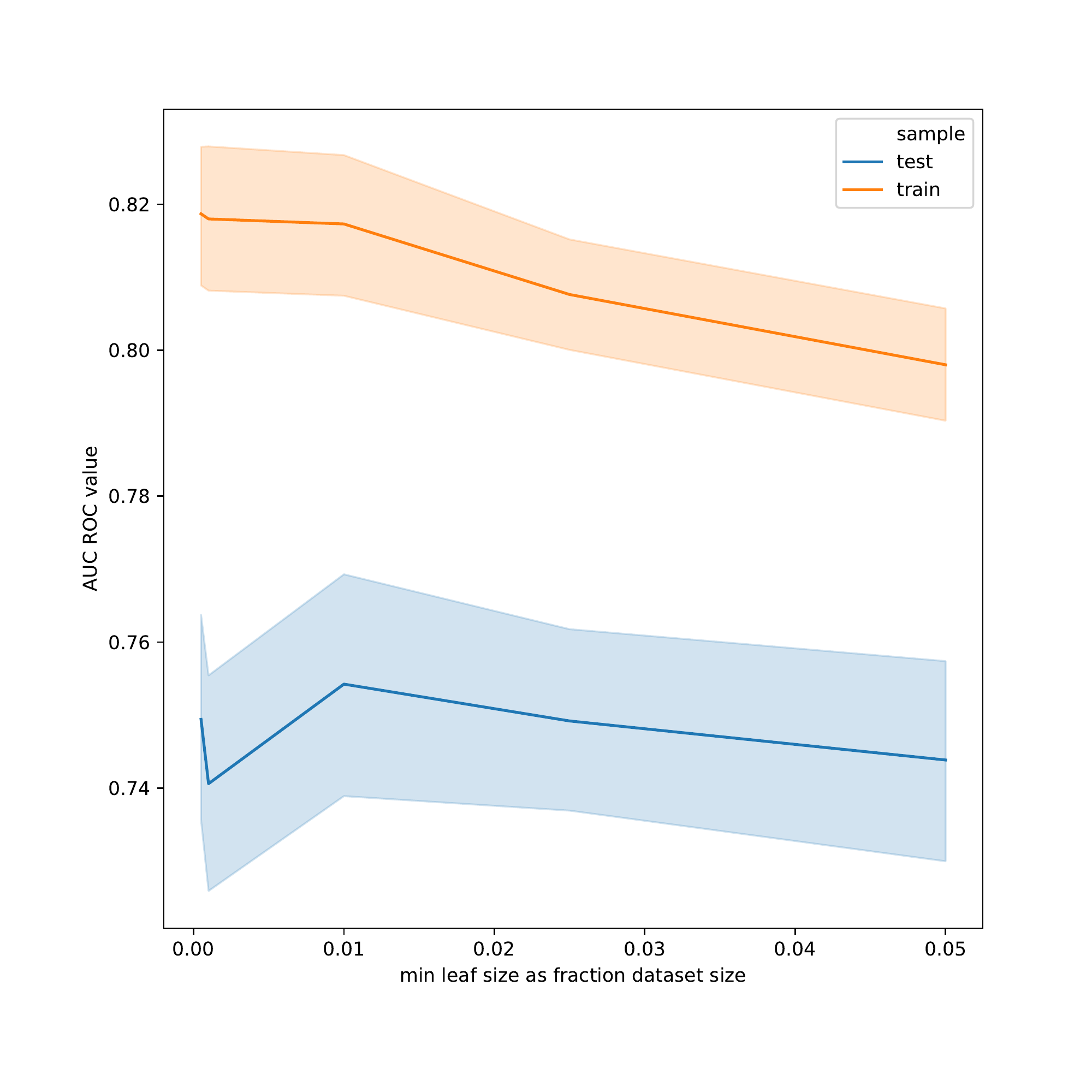}
        \includegraphics[width=0.33\linewidth,clip=]{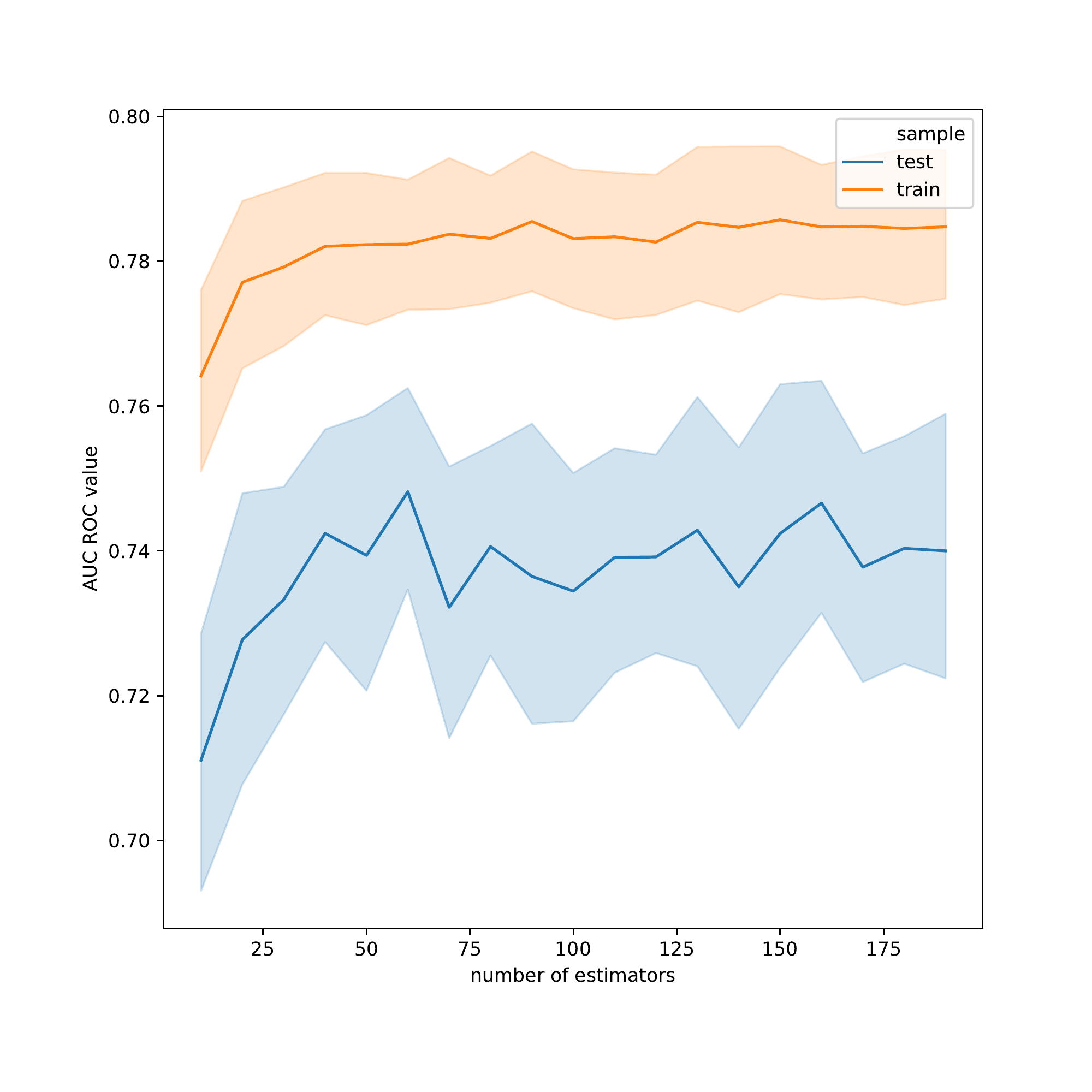}}
\caption{Claims model selection. GeneWays (top row), Literome (bottom row).
    Left: the distribution of ROC AUC as a function of depth of random forest.
    Center: the distribution of ROC AUC as a function of minimum number of samples in a decision tree leaf.
    Right: the distribution of ROC AUC as a function of the number of trees in a random forest.
    }
\label{fig:complexity_claims_auc}
\end{figure}

\clearpage
\begin{figure}
\mbox{
\includegraphics[width=0.5\textwidth]{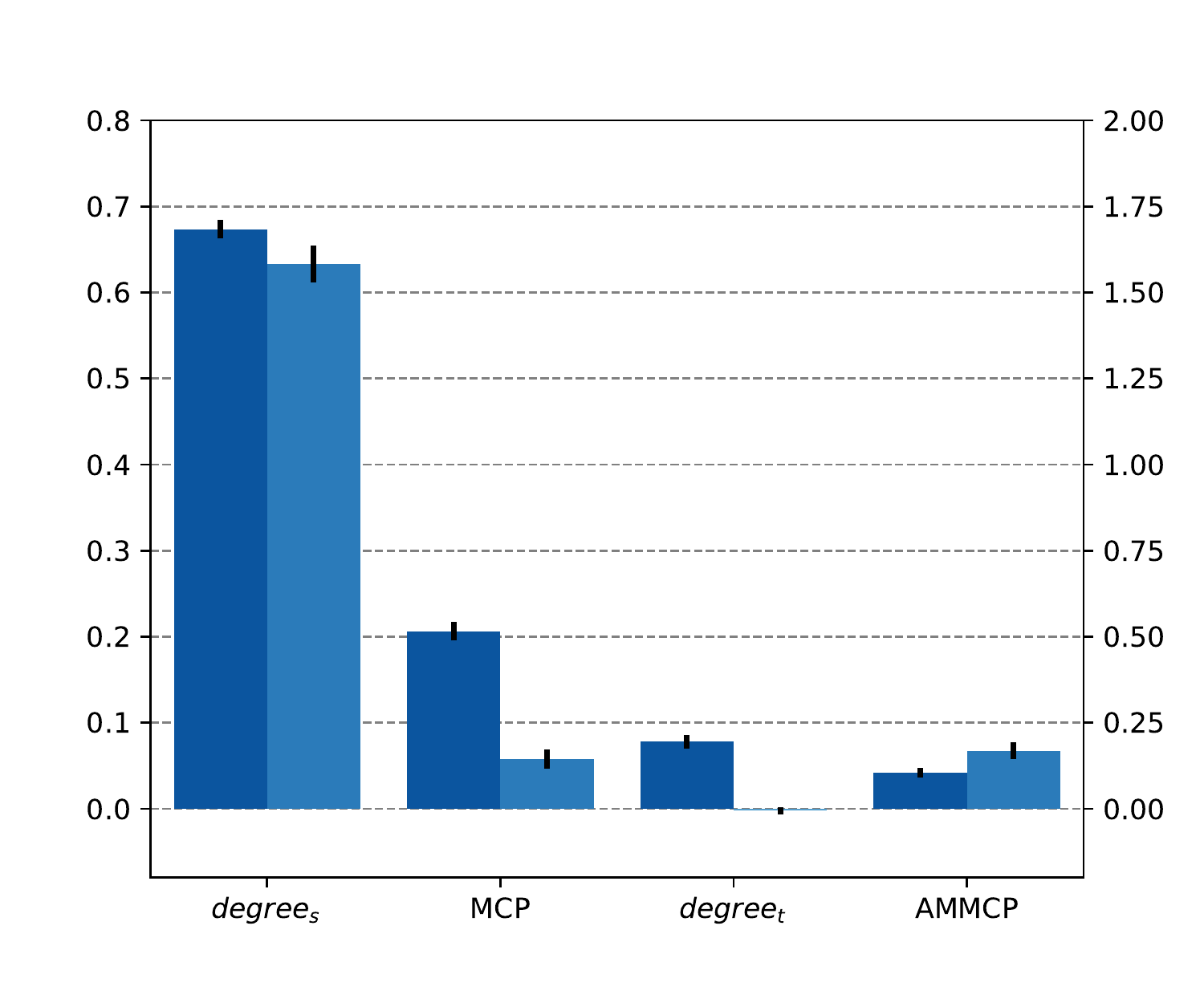}\\
\includegraphics[width=0.5\textwidth]{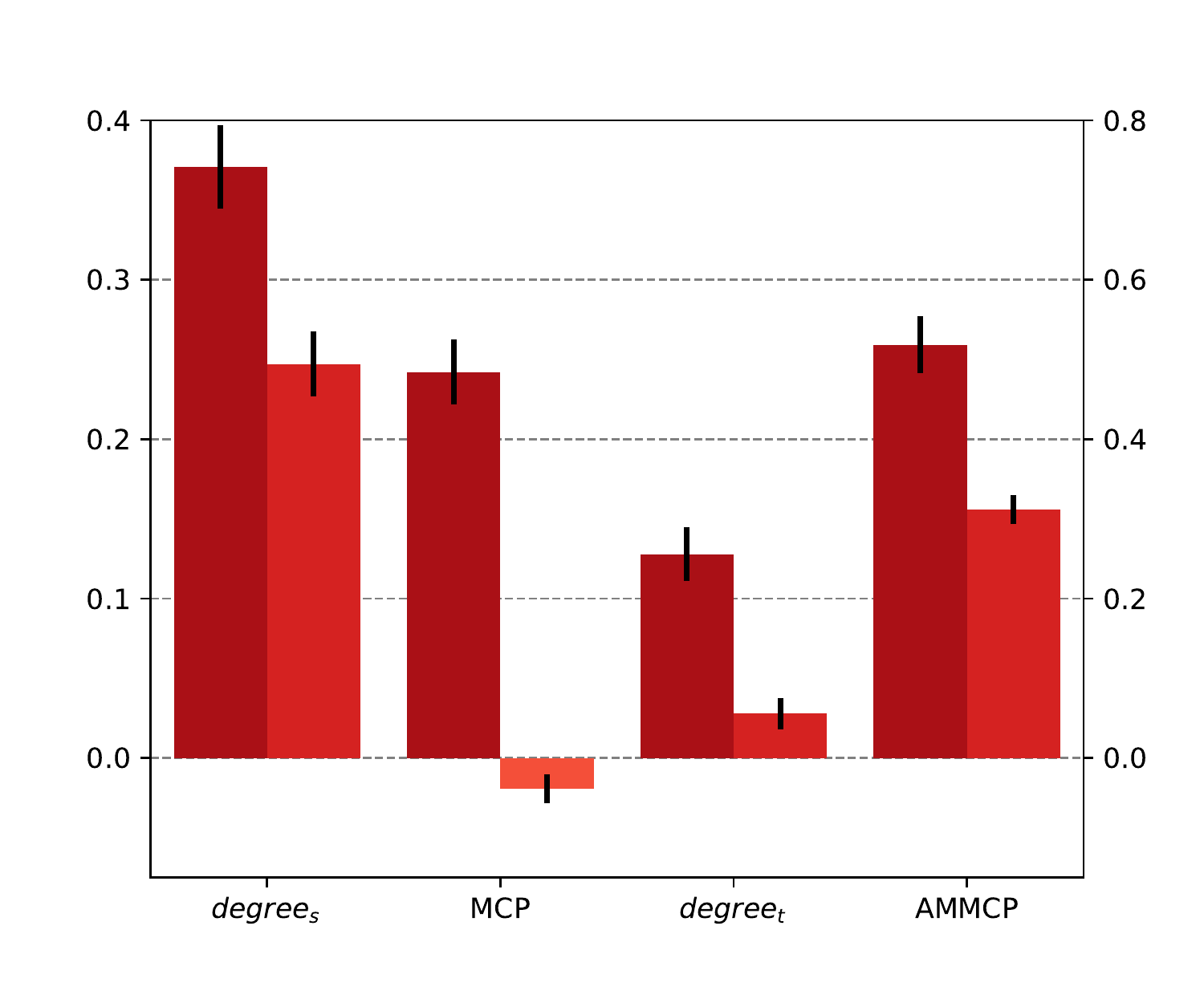}
}
\caption{Family importances of random forest model (left, darker shade) and logistic regression coefficients (right, lighter shade) for the model of classification of neutral interactions for GeneWays and Literome. Vertical centered lines show 95\% confidence level on the mean of the corresponding importance/coefficient.
}
\label{fig:imps_neut}
\end{figure}

\clearpage
\begin{figure}
\mbox{
\includegraphics[width=0.5\textwidth]{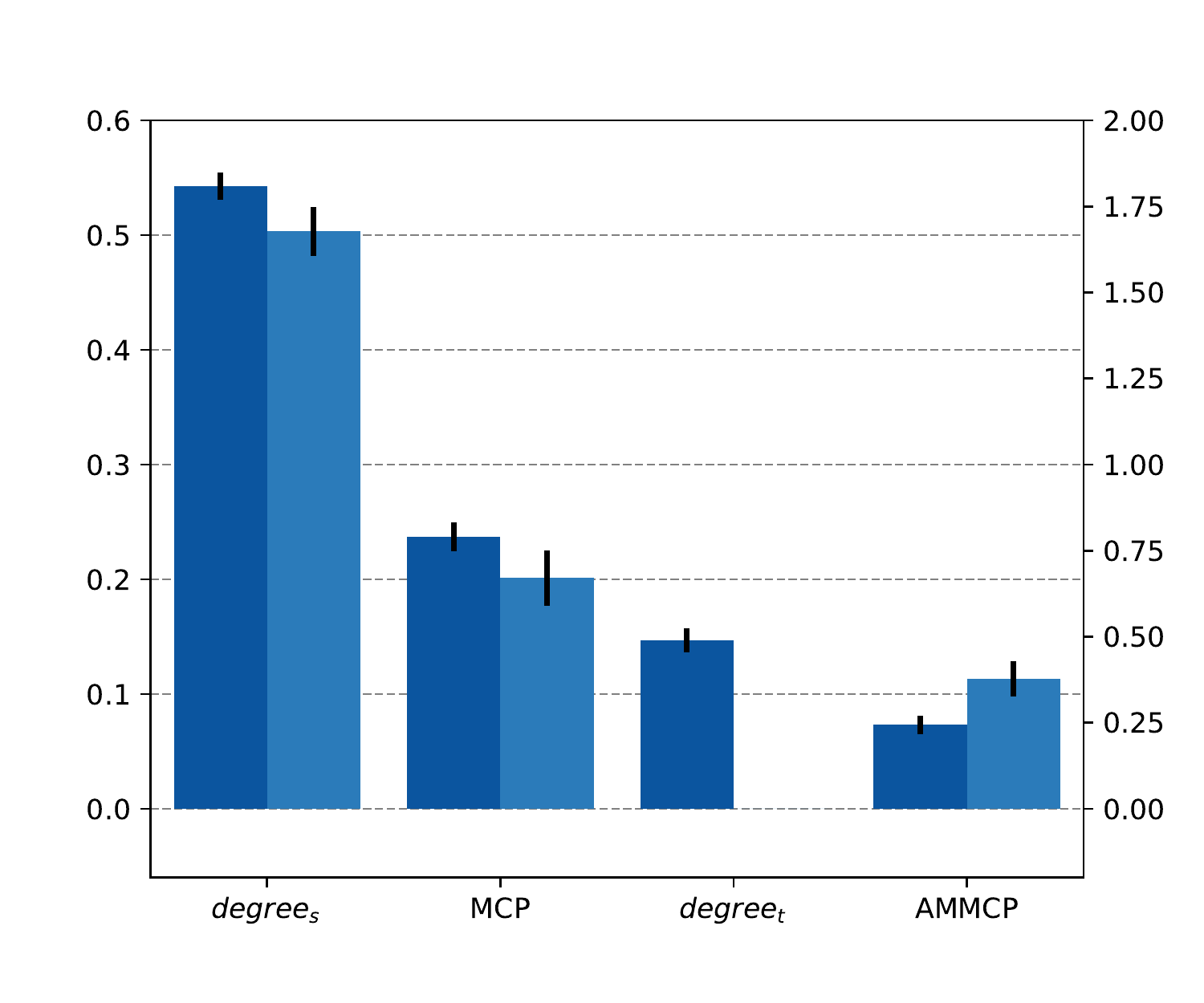}\\
\includegraphics[width=0.5\textwidth]{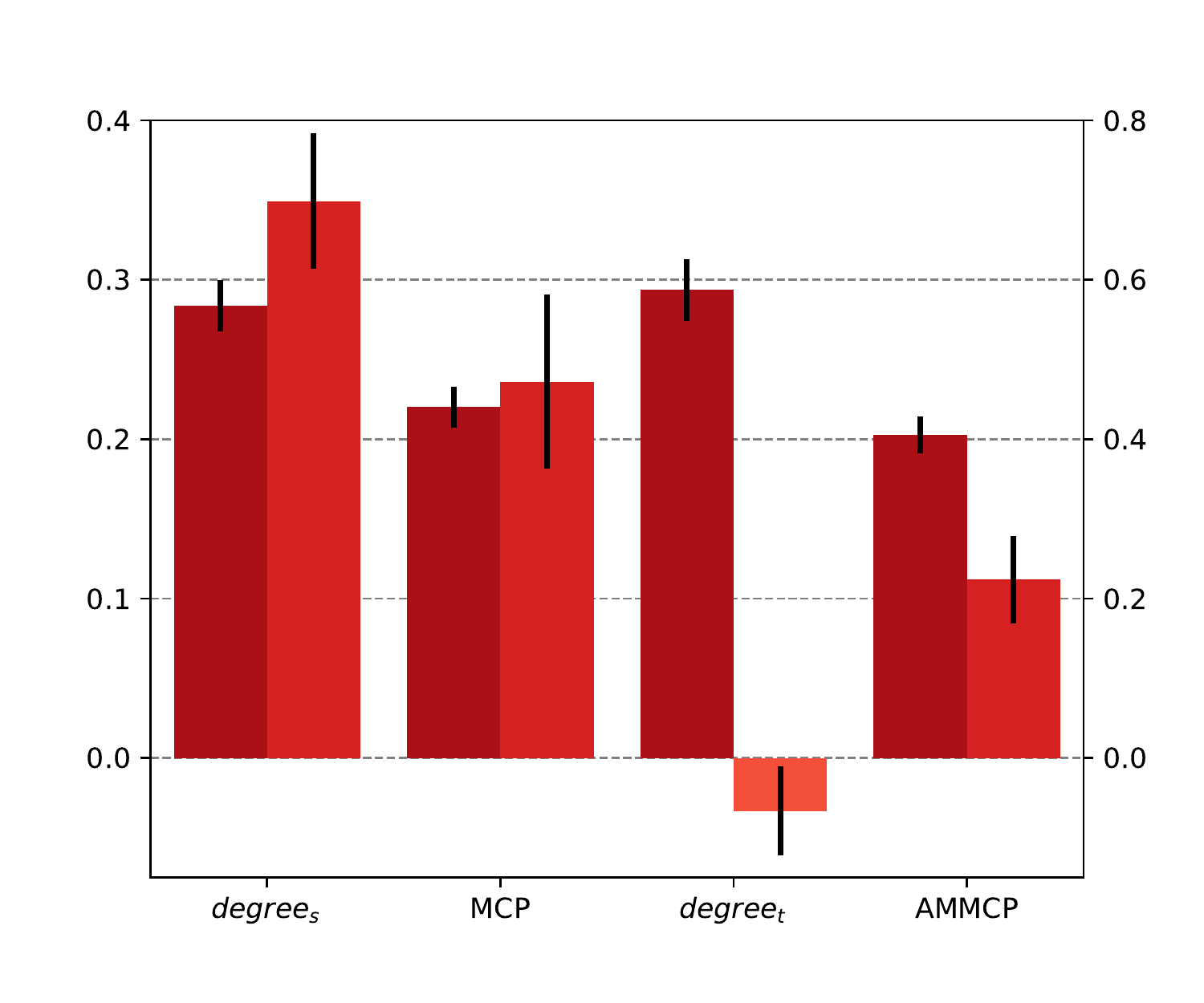}}
\caption{Family importances of random forest model (left, darker shade) and logistic regression coefficients (right, lighter shade) for the model of classification of positive interactions for GeneWays and Literome.
Vertical centered lines show 95\% confidence level on the mean of the corresponding importance/coefficient.
}
\label{fig:imps_posneg}
\end{figure}

\clearpage
\begin{figure}
\mbox{
\includegraphics[width=0.5\textwidth]{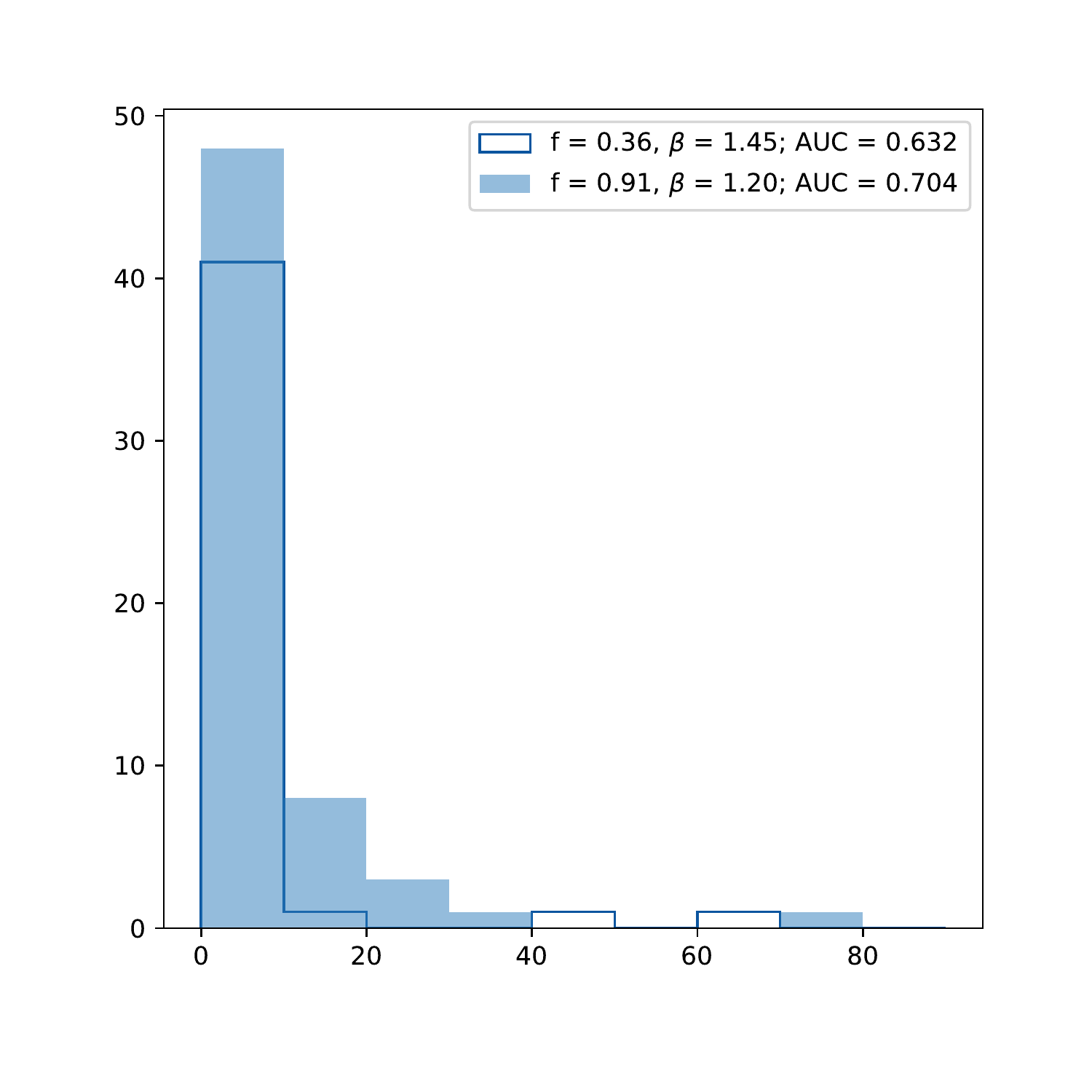}\\
\includegraphics[width=0.5\textwidth]{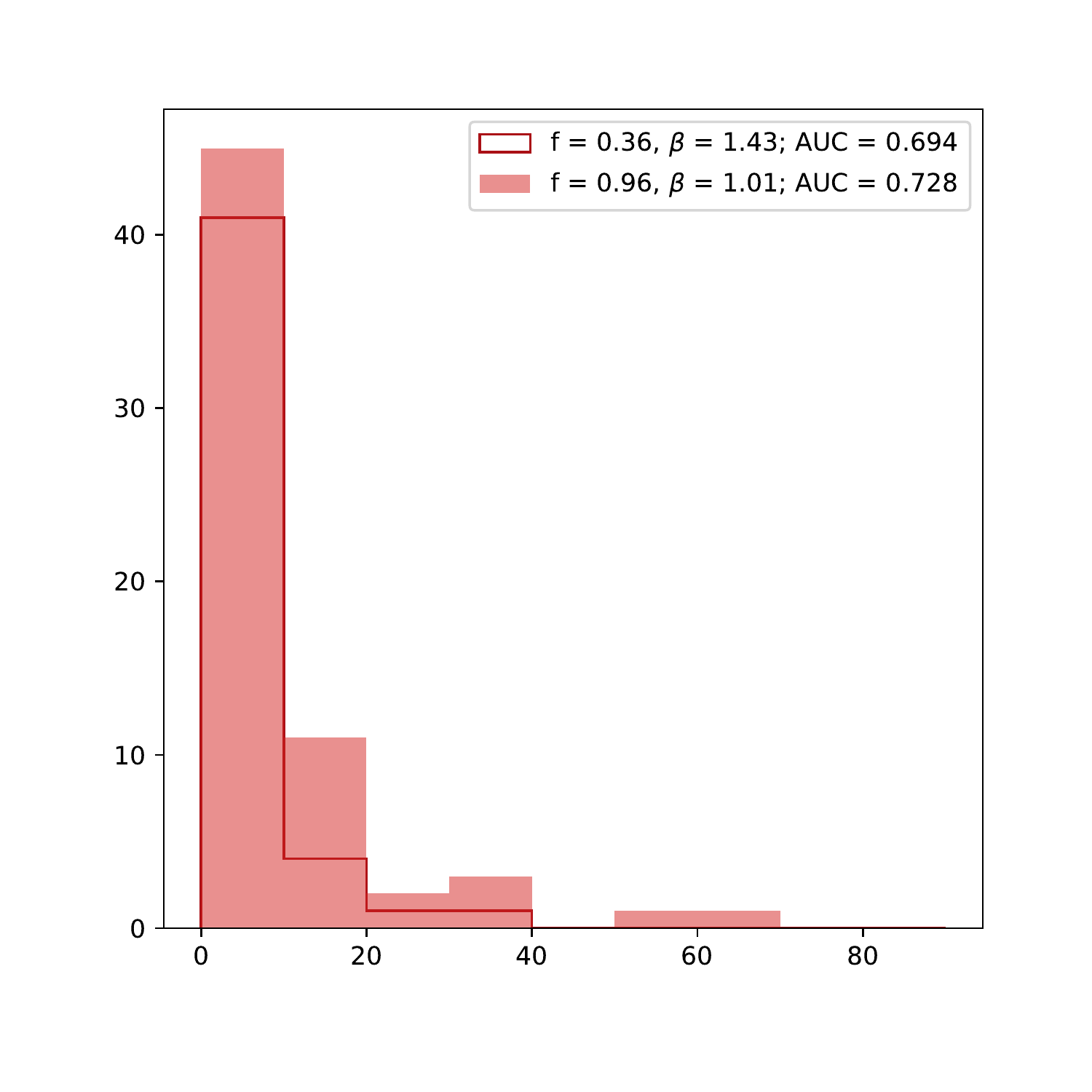}}

\caption{Typical examples of distributions of claim number $\rho(n_\alpha)$ per interaction for test subsamples. Left: GeneWays, right: Literome. 
}
\label{fig:len_distr}
\end{figure}

\clearpage
\begin{figure}
\mbox{
\includegraphics[width=0.5\textwidth]{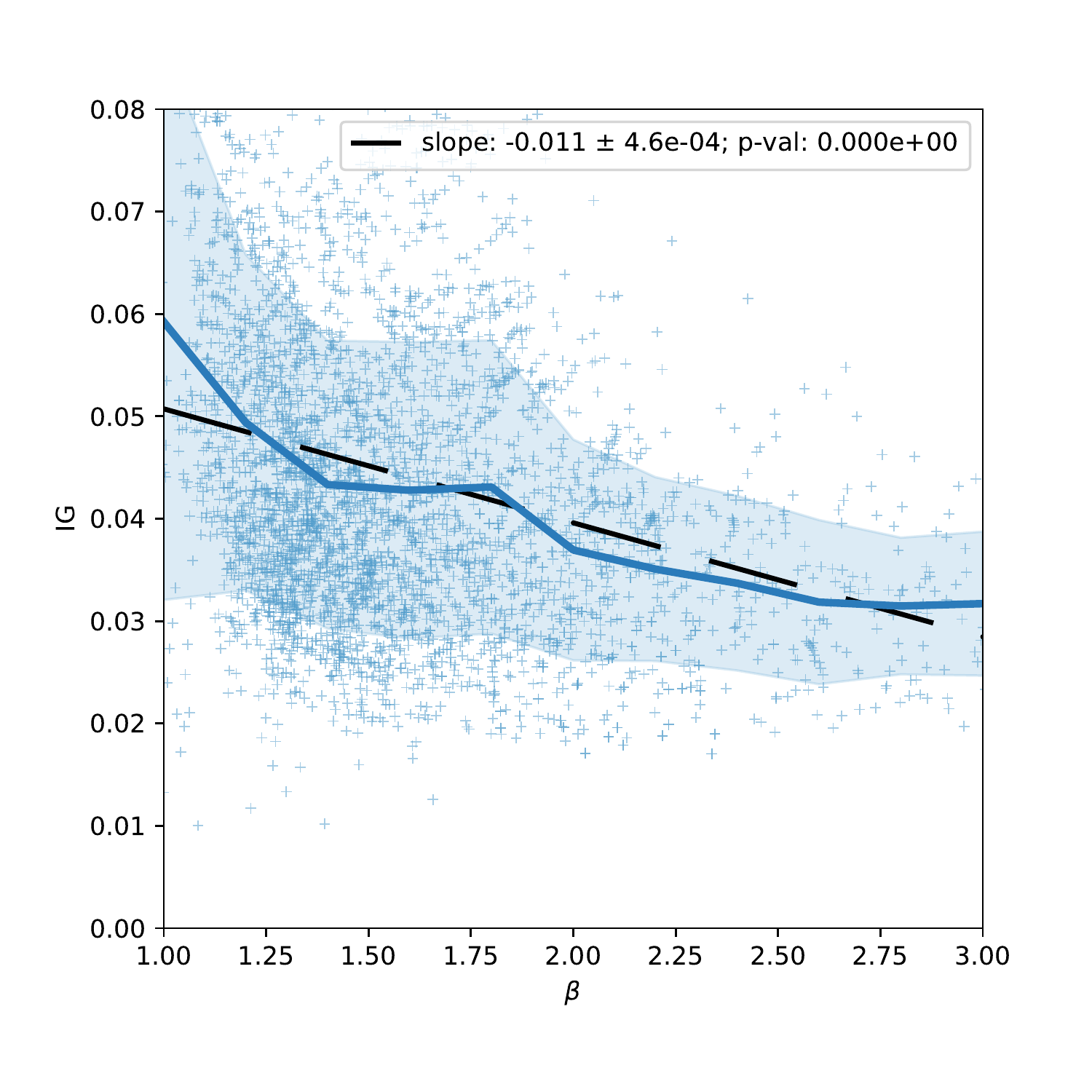}\\
\includegraphics[width=0.5\textwidth]{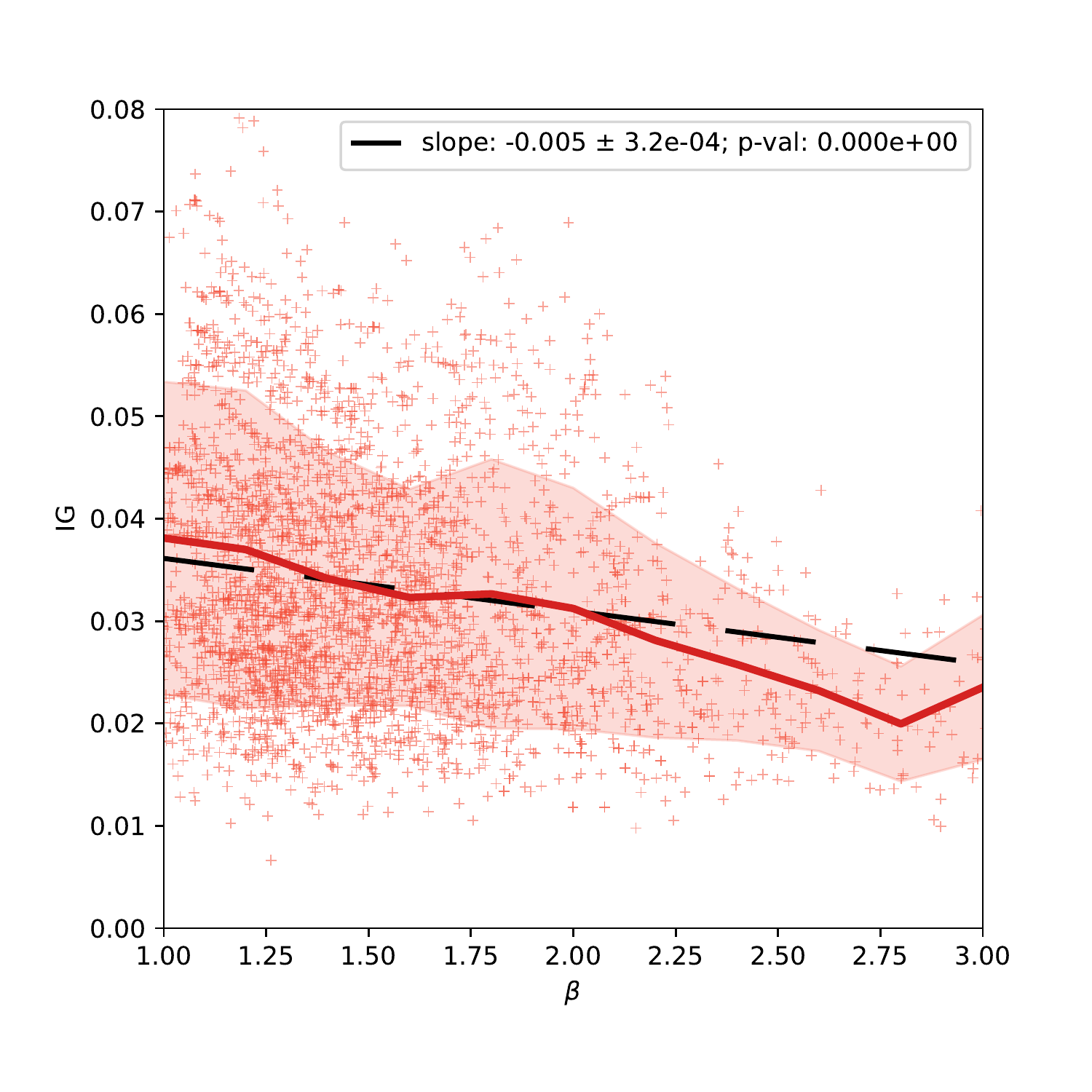}}
\caption{Information gain as a function of the slope of length distribution $\beta$ for the length policy. Solid lines correspond to binned averages and the shaded region to the binned region within one standard deviation. Left: GeneWays, Right: Literome.
}
\label{fig:len_distr_ig}
\end{figure}


\clearpage
\begin{thebibliography}{99}
\bibitem{Rzhetsky2004}
A.Rzhetsky et al, GeneWays: A System for Extracting, Analyzing, Visualizing, and Integrating Molecular Pathway Data, J. of Biomedical Informatics, 37, 2004.
\bibitem{Rodriguez2006}
R. Rodriguez-Esteban, I. Iossifov and A. Rzhetsky, Imitating Manual Curation of Text-Mined Facts in Biomedicine, PLOS Computational Biology, 2, 2006.

\bibitem{Poon2014}
H.Poon, C.Quirk, C.DeZiel and D.Heckerman, Literome: PubMed-scale genomic knowledge base in the cloud, Bioinformatics, 30, 2014.

\bibitem{lincs2017}
A. Subramanian et al, A Next Generation Connectivity Map: L1000 Platform and the First 1,000,000 Profiles, 
Cell, 171, 2017.
\bibitem{Belikov2020}
A. Belikov, J.Evans, Hierarchical string distance applied to disambiguation, in preparation.
\bibitem{Babuji2016}
Y. Babuji, K. Chard, A. Gerow, E.Duede,
Cloud Kotta: Enabling Secure and Scalable Data Analytics in the Cloud, IEEE International Conference on Big Data, 302, 2016.

\bibitem{Rosvall2009}
M. Rosvall, D. Axelsson, C. Bergstrom, The map equation, European Physical Journal Special Topics, 13, 2009.
\bibitem{igraph}
G. Csardi, T. Nepusz, The igraph software package for complex network research, InterJournal, 1695, 2006.
\bibitem{West2010-vn}
J.West, T.Bergstrom, C.Bergstrom, The Eigenfactor Metrics: A Network Approach to Assessing Scholarly Journals, Coll. Res. Libr., 236, 2010.
\end{thebibliography}
\end{document}